\documentclass[a4paper,11pt]{article}
\pdfoutput=1 

\usepackage{jinstpub} 
\usepackage[]{subfigure}
\usepackage[per-mode=fraction,separate-uncertainty,table-number-alignment = right]{siunitx}
\DeclareSIUnit[number-unit-product = \,, per-mode=fraction]\mip{MIP}
\DeclareSIUnit[number-unit-product = \,]\pixel{px}
\DeclareSIUnit[number-unit-product = \,]\Xnull {\ensuremath{\text{X}_0}}
\DeclareSIUnit[number-unit-product = \,]\RMoliere {\ensuremath{\text{r}_\text{M}}}
\DeclareSIUnit[number-unit-product = \,]\LambdaN {\ensuremath{\lambda_\text{n}}}
\DeclareSIUnit[number-unit-product = \,]\adc {ADC}
\DeclareSIUnit[number-unit-product = \,]\bit {bit}
\usepackage{multirow}
\usepackage{booktabs}
\usepackage{todonotes}
\usepackage{lineno}
\usepackage{textcomp}
\usepackage{nicefrac}
\usepackage{xspace}
\usepackage{microtype}

\newcommand{\ra}[1]{\renewcommand{\arraystretch}{#1}}

\newcommand\piminus{\(\mathrm{\pi^-}\)}
\newcommand\muminus{\(\mathrm{\mu^-}\)}
\newcommand\eminus{\(\mathrm{e^-}\)}
\newcommand\eplus{\(\mathrm{e^+}\)}
\newcommand\geant{\textsc{Geant\,4}\xspace}

\title{Hadronic Energy Resolution of a Combined High Granularity Scintillator Calorimeter System}

\emailAdd{oskar.hartbrich@desy.de}
\collaboration[]{The CALICE Collaboration}

\author[a]{J.\,Repond}
\author[a]{L.\,Xia}
\affiliation[a]{Argonne National Laboratory,\\9700 S.\ Cass Avenue, Argonne, IL 60439-4815, USA}

\author[b]{J.\,Apostolakis}
\author[b]{G.\,Folger}
\author[b]{V.\,Ivantchenko}
\author[b]{A.\,Ribon}
\author[b]{V.\,Uzhinskiy}
\affiliation[b]{CERN,\\1211 Gen\`{e}ve 23, Switzerland}

\author[c]{D.\,Boumediene}
\author[c]{V.\,Francais }
\affiliation[c]{Universit\'e Clermont Auvergne, Universit\'e Blaise Pascal, CNRS/IN2P3, LPC,\\4 Av. Blaise Pascal, TSA/CS 60026, F-63178 Aubi\`ere, France }

\author[d]{G. C.\,Blazey}
\author[d]{A.\,Dyshkant}
\author[d]{K.\,Francis}
\author[d]{V.\,Zutshi }
\affiliation[d]{NICADD, Northern Illinois University,\\Department of Physics, DeKalb, IL 60115, USA }

% ...

\abstract{This paper presents results obtained with the combined CALICE Scintillator Electromagnetic Calorimeter, Analogue Hadronic Calorimeter and Tail Catcher \& Muon Tracker, three high granularity scintillator-SiPM calorimeter prototypes. The response of the system to pions with momenta between \SI{4}{\GeV/c} and \SI{32}{\GeV/c} is analysed, including the energy response, resolution, and longitudinal shower profiles. The results of a software compensation technique based on weighting according to hit energy are compared to those of a standard linear energy reconstruction. The results are compared to predictions of the \geant physics lists QGSP\_BERT\_HP and FTFP\_BERT\_HP.}


\begin{document}
\maketitle
\flushbottom

\section{Introduction}
Experiments at future \eplus\eminus~colliders require unprecedented jet energy resolutions of \SIrange{3}{4}{\percent} across the full expected jet energy range up to hundreds of \si{\GeV} \cite{ILD, CLiC}. One concept to achieve such resolutions are Particle Flow Algorithms (PFAs) which aim to combine reconstructed tracks from the tracking system with calorimeter deposits, requiring exceptionally granular calorimeters \cite{Brient, Morgunov, PandoraPFA}. The CALICE collaboration develops, builds and tests different calorimeters that aim to fulfil the requirements for the optimal application of PFAs. One concept consists of scintillator tiles or strips of a few \si{\cm\squared} size, individually read out by a Silicon Photomultiplier (SiPM). Several such prototypes with different absorbers, granularities and sampling structures have been constructed.  

In a common beam test at Fermi National Accelerator Laboratory (FNAL) in 2009 the Scintillator Electromagnetic Calorimeter (ScECAL) \cite{ScECALPaper}, Analogue Hadronic Calorimeter (AHCAL) \cite{CommPaper} and Tail Catcher \& Muon Tracker (TCMT) \cite{TCMTPaper} comprised a combined scintillator-SiPM calorimeter system exposed to beams of muons, electrons and pions in the momentum range \SIrange{1}{32}{\GeV/c}. 
Studying the characteristics of single pion events is of special interest as significant fractions of the full particle energy are typically deposited in each calorimeter subsystem. 
The single pion energy resolution of a calorimeter system is expected to be a significant contribution to its performance in both PFA schemes and classic jet energy reconstructions. 
As well as PFA reconstruction, the high granularity of these calorimeters, on the scale of electromagnetic shower development ($\approx$\num{1} radiation length \si{\Xnull} longitudinal sampling, $\approx$\num{1} Moli\`ere radius \si{\RMoliere} transverse segmentation), should enable the statistical identification of electromagnetic subshowers within hadronic showers. This allows the application of software compensation (SC) techniques to improve the energy resolution and response linearity of the calorimeter system. This has been demonstrated in the CDHS calorimeter \cite{SC_CDHS} and was successfully used in collider experiments such as H1 \cite{SC_H1} and ATLAS \cite{SC_ATLAS}. The concept has already been applied to data taken with the CALICE AHCAL and was shown to significantly improve its single pion energy resolution \cite{SCPaper}. This study extends the software compensation scheme to a combined calorimeter system.

\section{Testbeam setup and simulation model}
The datasets used in this note were acquired during a testbeam campaign in May 2009 at the Fermilab Testbeam Facility (FTBF). The MTest beamline delivers secondary particle beams with particle momenta in the range \SIrange{1}{32}{\GeV/c}. The beam line is equipped with a two channel differential gas Cherenkov counter as well as various scintillator triggers. The calorimeter prototypes are installed in the order ScECAL, AHCAL, TCMT downstream of the beam instrumentation. More detailed descriptions of the beamline setup are available in Refs. \cite{Hartbrich, ScECALPaper}.

Each ScECAL layer consists of \SI{3.5}{\mm} tungsten-based absorber and 72 scintillator strips of \\\SI{10x45x3}{\mm} each for a total area of \SI{18x18}{\cm}. Thirty layers are stacked, giving a total depth of around \SI{20}{\Xnull} (\num{\approx0.9} nuclear interaction lengths \si{\LambdaN}) and 2160 readout channels. The SiPMs used in the ScECAL are Hamamatsu MPPC-11-025M models with \num{1600} pixels on an active area of \SI{1x1}{\mm} \cite{ScECALPaper}.
The AHCAL is a \SI{\approx1}{\m^3} hadron calorimeter prototype with steel absorber plates, instrumented with 7608 \SI{5}{\mm} thick scintillator tiles with area between \SI{3x3}{\cm} to \SI{12x12}{\cm} in 38 layers, giving a full depth of \SI{5.3}{\LambdaN} \cite{CommPaper}. The TCMT covers \SI{1x1}{\m} with steel absorber plates and a total of 320 scintillator strips of \SI{5x100}{\cm} each in 16 layers, instrumenting a depth of \SI{5.2}{\LambdaN} \cite{TCMTPaper}. The AHCAL and TCMT use the same type of SiPMs produced by MEPhI/PULSAR with 1156 pixels in an active area of \SI{1.1x1.1}{\mm}.

Details of the simulation models of the ScECAL, AHCAL and TCMT are discussed in Refs. \cite{ScECALPaper, Feege, Guenter}. The right-handed coordinate system is laid out such that the Z-axis is pointing in the beam direction and the Y-axis is pointing upwards. The beam instrumentation present in the MTest beamline is not included in the simulation model apart from the main trigger scintillators. Extra material upstream of the calorimeters associated with about \SI{0.1}{\Xnull} from beam instrumentation is  simulated by adding an Al plate in front of the ScECAL \cite{Hartbrich, Feege}. The beam momentum spread is set to \SI{2.7}{\percent} for samples with beam momentum \SI{<=4}{\GeV/c} and \SI{2.3}{\percent} for samples with higher beam momenta \cite{ScECALPaper, FNALSpread}. The beam profile is extracted from the calorimeter data and transferred into the simulation individually for each sample, as described in Ref. \cite{Hartbrich}.

The sensor digitisation implemented into the simulation accurately models the statistical effects of the scintillator-SiPM readout, using a computationally efficient procedure. Where available, individual sensor parameters extracted from data or separate lab measurements are used. To include the effects of sensor noise into the simulation, noise samples extracted from random trigger events taken during each data run are overlaid onto their respective simulated samples \cite{Hartbrich}. After applying the digitisation procedure to the simulated events, the identical software chain is used for the reconstruction, selection and analysis for real and simulated data.

\begin{figure}[htbp]
{\includegraphics[width=0.50\textwidth]{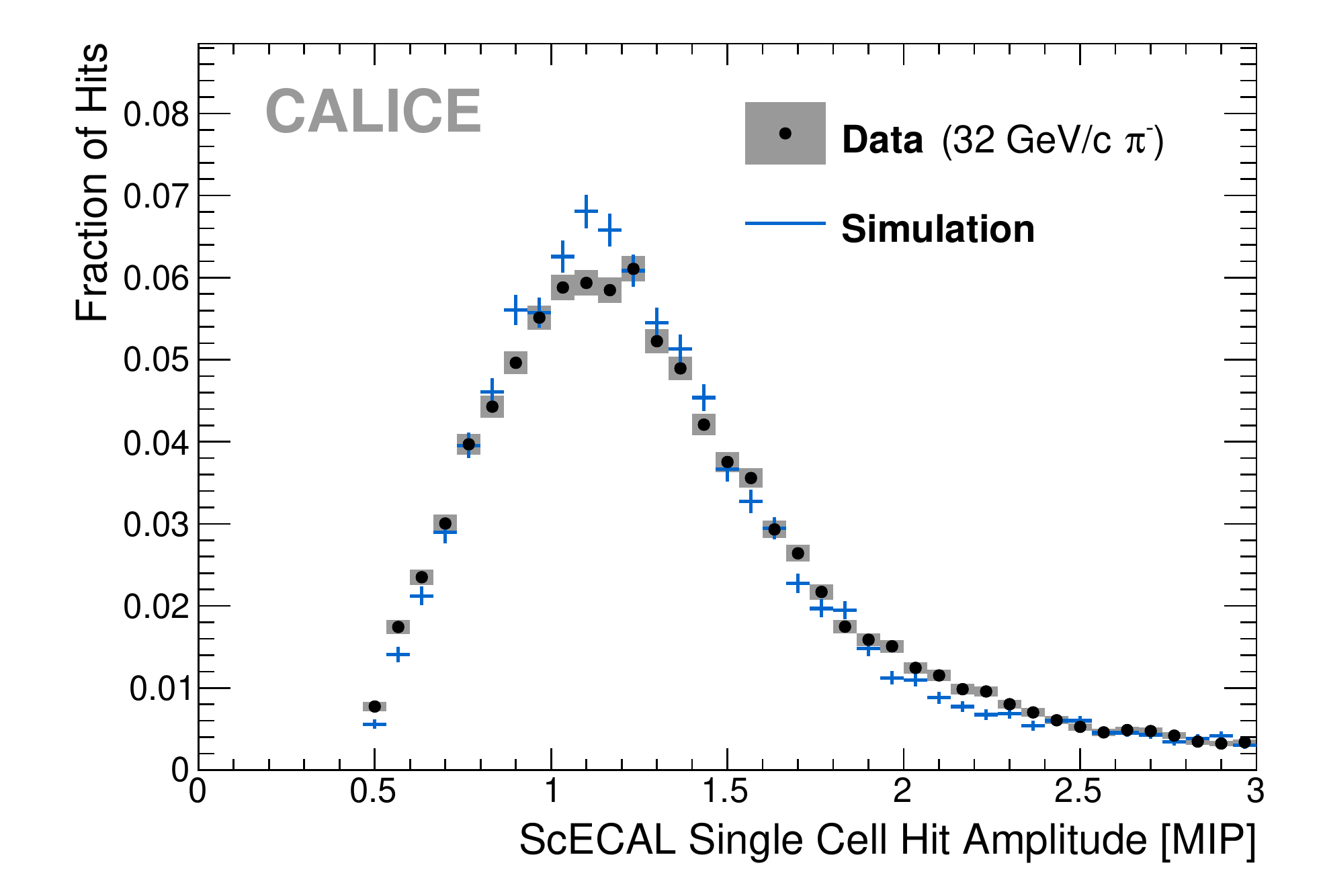}}\hfill
{\includegraphics[width=0.50\textwidth]{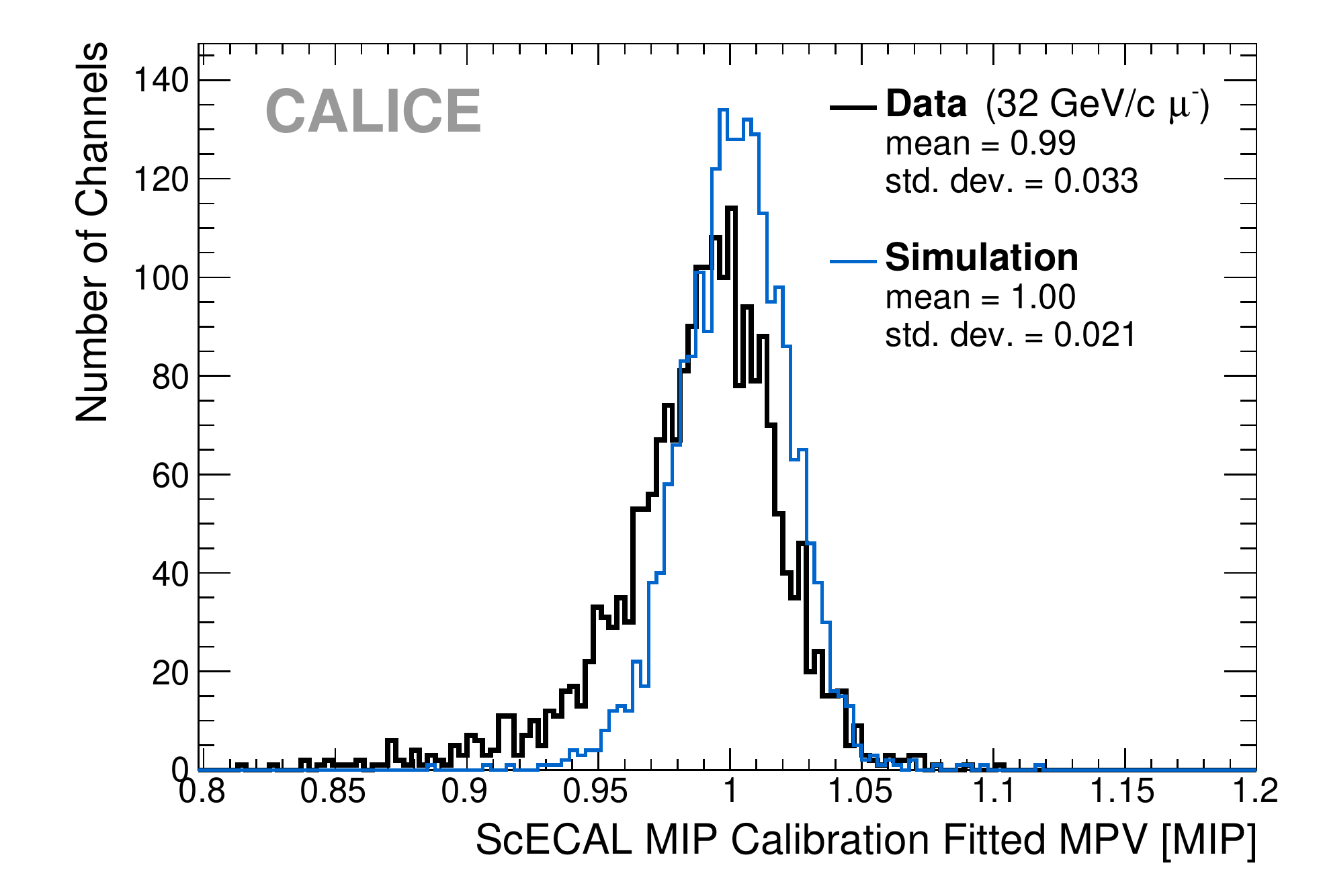}}
	\caption[]{Comparison of MIP-like particles in data and simulation. Normalised hit energy spectrum of a single ScECAL cell from MIP-like tracks deposited by \SI{32}{\GeV/c}~\piminus~(left). Distribution of most probable values of the MIP spectra in single ScECAL cells from \SI{32}{\GeV/c}~\muminus~(right).}
	\label{fig:MIP}
\end{figure}

While sensor effects only have a minimal impact on the calorimetric energy resolution, they significantly widen the response to single MIP-like particles traversing individual cells, which is used to calibrate the whole detector to the MIP scale. The shape of single MIP spectra from an individual ScECAL cell shown in \autoref{fig:MIP} (left) is well reproduced by the simulation, indicating that sensor effects are properly modelled by the digitisation procedures.
The most probable value (MPV) of each MIP spectrum is extracted for each individual detector cell from fitting these MIP spectra with the convolution of a Landau function and a Gaussian. The distribution of ScECAL MIP MPVs from data and simulated samples are shown in \autoref{fig:MIP} (right). The data sample shows a mean close to unity, demonstrating a MIP calibration accuracy at the few percent regime. The data distribution is only slightly wider than the simulated sample, indicating that the calibration quality is close to the systematic limits.

%

Further comparisons to data show a typically very good description of electromagnetic shower variables in the ScECAL simulation, with only a slight underestimation of the effective shower radius in simulation, and a resulting slight overestimation of electromagnetic hit energy densities \cite{Hartbrich, ScECALPaper}. The AHCAL and TCMT simulation models have been well validated in previous studies \cite{SCPaper}.

\geant 10.1p2 was used to simulate all samples used in this analysis, using the QGSP\_BERT\_HP and FTFP\_BERT\_HP physics list, two well validated and actively maintained models also currently used by the LHC experiments \cite{Geant4, PhysLists}. In order to estimate selection efficiencies and biases, samples of \num{100000} electron, pion, proton and muon events were simulated for each data run used in this analysis.

\section{Run and event selection}\label{sec:pionselection}
During the FNAL testbeam period in 2009, \piminus-runs were taken with beam momenta ranging from \SIrange{2}{32}{\GeV/c}. The \SI{2}{\GeV/c} momentum point is omitted in this analysis due to a large admixture of electrons and multi-particle events, a very wide beam profile
as well as inefficient and imprecise determination of the layer of first hadronic interaction, leading to an inefficient and impure selection of single pion events for further analysis  \cite{Hartbrich}. This analysis uses one data run per available beam momentum of \num{4}, \num{12}, \num{15}, \num{20} and \SI{32}{\GeV/c}.
Each run contains between \numrange{180000}{250000} recorded events. All used data runs were recorded within the same beam period without major changes to the system setup between runs.

As the MTest beamline at FTBF does not offer a direct selection of the particle type apart from its charge, the delivered particle beam is a mixture of mostly electrons, pions and muons in varying fractions depending on the beam momentum. Especially at lower beam momenta there is also a large fraction of events with multiple particles hitting the calorimeters in the same event. The goal of the selection described in the following is to efficiently retain events containing a single contained pion shower in the detector system without biasing the composition of the selected pion showers (e.g. the fraction of the shower energy deposited in electromagnetic subshowers). All applied event selection criteria as well as the reconstruction algorithms and their efficiencies are discussed in detail in Ref. \cite{Hartbrich}. 

\paragraph{Raw}{A preselection requires signals in the primary beam triggers in order to exclude pedestal and gain calibration events recorded during beam runs from the analysis. Likewise, not all simulated initial particles pass through the beam trigger as the beam profiles for the lower beam momentum samples are slightly wider than the used scintillator counter.}

\paragraph{Event quality}{The first selection step evaluates readouts from the external beam instrumentation, differential Cerenkov counters, the multi-particle counter and other trigger scintillators \cite{Hartbrich, ScECALPaper}. Events that do not show summed energy deposits of at least \SI{3.5}{\mip} in the first five ScECAL layers are excluded. }

\paragraph{Pion selection}{The pion selection is supposed to suppress electron and muon events in the sample. Single muons and punch-through pions are rejected based on the number of hits and their centre of gravity along the beam axis. 
Electrons are suppressed by reconstructing the layer of first hadronic interaction (FHI layer). The FHI layer is the layer with the first large deposit of energy in the detector, which is taken as an indication of a hadronic interaction of the pion with the detector. The FHI layer reconstruction in the combined system is based on the AHCAL Primary Track Finder algorithm \cite{MarinaFHI}, using optimised thresholds to seamlessly extend into the ScECAL \cite{Hartbrich}. The distribution of reconstructed FHI layers in data and simulation is discussed further in \autoref{sec:profiles}. The reconstructed FHI layer is required to be no earlier than the fifth layer of the ScECAL, which also removes events that have started showering upstream of the calorimeters.}

\paragraph{Multi particle suppression}{Due to the high granularity of the calorimeter system, it is possible to reconstruct the primary MIP-like track a pion leaves in the detector before its first hadronic interaction. Events with multiple beam particles in the detector are suppressed by requiring exactly one such isolated primary track in the event. To suppress events with additional muons entering the AHCAL outside of the coverage of the ScECAL, all events with tracks longer than five layers parallel to the beam axis in the outer parts of the AHCAL, reconstructed with the track segment finder algorithm described in Ref. \cite{TrackSegments}, are also rejected.}

\paragraph{Shower start}{Finally, showers are selected for their reconstructed shower start position. Events must have the primary pion track within the central \SI{10x10}{cm} of the ScECAL and the FHI layer must be reconstructed no later than the fifth AHCAL layer.}

\begin{figure}[htbp]

{\includegraphics[width=0.5\textwidth]{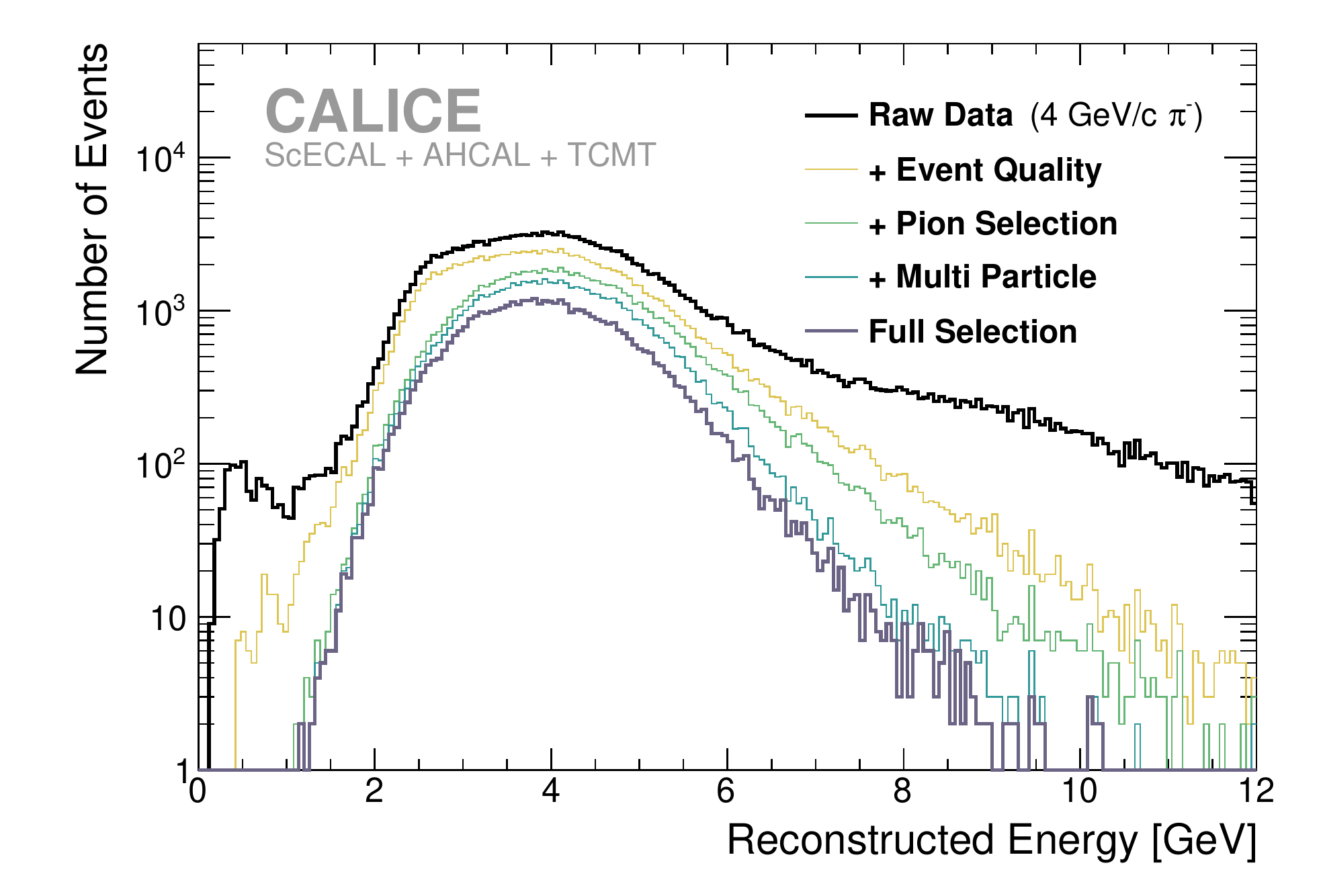}}\hfill
{\includegraphics[width=0.5\textwidth]{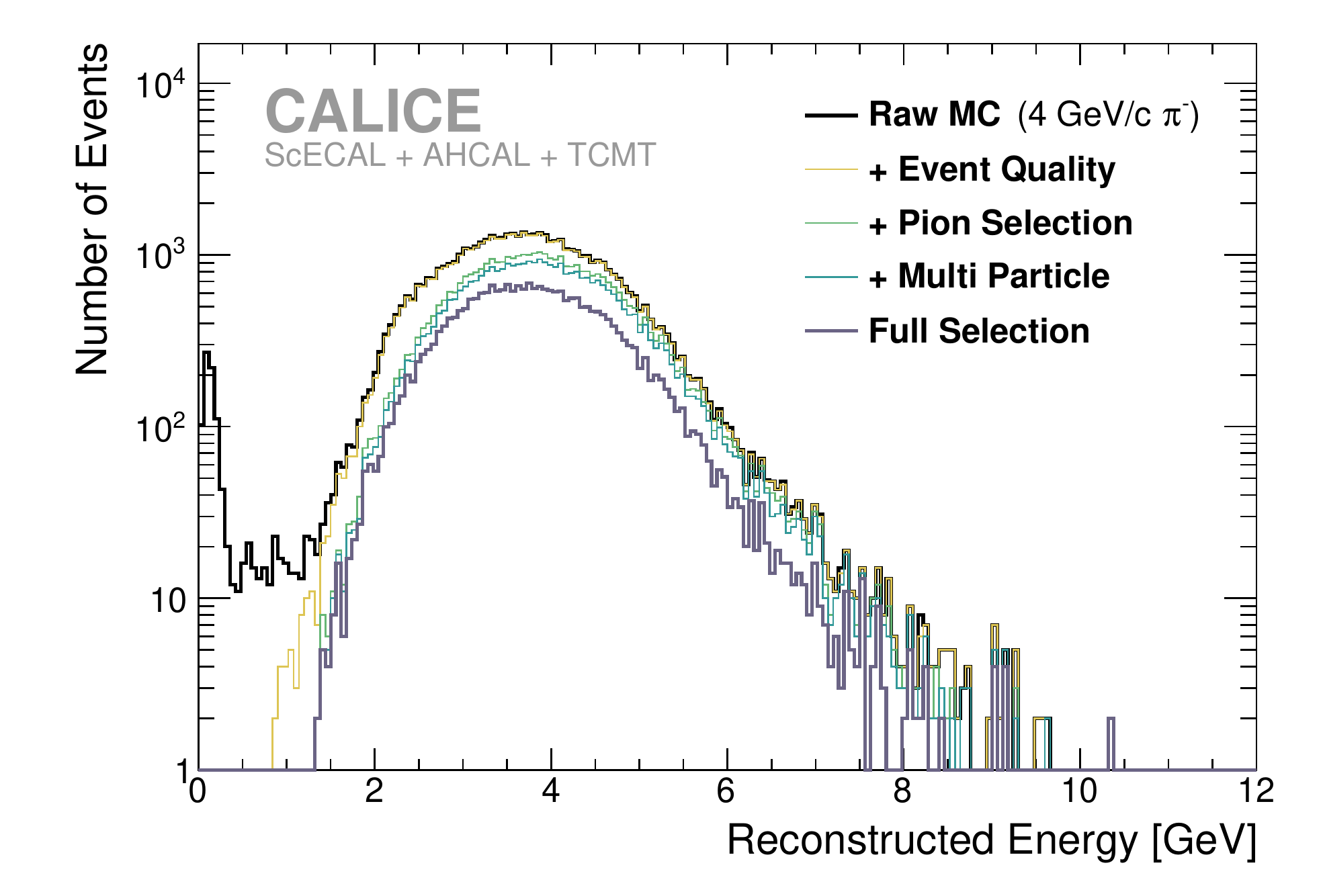}}

{\includegraphics[width=0.5\textwidth]{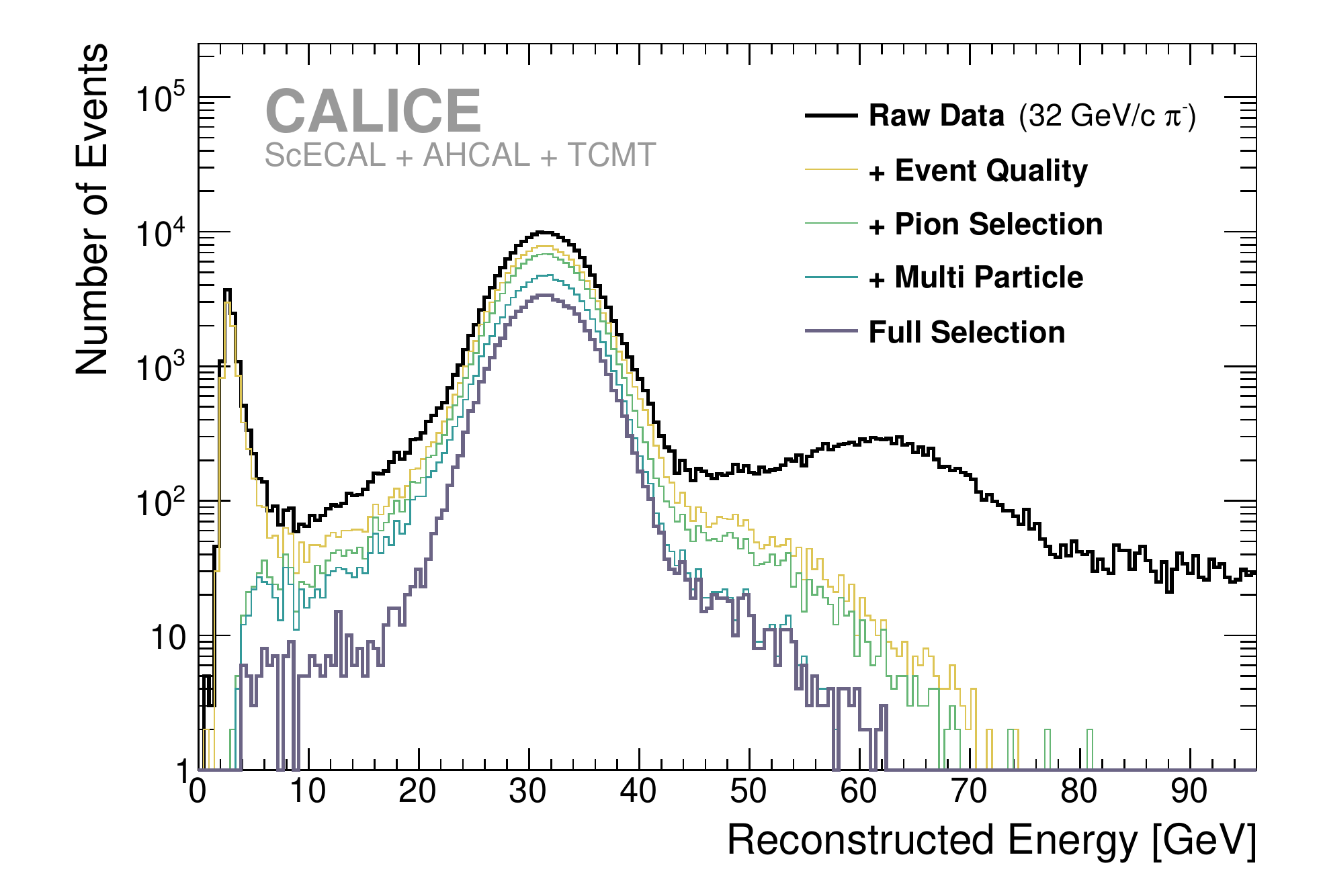}}\hfill
{\includegraphics[width=0.5\textwidth]{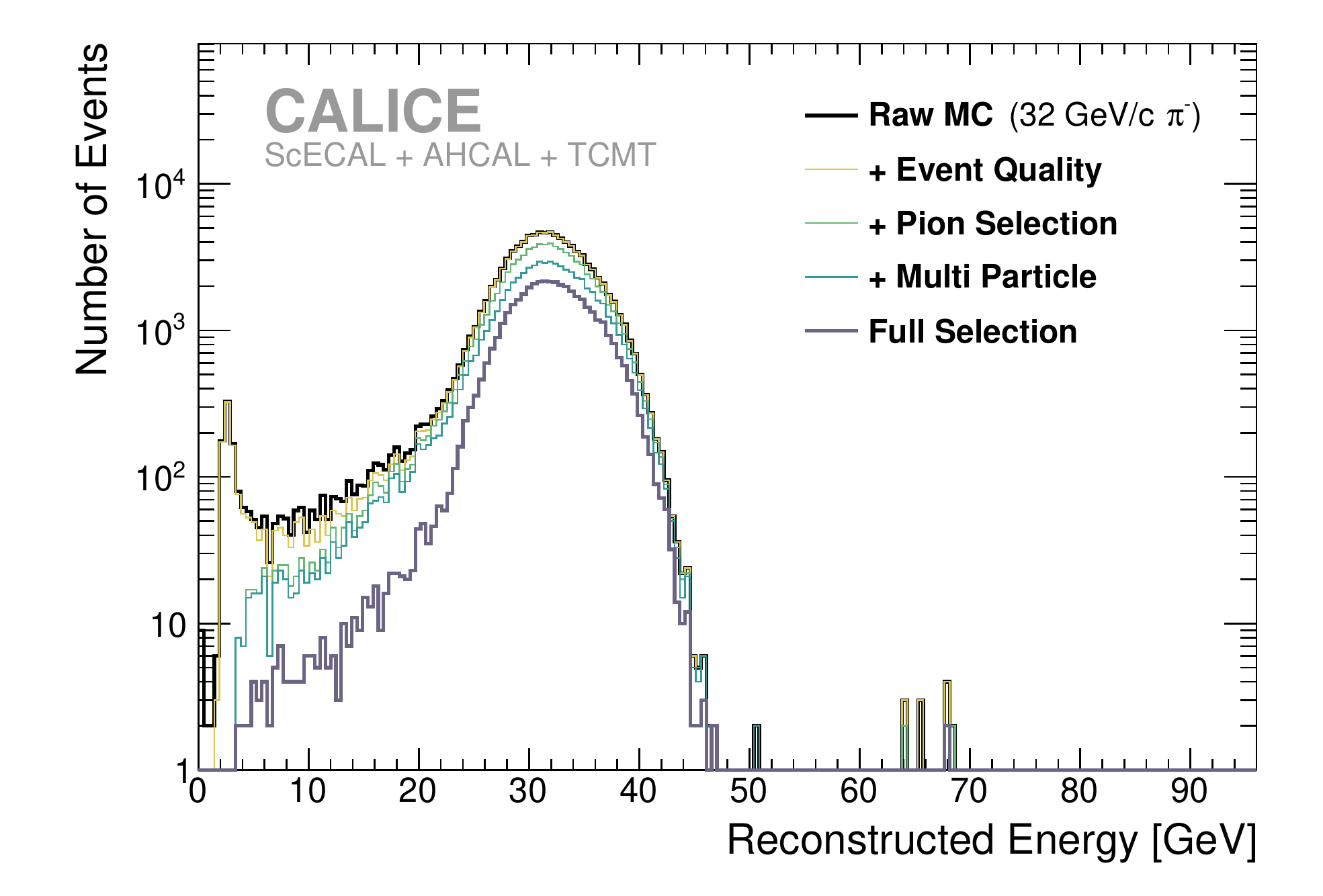}}
	
	\caption{Reconstructed energy spectra of \piminus events in data (left) and FTFP\_BERT\_HP (right) for \SI{4}{\GeV/c} events (top) and \SI{32}{\GeV/c} events (bottom) at different steps of the applied event selection.}
	\label{fig:cutflow}
\end{figure}

\paragraph{}{Examples of the full step-by-step event selection in data and simulation are shown for two beam momenta in \autoref{fig:cutflow}. The event selection suppresses the clearly visible multi-particle peaks in the data distributions without visibly biasing the spectrum of the simulated samples. Especially for the higher beam momentum samples, some entries clearly above the main reconstructed energy peak remain. The efficiencies and biases of the event selection as well as the potential influence of the remaining data sample impurities are discussed in the following sections.}

\subsection{Event selection efficiencies}
Selection efficiencies and biases are studied using simulated event samples for different physics lists and particle types. \autoref{table:pion_selection_efficiency} shows that \SIrange{45}{50}{\percent} of simulated single pion events pass the dull pion event selection in both examined physics lists. Most of the excluded pion events are rejected by the FHI layer requirements. Applying the pion event selection on simulated samples of electrons shows that the combination of FHI layer and isolated primary track cuts provides an excellent electron suppression of $\gtrsim99.9$\% with a slightly worse efficiency for \SI{4}{\GeV/c} beams.

\subsection{Remaining sample impurities}\label{sec:selection_impurities}
To estimate the remaining fraction of electrons in data events passing the pion selection, the distribution of the energy fraction reconstructed in the ScECAL $f = \frac{E_\text{rec}^\text{ScECAL}}{E_\text{rec}}$, using the standard energy reconstruction as discussed in \autoref{sec:energyreco}, is fitted with a linear combination of templates extracted from simulated electron and pion events. Electron showers are typically fully contained in the ScECAL and thus result in values of $f$ close to unity. Pion showers can deposit variable fractions of energy in all parts of the detector systems and can thus generate values of $f$ in the full range of 0 to 1.

\begin{figure}[htbp]
	{\includegraphics[width=0.50\textwidth]{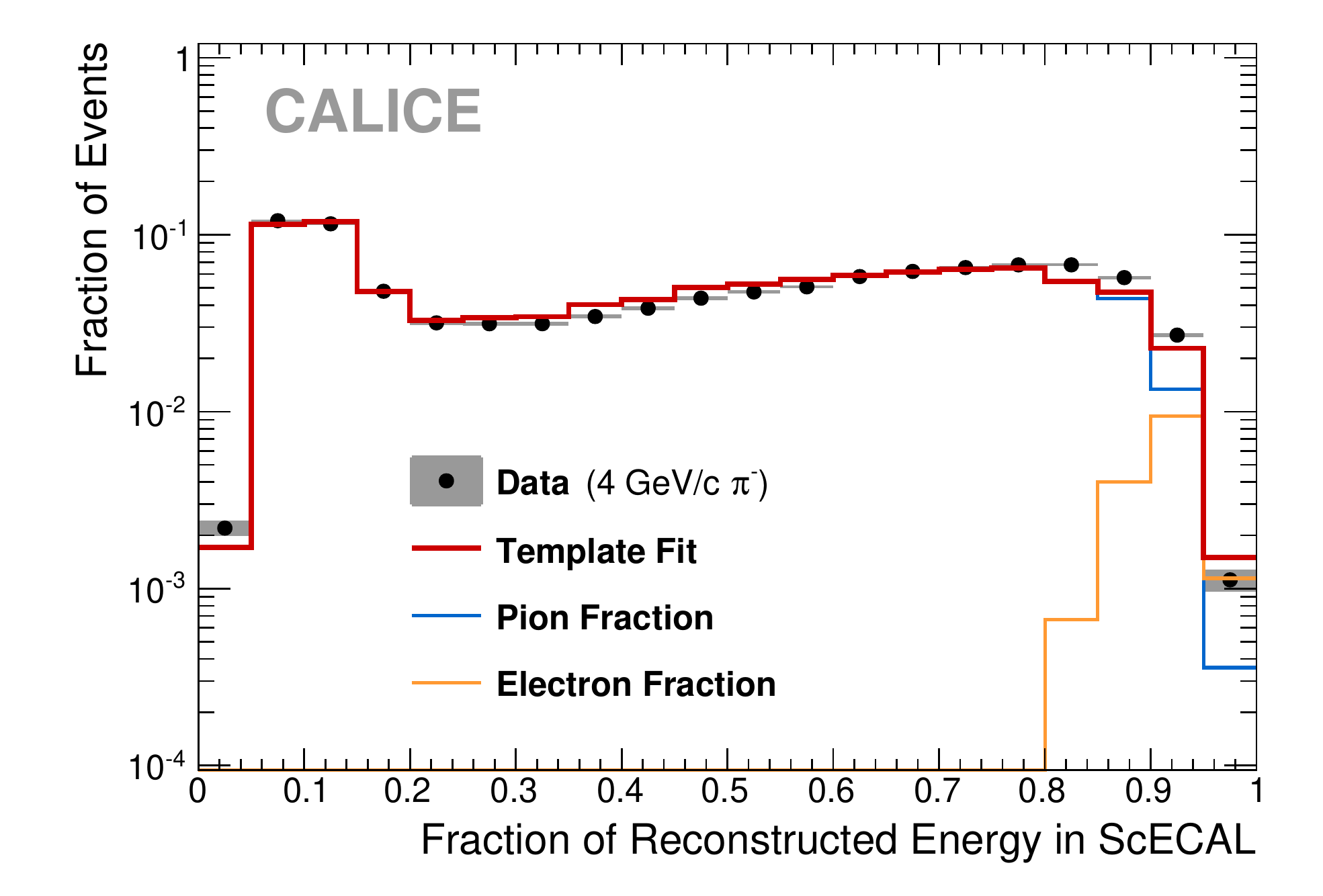}}\hfill
	{\includegraphics[width=0.50\textwidth]{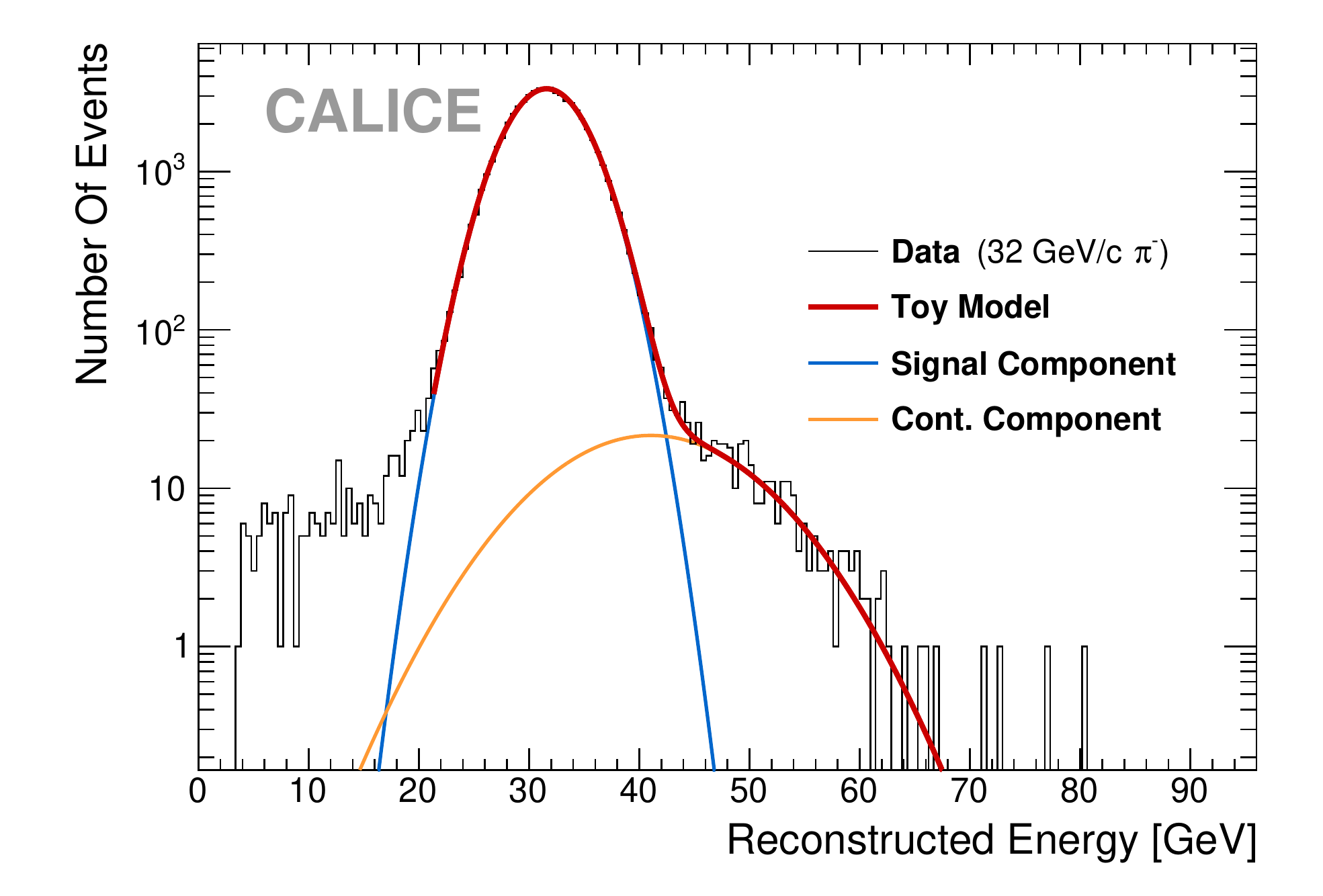}}
	
	\caption[]{Estimations of remaining beam contaminations after the pion selection. Fraction of the total event energy reconstructed in the ScECAL in the 4\,GeV/c pion data sample, including a fit consisting of templates for the pion fraction and electron fraction (left). Reconstructed energy spectrum of the 32\,GeV/c pion data sample, including simple signal and contamination models fitted to the data distribution (right).}
	\label{fig:contaminations}
\end{figure}

\autoref{fig:contaminations} shows the result of this template fit applied to the data sample taken at \SI{4}{\GeV/c} beam momentum. The fit result varies between \SI{1.70}{\percent} and \SI{1.83}{\percent} depending on the used fit range. An electron contamination of \SI{1.8}{\percent} is thus assumed for the \SI{4}{\GeV/c} data. The electron contamination of the other samples used in this analysis is estimated to be negligible with \SI{0.2}{\percent} at \SI{12}{\GeV/c} and \SI{<0.1}{\percent} for beam momenta \SI{>=15}{\GeV/c}.  All these electron contaminations are to be interpreted as upper limits on the single electron contamination fraction, as other possible beam contaminations are not accounted for in the template fit.

The proton contamination in the data samples used for this analysis is expected to be minimal, as exclusively negative charge pion runs were recorded. Thus, only anti-protons could possibly contaminate the beam, which are produced very rarely. The anti-proton fraction of the used data samples was estimated from comparing the fitted slopes of the reconstructed FHI layer spectra (see \autoref{sec:profiles}) from data to simulated proton and pion events.  The reconstructed FHI layer spectrum is sensitive to the proton beam content since the hadronic interaction length of protons and pions in steel differs by about \SI{20}{\percent}. The fit results support no significant anti-proton contamination, as expected \cite{Hartbrich}. 

A general estimate of the remaining fraction of contaminated data events after the pion selection is obtained from the distribution of reconstructed energies in data as shown for the \SI{32}{\GeV/c} sample in \autoref{fig:cutflow}. Even after applying the full event selection, a number of entries clearly above the main reconstructed energy peak remain visible in the distribution. Assuming a simple toy model as shown in \autoref{fig:contaminations}, consisting of two Gaussian curves modelling the main signal and contamination contributions respectively, the fraction of contaminated events can be conservatively estimated to be \SIrange{<=2}{<=3}{\percent} depending on the assumed mean contamination energy for all data samples with beam momenta \SI{>=12}{\GeV/c}. As the reconstructed energy spectrum of the \SI{4}{\GeV} sample is asymmetric with a tail to higher energies (see \autoref{fig:cutflow}), the fraction of contamination events is well hidden below the signal and cannot be estimated in this way at this beam energy. 

It is known that the MTest beamline delivers a significant fraction of events with more than one beam particle, which are not perfectly suppressed in this analysis. The remaining contaminations are unlikely to be events containing two pions or one pion and one electron which both carry the full beam momentum each, as such events would form a peak around twice the reconstructed energy, which is not seen. The distribution of reconstructed energies rather shows a continuous contamination tail towards higher reconstructed energies. This could be caused by additional contaminating particles with energy lower than that of the beam, e.g. particles which punch through the final beam collimator. The observed behaviour cannot be explained by additional muons in the beam halo, as such muons would deposit far less energy than required to explain the observed contamination. The visible tail towards lower reconstructed energies is part of the signal, caused by pion showers that are not fully contained in the calorimeter system.

\section{Pion energy reconstruction}\label{sec:energyreco}
The first step of the energy reconstruction consists of converting the raw measured ADC inputs of each channel in a given event to a common energy scale per subdetector, calibrated with respect to the most probable deposited energy of a MIP-like particle crossing a single detector scintillator cell. This calibration includes the correction of saturation effects in the sensors, temperature corrections, correction of electronics effects (etc.) with calibration data obtained in lab tests, dedicated calibration runs and {\it in situ} methods with down to channel-wise granularity \cite{ScECALPaper, CommPaper, TCMTPaper}.

The incident particle energy can be reconstructed from these calibrated hit amplitudes by summing all energy deposits $\mathbf{D}$ weighted by any number of measured event quantities and external parameters $\mathbf{X}$. In the simplest case, they are directly weighted by their inverse sampling fraction. For any chosen energy reconstruction algorithm, external parameters can be optimised by minimising the sum of quadratic distances of the reconstructed event energy to the known beam energy, resembling a $\chi^2$ function:
\begin{equation}\label{eq:chi2}
\chi^2 = \sum_{\text{events}}\frac{\left( E_\text{rec}^\text{event}(\mathbf{D},\mathbf{X})\left[\si{\GeV}\right]-E_\text{beam}^\text{event}\left[\si{\GeV}\right]\right)^2}{{\left(\SI{55}{\percent}\right)}^2\cdot E_\text{beam}^\text{event}\left[\si{\GeV}\right]}
\end{equation}
In this formalism, parameters can be estimated for multiple samples at multiple beam momenta in a single optimisation. 
In the energy range considered here, the variance of the reconstructed energy spectrum is expected to scale with the beam energy according to a $\nicefrac{1}{\sqrt{E}}$ stochastic term in the calorimetric energy resolution dependence. 
To normalise contributions to $\chi^2$ between different beam momenta, each event is thus weighted with the inverse of its known beam energy. The constant factor of \SI{55}{\percent} in the denominator is approximating the expected stochastic term of the calorimeter to enable the correct parameter uncertainty estimation by the optimisation algorithm, but does not influence the estimated parameter values. In order not to bias the parameter estimation towards a specific beam energy, the same number of events are added to the $\chi^2$ for each beam momentum. The first 40,000 selected events of each data run (20,000 in simulated samples) are used in the parameter optimisation. No significant difference in reconstructed energy resolution is observed whether the events used for parameter optimisation are excluded in the reconstruction or not for both energy reconstruction schemes discussed here.

\subsection{Standard weighting}\label{sec:energyrecoclassic}
In an idealised sampling calorimeter the reconstructed energy for an incoming particle is directly proportional to the measured energy deposits. For a calorimeter system combining different sampling ratios the reconstructed energy is the sum of all hit energies weighted by a constant factor for each calorimeter. As the sampling fraction within the ScECAL is constant and the AHCAL and first eight layers of the TCMT have identical sampling fractions, the standard energy reconstruction only has two parameters $w_{\text{ECAL}}$ and $w_{\text{HCAL}}$:  
\begin{equation}
 E_\text{rec}^\text{classic}=w_{\text{ECAL}}\cdot E^\text{ScECAL}_\text{sum} + w_{\text{HCAL}}\cdot \left( E^\text{AHCAL}_\text{sum} + E^\text{TCMT}_\text{sum}\right) 
\end{equation}
The $\chi^2$-optimisation described above inherently assumes Gaussian distributions. The reconstructed energy spectra obtained from the used calorimeter setup can exhibit non-Gaussian features from fluctuations in the electromagnetic fraction of pion showers, leakage effects and sample impurities remaining in data. The optimisation of weights is thus performed iteratively on the central 90\% of reconstructed energies. 

\begin{figure}[htbp]
\begin{center}
\includegraphics[width=0.8\textwidth]{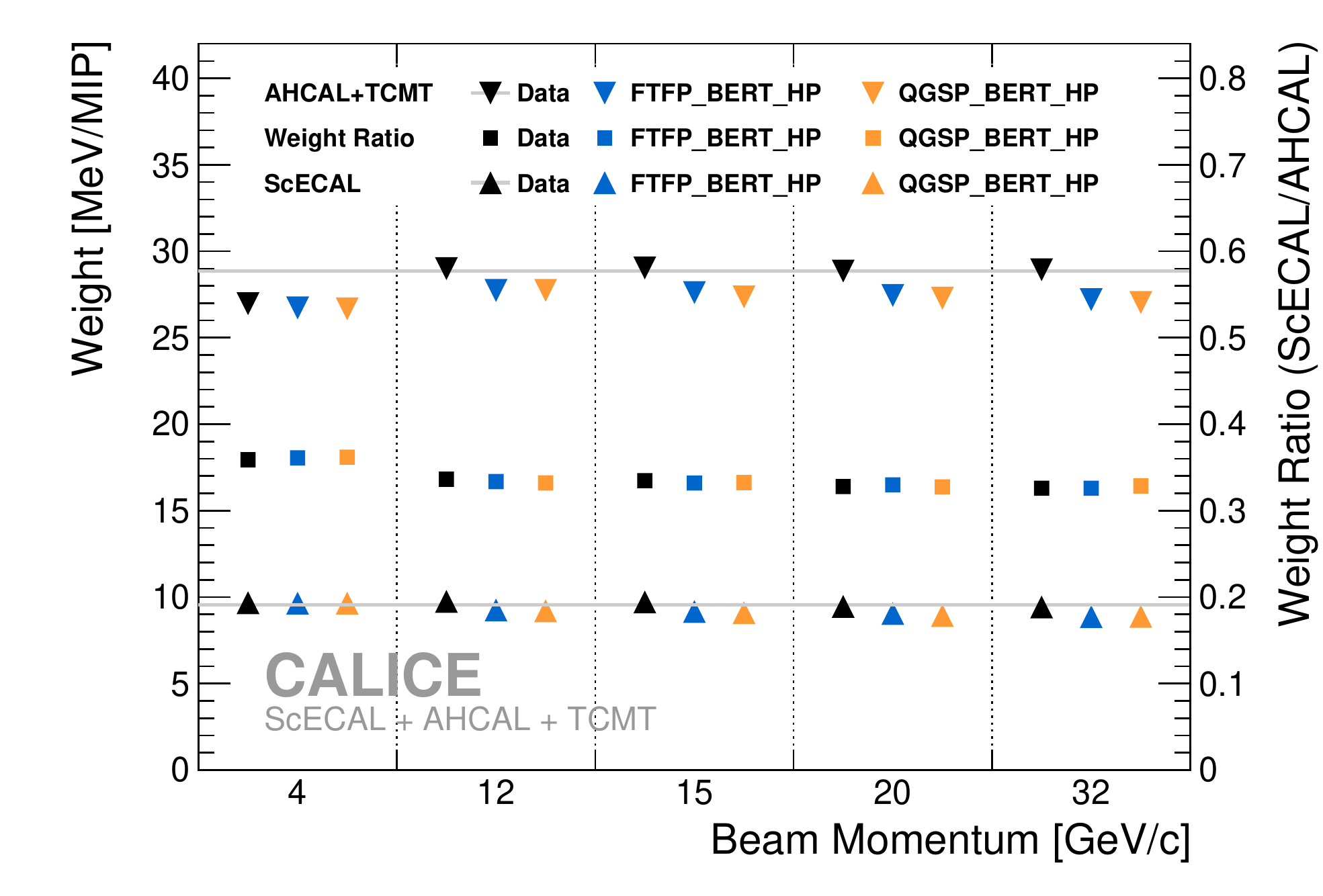}
\caption{Standard energy reconstruction weights obtained from data and simulations. 
Weights are optimised for single beam momenta individually (markers) or for all beam momenta simultaneously (lines). Square markers indicate the ratio between the ScECAL and AHCAL+TCMT weights. All statistical uncertainties are smaller than the used markers.}
\label{fig:weights_classic}
\end{center}
\end{figure}

\autoref{fig:weights_classic} lists the weights optimised for each beam momentum individually as well as all momenta at once. Variations of the weight with the beam momentum are rather small, with only the lowest beam momentum point at \SI{4}{\GeV/c} preferring slightly different weights. The two used physics lists produce very similar weights. The weights obtained from data for beam momenta \SI{>=12}{\GeV/c} have slightly (\SI{\approx 5}{\percent}) higher values than those obtained from simulations, hinting at a general overestimation of energy deposits in simulations, as is also  observed in the longitudinal shower profile as shown in \autoref{sec:profiles}.

The reconstructed resolution and linearity only depend on the ratio of weights, which is very similar between data and simulations. Using the standard reconstruction weights obtained from simulation to reconstruct data events (or vice versa) would thus not notably influence the energy resolution, but only result in a shifted energy scale. 
\subsection{Software compensation}\label{sec:energyrecosc}
The measured response of an undercompensating calorimeter to a hadron shower is typically smaller than the response to an electromagnetic shower of the same initial particle energy. Within hadronic showers, energy can be lost to invisible processes such as recoil, excitation and fragmentation of absorber nuclei and, depending on the active material, neutron emission, leading to a lower response compared to purely electromagnetic showers. Additionally, inelastic interactions in hadron showers can develop purely electromagnetic subshowers from $\pi^0$/$\eta$ production and their subsequent decay to two photons, yielding the full electromagnetic response for the two photons. \cite{Wigmans}

Both the loss of measurable energy to invisible mechanisms and the creation of $\pi^0$/$\eta$ are stochastic processes and thus subject to statistical fluctuations from event to event, leading to large fluctuations in measured energy deposits and degraded energy resolution. Furthermore, the number of generated $\pi^0$/$\eta$ depends on the number of inelastic interactions within a hadron shower which scales with initial particle energy, degrading the linearity of the calorimeter response. If it were possible to identify electromagnetic subshower contributions within hadron showers, reweighting them to the purely hadronic scale should lead to an improvement in both the resolution and linearity of the energy measurement.

With the materials used in this setup, the length scales of hadronic and electromagnetic showers are notably different ($\nicefrac{X_0}{\text{layer}}\gg\nicefrac{\lambda_{\pi}}{\text{layer}}$ in both ScECAL and AHCAL). In combination with the high transverse granularity similar to the Moli\`ere radius of electromagnetic showers, this enables the discrimination of electromagnetic and hadronic shower components by means of their deposited energy density. Each measured cell deposit is thus weighted as a function $w(\rho, E_\text{est})$ of the local deposition density $\rho$ and an estimate of the full shower energy $E_\text{est}$. 

Instead of a continuous two-dimensional parametrisation of $w(\rho, E_\text{est})$, the full range of $\rho$ is divided into fixed bins, while the dependence on $E_\text{est}$ is parametrised over the used energy range for each such bin. This scheme differs from the local software compensation implementation in \cite{SCPaper}, in which $w$ is iteratively parametrised as a continuous function of both $\rho$ and $E_\text{est}$. The scheme presented here leads to more free parameters and thus more degrees of freedom but maintains a stable convergence of the optimisation. 

This analysis uses eight bins in deposition density, individually chosen to be logarithmically distributed in the range of occuring depositions for both the ScECAL and the AHCAL, respectively. The obtained results do not critically depend on the number of bins or exact bin boundaries. For the two lowest deposition density bins, instead of summing up hit energies, only the number of hits falling into these bins are counted to suppress Landau fluctuations from low particle multiplicity hits (similar to the reconstruction scheme applied in the CALICE SDHCAL \cite{SDHCAL}), slightly improving the resolution of the algorithm. 

Instead of using the deposition density as the hit amplitude divided by the cell size, the hit energy, the energy deposition of each individual scintillator cell measured in photo-electrons and then converted to the MIP scale, is used directly. This is disregarding the differently sized tiles in the AHCAL, which slightly improves the performance of the full algorithm. All TCMT energy deposits are treated as falling into the same hit energy bin, effectively parametrising the relative TCMT weight as a function of only the first energy estimate.

The lowest hit energy bin has significant contributions from the primary pion track before the first hadronic interaction, which show nearly no dependence on beam momentum. To avoid biasing of the parameter optimisation towards weighting up the primary track hits to the full beam energy, hits on the primary track are excluded from the software compensation weighting. All hits on the axis of the reconstructed isolated primary track (as described in \autoref{sec:pionselection}) from the first ScECAL layer up to two layers before the reconstructed FHI layer are included into the energy reconstruction without hit energy or shower energy dependent weighting. To exclude Landau fluctuations, only the number of such hits is used and multiplied by the mean energy deposit of a MIP-like particle in a single cell of the given calorimeter section.

\begin{figure}[htbp]
\includegraphics[width=0.5\textwidth]{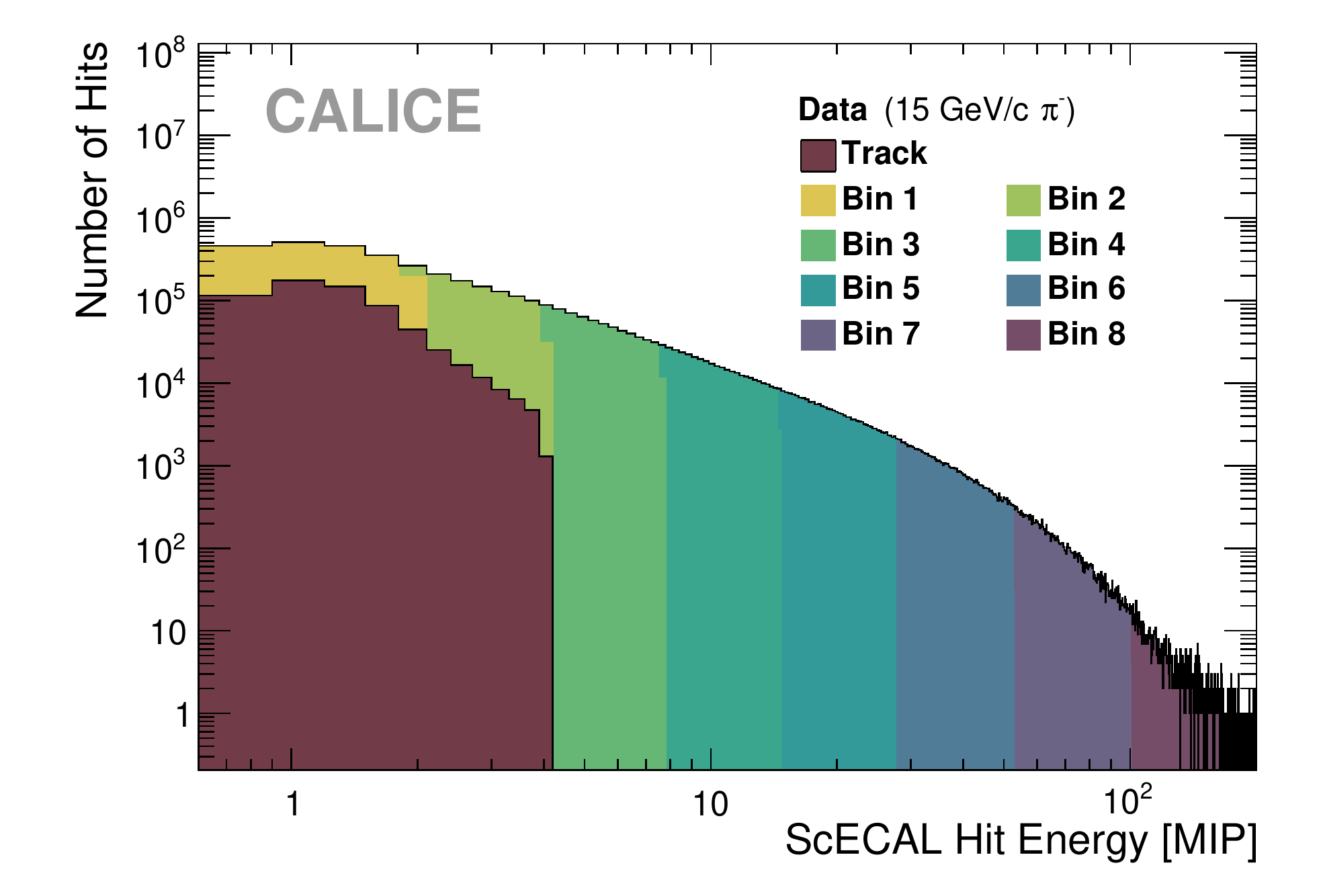}\hfill
\includegraphics[width=0.5\textwidth]{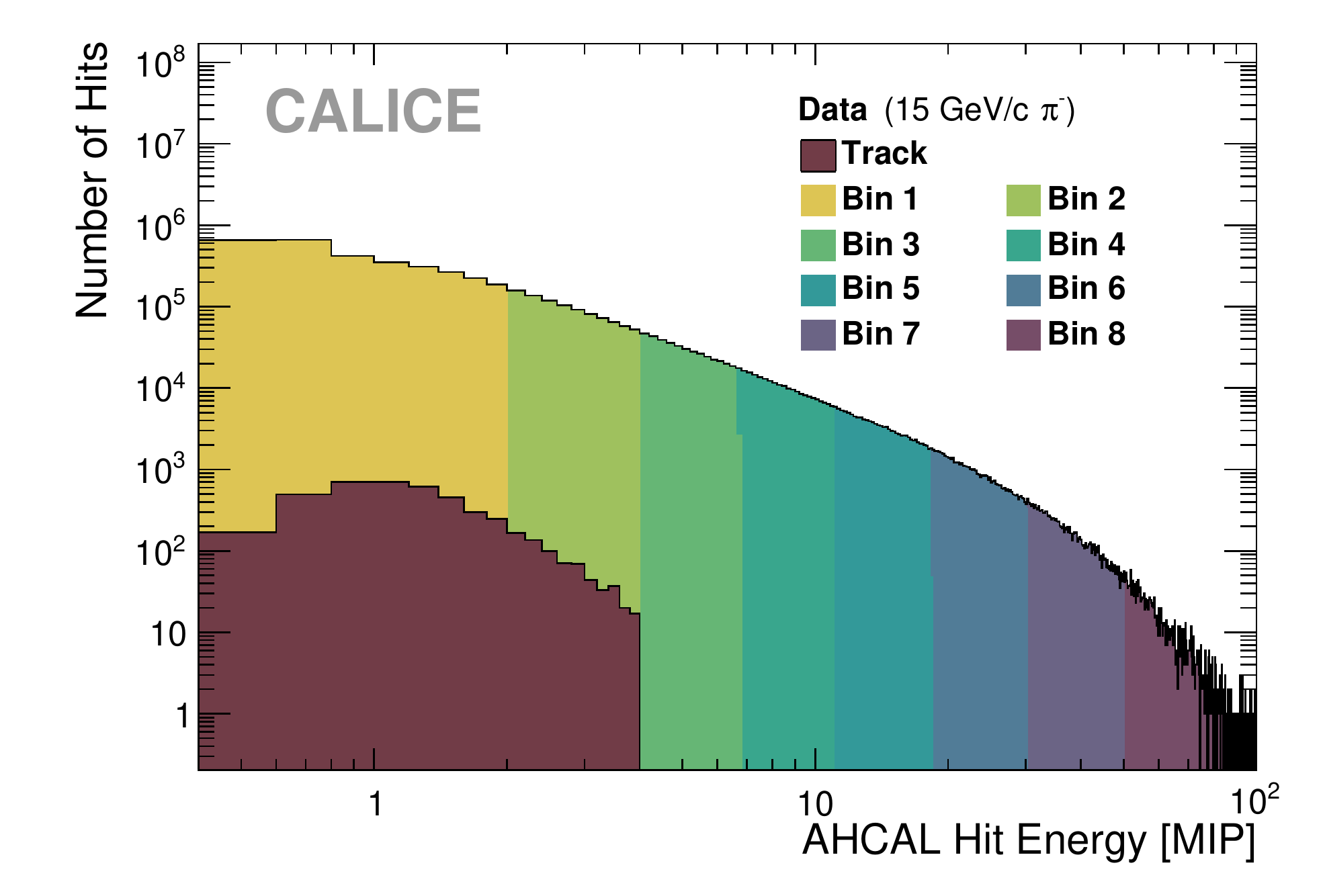}
	
	\caption{Hit energy spectra of the ScECAL (left) and the AHCAL (right) for 15\,GeV/c \piminus~data. Colours are assigned to hits reconstructed on the primary pion track and by software compensation bin.}
	\label{fig:hitEnergySCbinsData}
\end{figure}

An example of the distribution of hits into hit energy bins is given in \autoref{fig:hitEnergySCbinsData}. In the lowest hit energy bin in the ScECAL, around one quarter of all contributions would originate from the primary track if not identified and excluded. The contribution of primary track hits to the AHCAL hit energy spectrum is small, as around 70\% of selected events start showering in the ScECAL.

In this analysis the weights, $\alpha_i, \beta_i$, for the $i$th hit energy bin as well as the TCMT weight $\gamma$ are parametrised as second order Chebyshev polynomials of the estimated particle energy $E_\text{est}$.
The full formula to reconstruct the energy in the combined system, with the sampling weights $w$ used from the standard energy reconstruction, the software compensation weights $\alpha_i, \beta_i$, $\gamma$, the sum (or count) of energy deposits in the $i$th hit energy bin $E_i$ and the energy deposits on the primary track $E_\text{track}$, is:
\begin{multline}
  E_\text{rec}^\text{SC}=w_{\text{ECAL}}\cdot\left( \sum_{i}^{\text{bins}}\alpha_i\left(E_\text{est}\right)\cdot E^\text{ECAL}_i+E^\text{ECAL}_\text{track}\right) \\ + w_{\text{HCAL}} \cdot\left( \sum_{i}^{\text{bins}}\beta_i\left(E_\text{est}\right)\cdot E^\text{HCAL}_i+E^\text{HCAL}_\text{track} + \gamma\left(E_\text{est}\right)\cdot E^\text{TCMT}_\text{sum}\right)
\end{multline}
The software compensation reconstruction is defined by a total of 51 parameters (8 bins in the ScECAL $\times$ 3 parameters per bin, $8\times3$ parameters for the AHCAL and $3$ parameters for the TCMT). The parameter values are obtained by minimising the $\chi^2$ function described in \autoref{eq:chi2}, using all available beam momentum samples in one common optimisation procedure. During the parameter optimisation, the known beam energy is used for $E_\text{est}$, while during reconstruction the standard reconstruction result is used as an estimate.

\begin{figure}[htbp]
	{\includegraphics[width=0.5\textwidth]{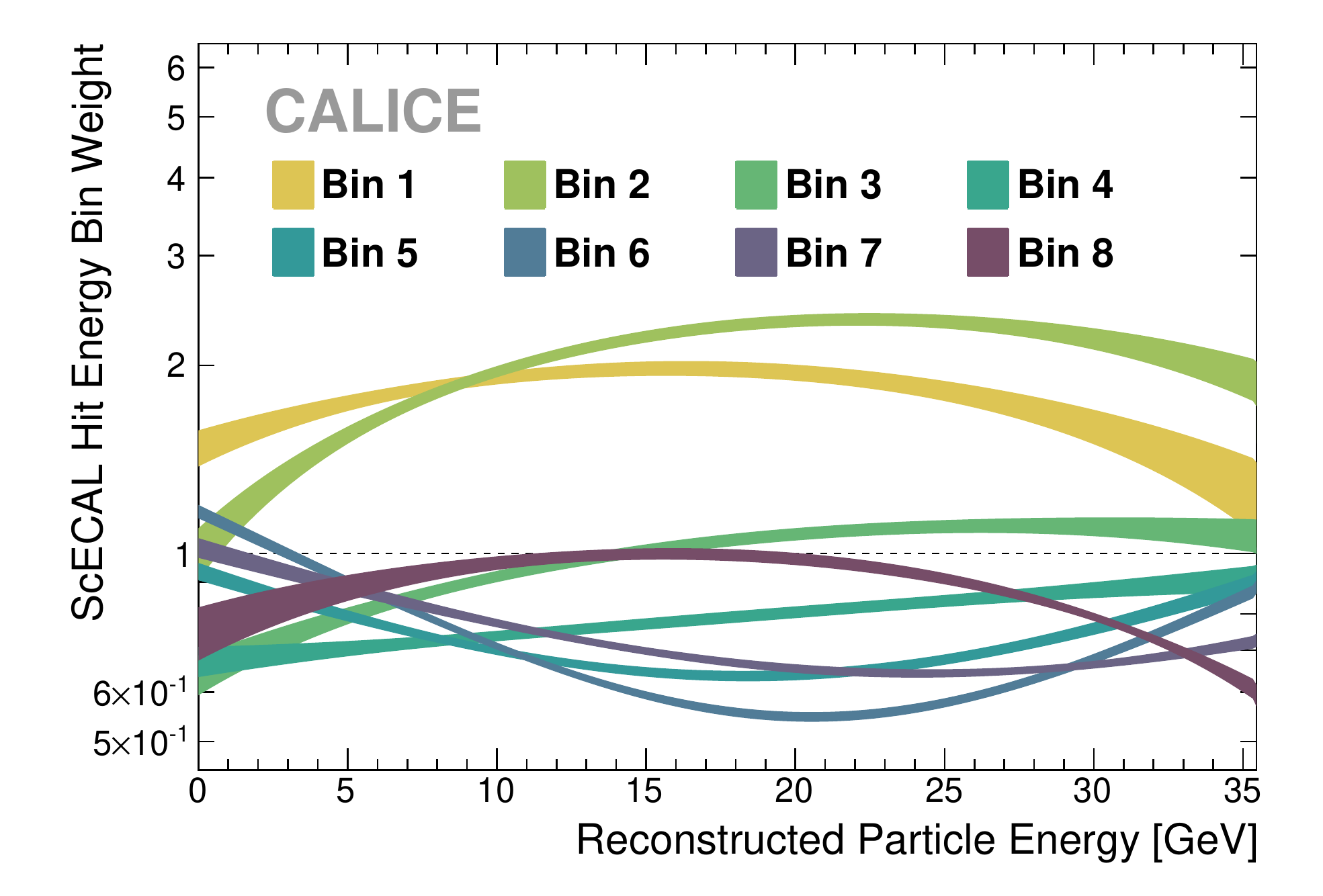}}\hfill
	{\includegraphics[width=0.5\textwidth]{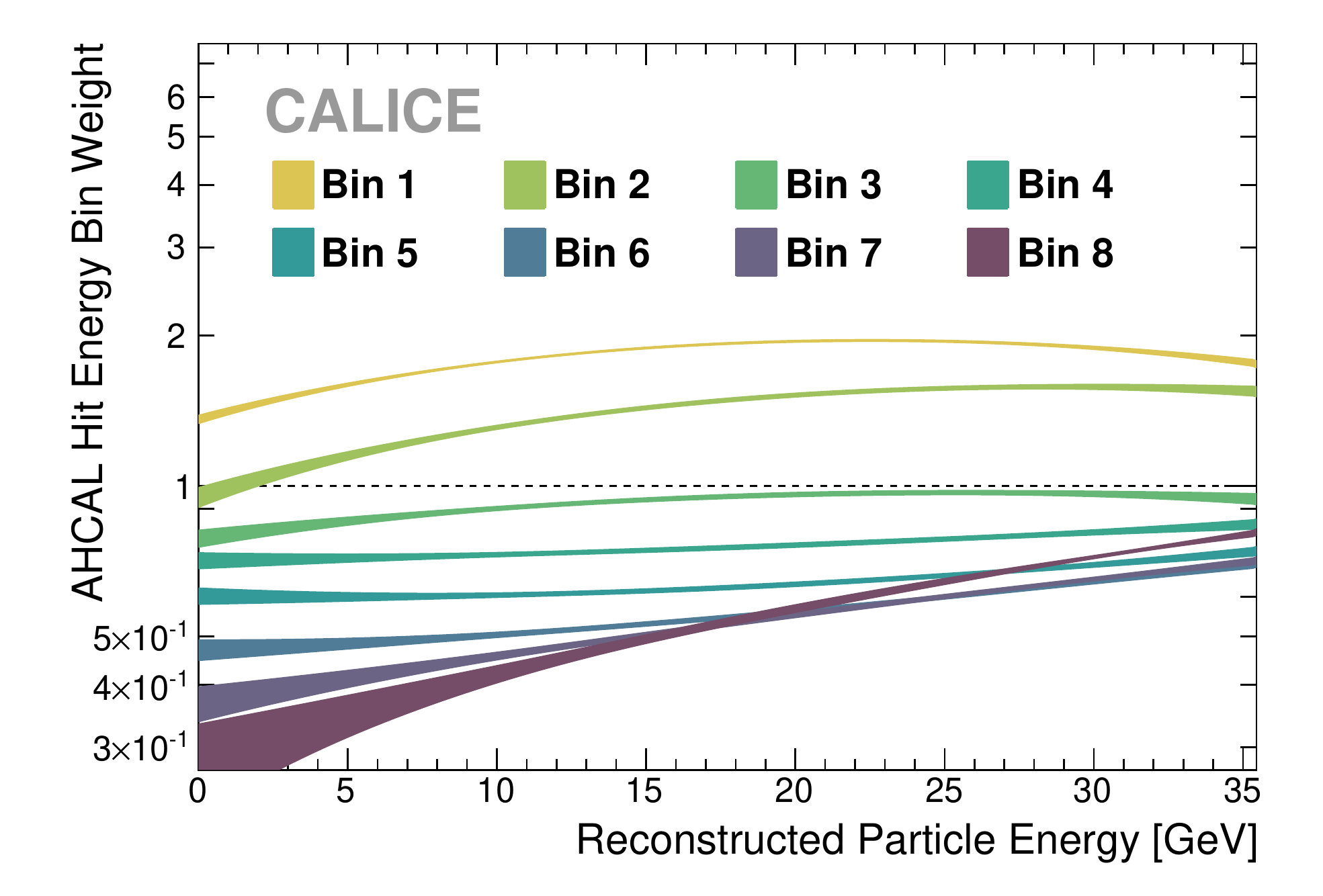}}
	{\includegraphics[width=0.5\textwidth]{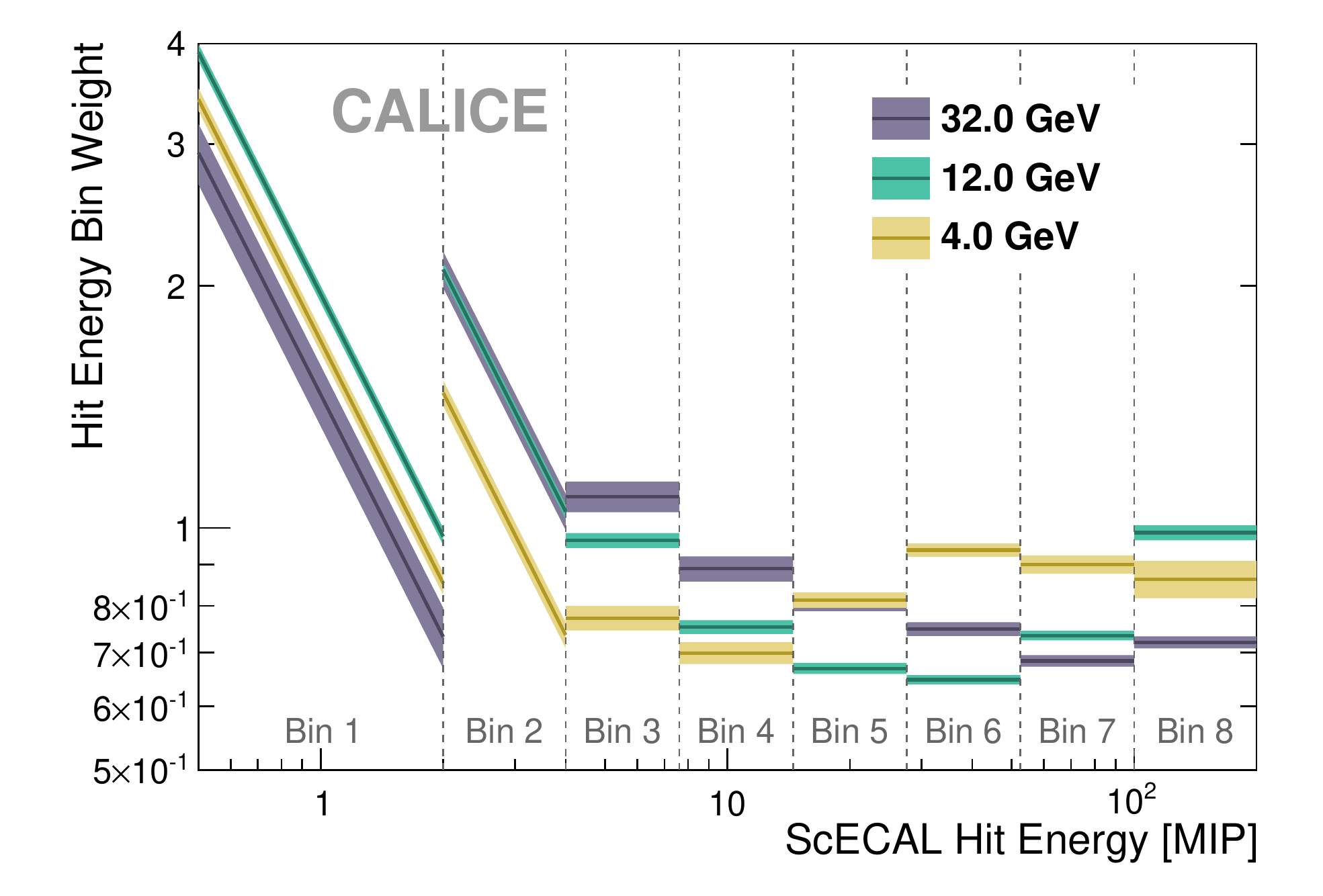}}\hfill
	{\includegraphics[width=0.5\textwidth]{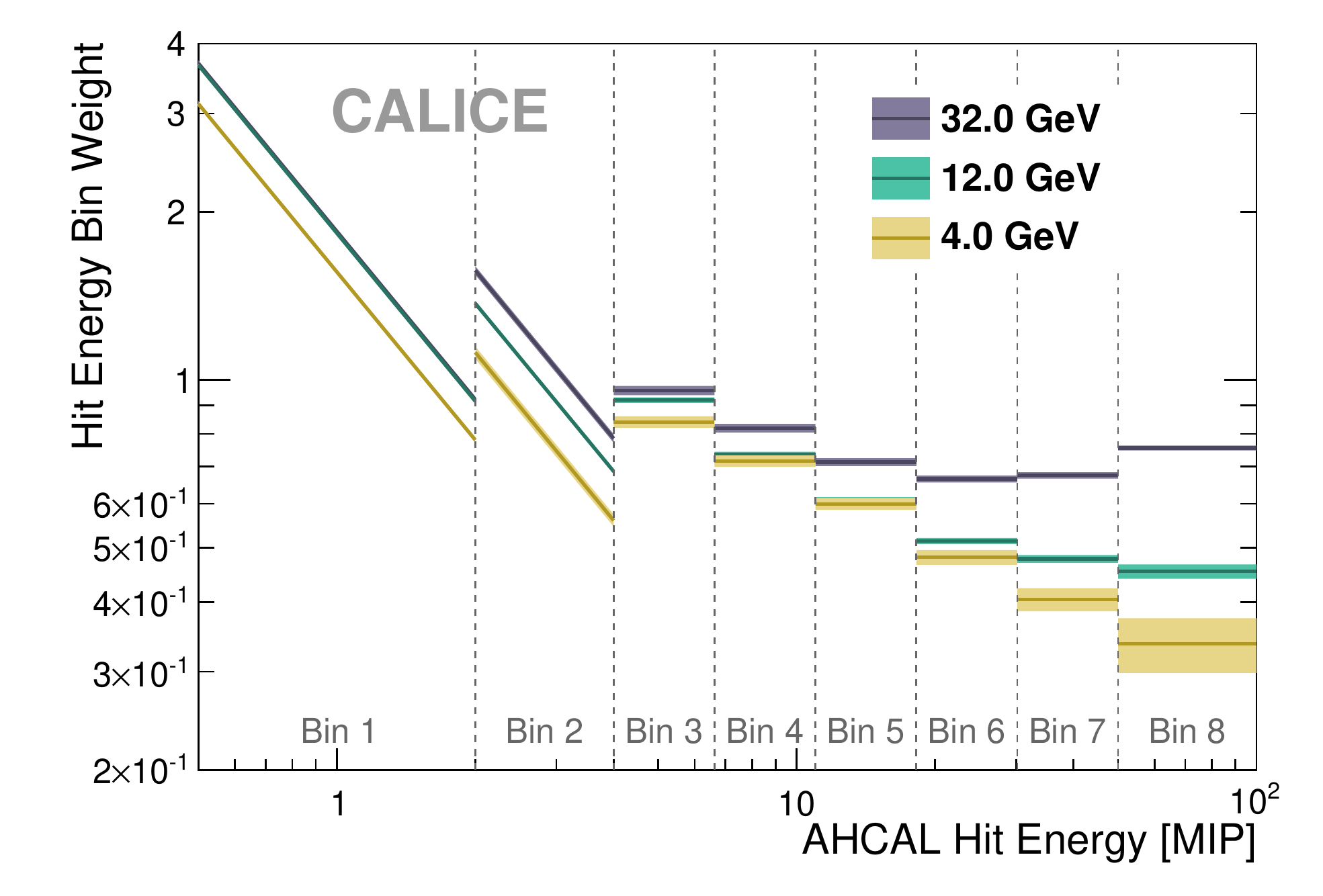}}
	
	\caption[]{Software compensation weights for the ScECAL (left) and AHCAL (right) optimised from data. The upper row shows the weights for each hit energy bin as a function of the estimated particle energy. The bottom row shows vertical slices through the weights shown in the upper plots. The hit energy dependent weights of the first two bins correspond to a $\nicefrac{1}{E}$ dependence and thus a counting of hits in these bins. The width of each line indicates the weight uncertainty propagated from the parameter errors.
	}
	\label{fig:SCweightsData}
\end{figure}

\autoref{fig:SCweightsData} shows the polynomial functions obtained for the energy dependence of the bin weights for ScECAL and AHCAL resulting from the parameter optimisation. The slopes in the first two bins of ScECAL and AHCAL in \autoref{fig:SCweightsData} (bottom) correspond to a $\nicefrac{1}{E}$ dependence and thus a constant contribution to the reconstructed energy of each hit in these bins, regardless of the hit energy. Assuming a shower of $E_\text{est} = \SI{4}{\GeV}$, a hit in the AHCAL with a measured hit energy $E_\text{hit} = \SI{1}{\mip}$ would be weighted with a factor of around $1.5$ (as given by the yellow lines in \autoref{fig:SCweightsData}, bottom right) for a contribution to the reconstructed shower energy of $1.5 \times \SI{1}{\mip} = \SI{1.5}{\mip}$. A hit of measured energy $E_\text{hit} = \SI{0.5}{\mip}$ would be weighted with the doubled weight, due to the $\nicefrac{1}{E}$ dependence of the first two hit energy bins, for the identical contribution of $3 \times \SI{0.5}{\mip} = \SI{1.5}{\mip}$ to the reconstructed shower energy. In hit energy bins $\geq 3$, two hits of different hit energy within the same hit energy bin would contribute to the reconstructed shower energy proportionally to their hit energy.

Higher hit energy bins tend to be weighted below unity, indicating that a high energy hit is more likely to belong to an electromagnetic subshower. Especially in the ScECAL, bin weights do not monotonically decrease for increasing hit energies, as would be enforced in the local software compensation scheme used in \cite{SCPaper}. However, the hit energy range which is assigned the lowest reconstruction weight increases with energy, indicating that the typical hit energy scale for electromagnetic subshowers increases with the incident pion energy.

\begin{figure}[htbp]
	{\includegraphics[height=0.4\textwidth]{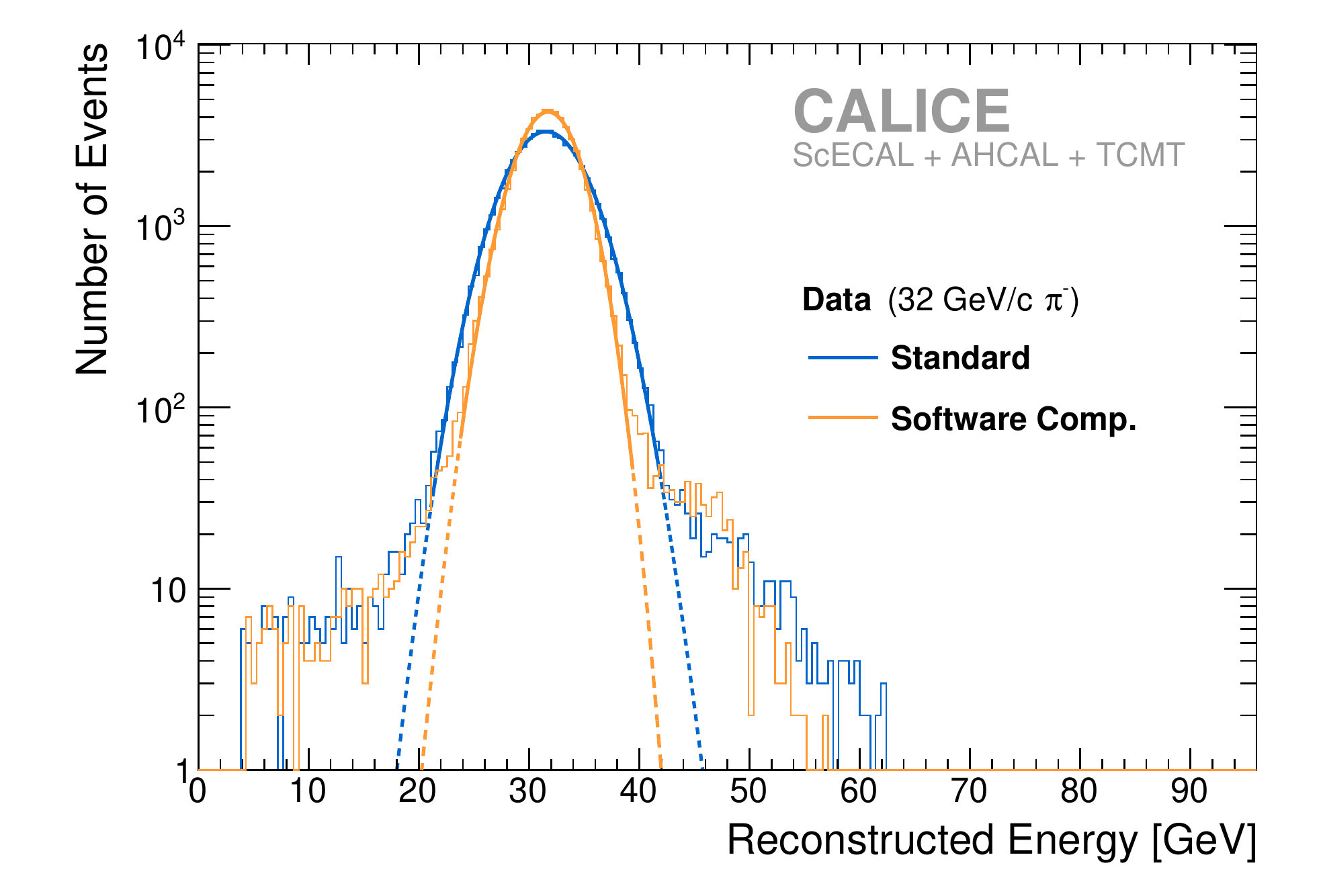}}\hfill
	{\includegraphics[height=0.4\textwidth]{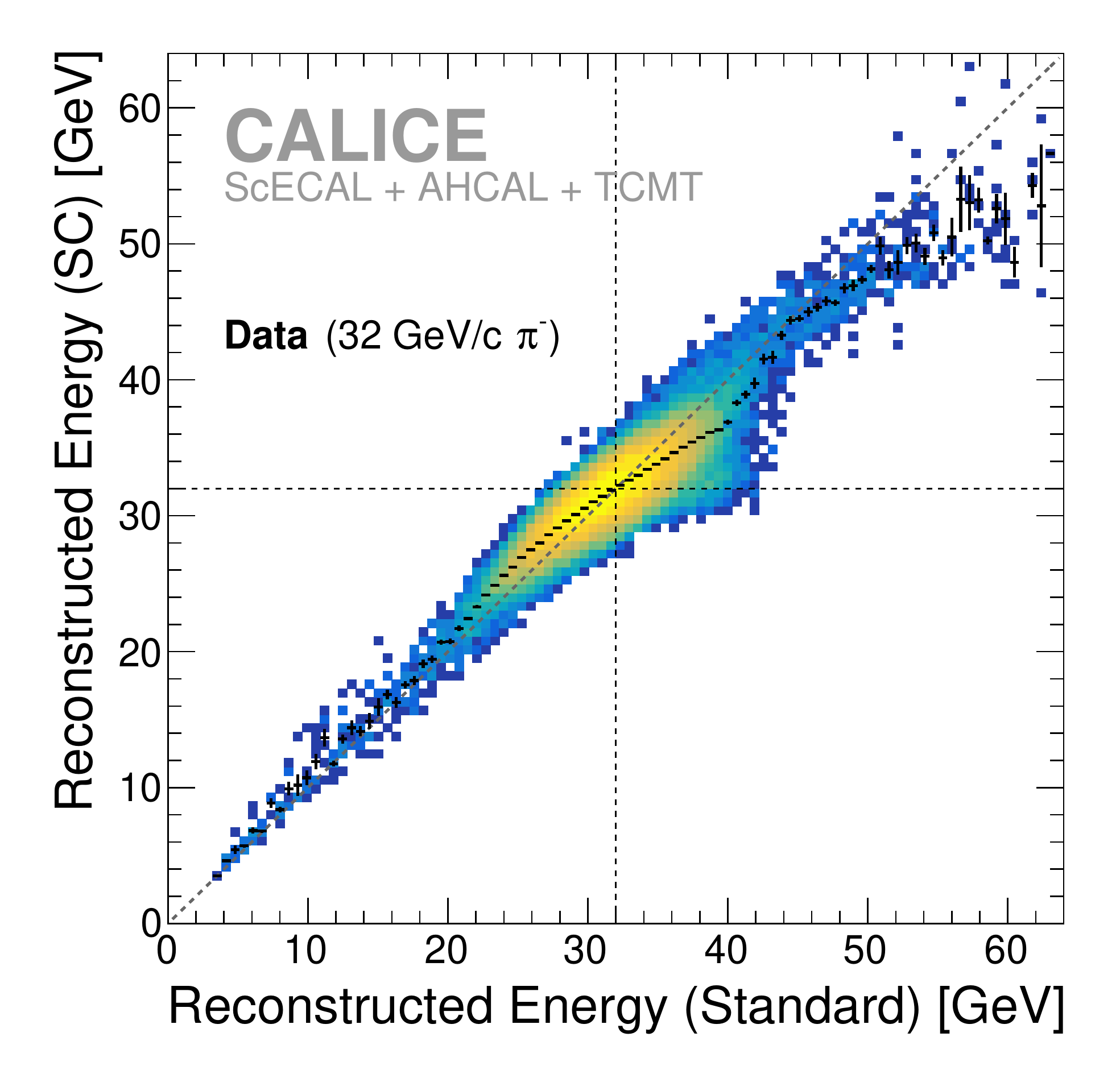}}
	
	\caption{Reconstructed energy distributions from standard and software compensation reconstructions for the 32\,GeV/c~\piminus data. The black markers in the correlation plot show the profile of the mean software compensation reconstructed energy for bins of the standard reconstruction energy (right). The black dashed lines in the correlation plot indicate the beam energy of the sample.}
	\label{fig:ClassicVsSC}
\end{figure}

Applying the weights shown in \autoref{fig:SCweightsData} to the dataset yields an improved energy resolution as shown in \autoref{fig:ClassicVsSC} (left) and \autoref{fig:additionalspectra}. Iterative applications of the software compensation reconstruction using the result of the previous iteration as $E_\text{est}$ do not further improve the energy resolution.  The correlation between standard and software compensation reconstruction in \autoref{fig:ClassicVsSC} (right) shows a clear non-linearity in the central part of the reconstructed energies, suggesting that events with a high hadronic fraction, and thus lower standard reconstructed energy, are shifted up in the software compensation reconstruction. Likewise, events with above average electromagnetic shower content, and thus too high standard reconstructed energy, are shifted down when reconstructed with the software compensation reconstruction.

\begin{figure}[htbp]
	{\includegraphics[width=0.5\textwidth]{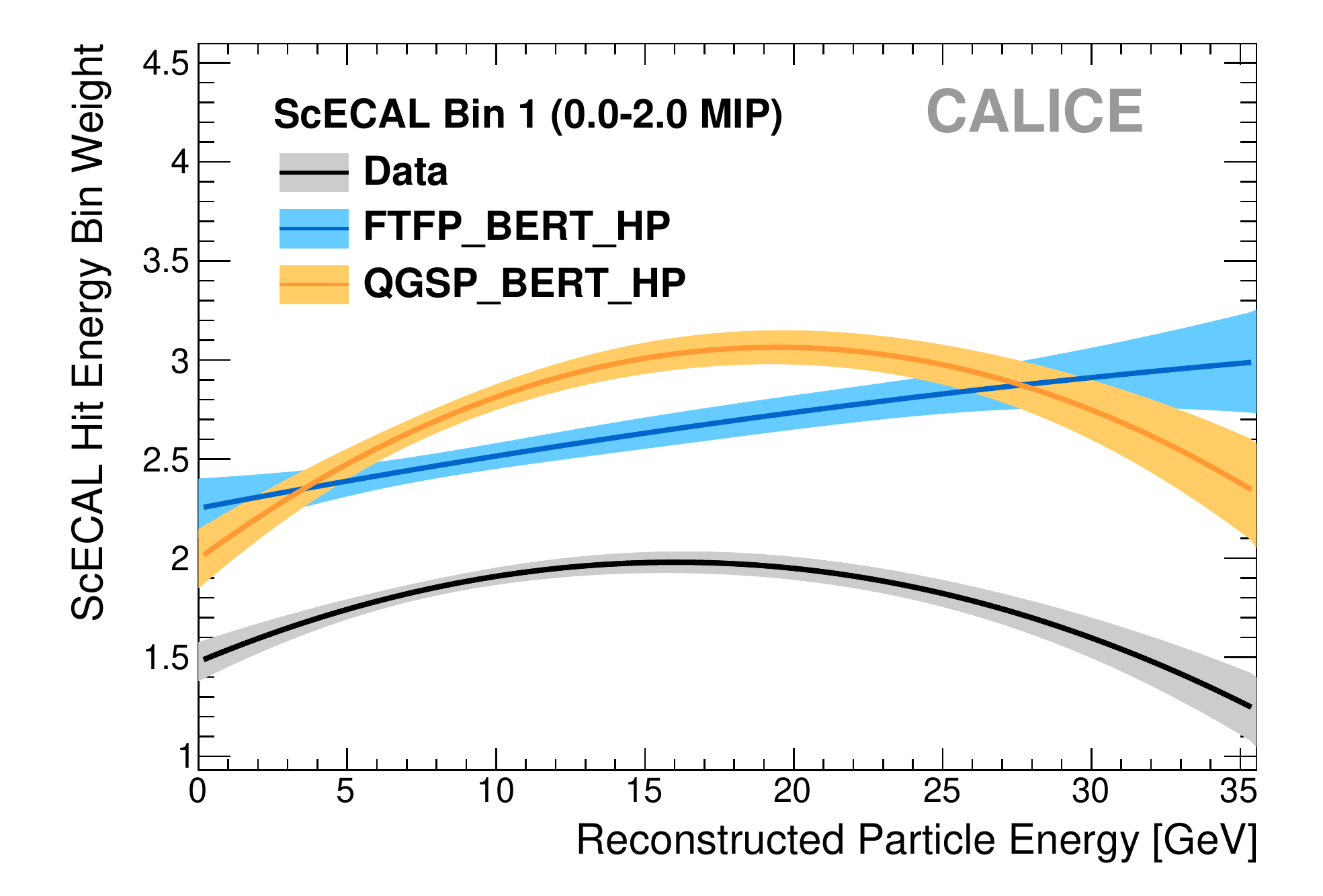}}\hfill
	{\includegraphics[width=0.5\textwidth]{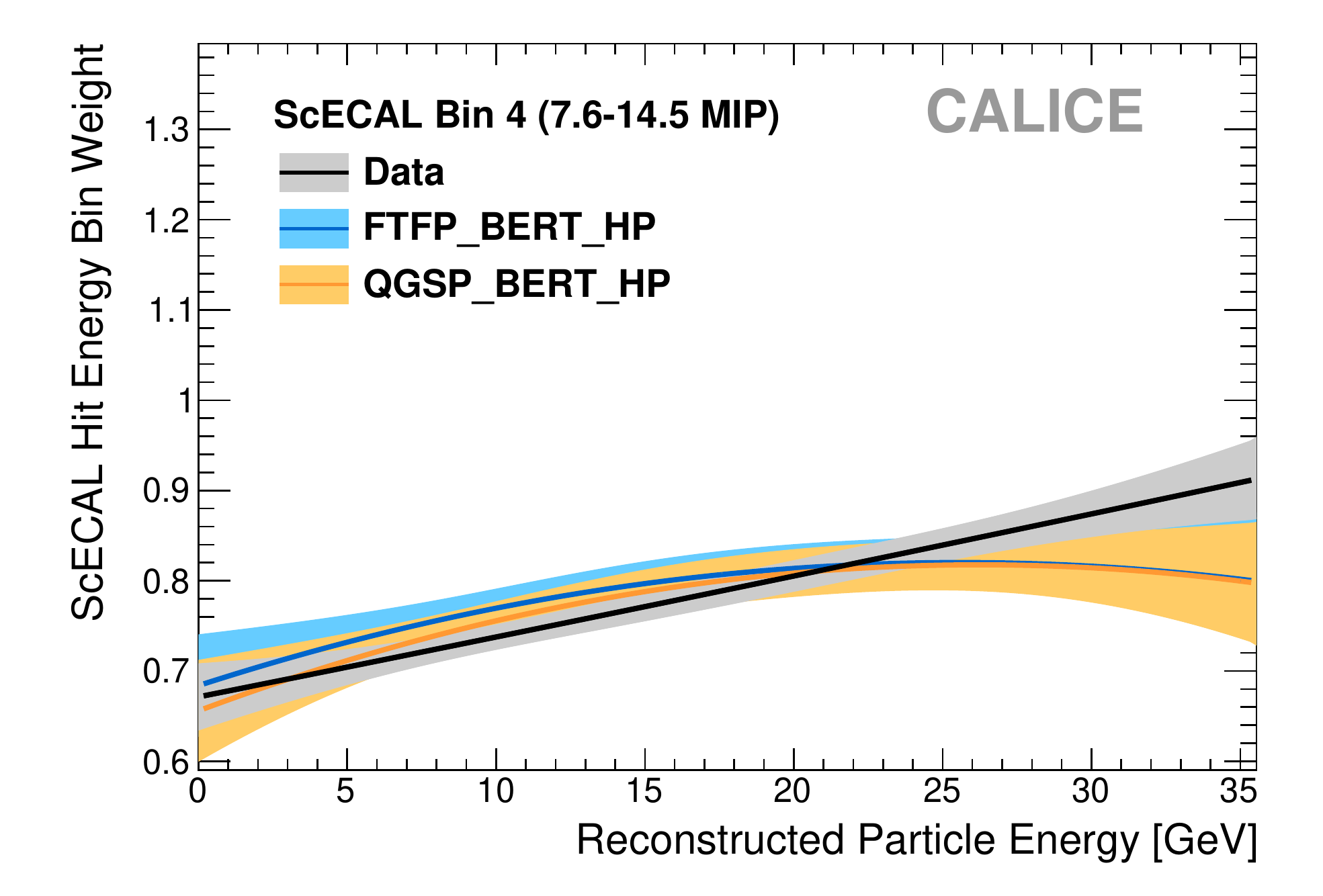}}
	{\includegraphics[width=0.5\textwidth]{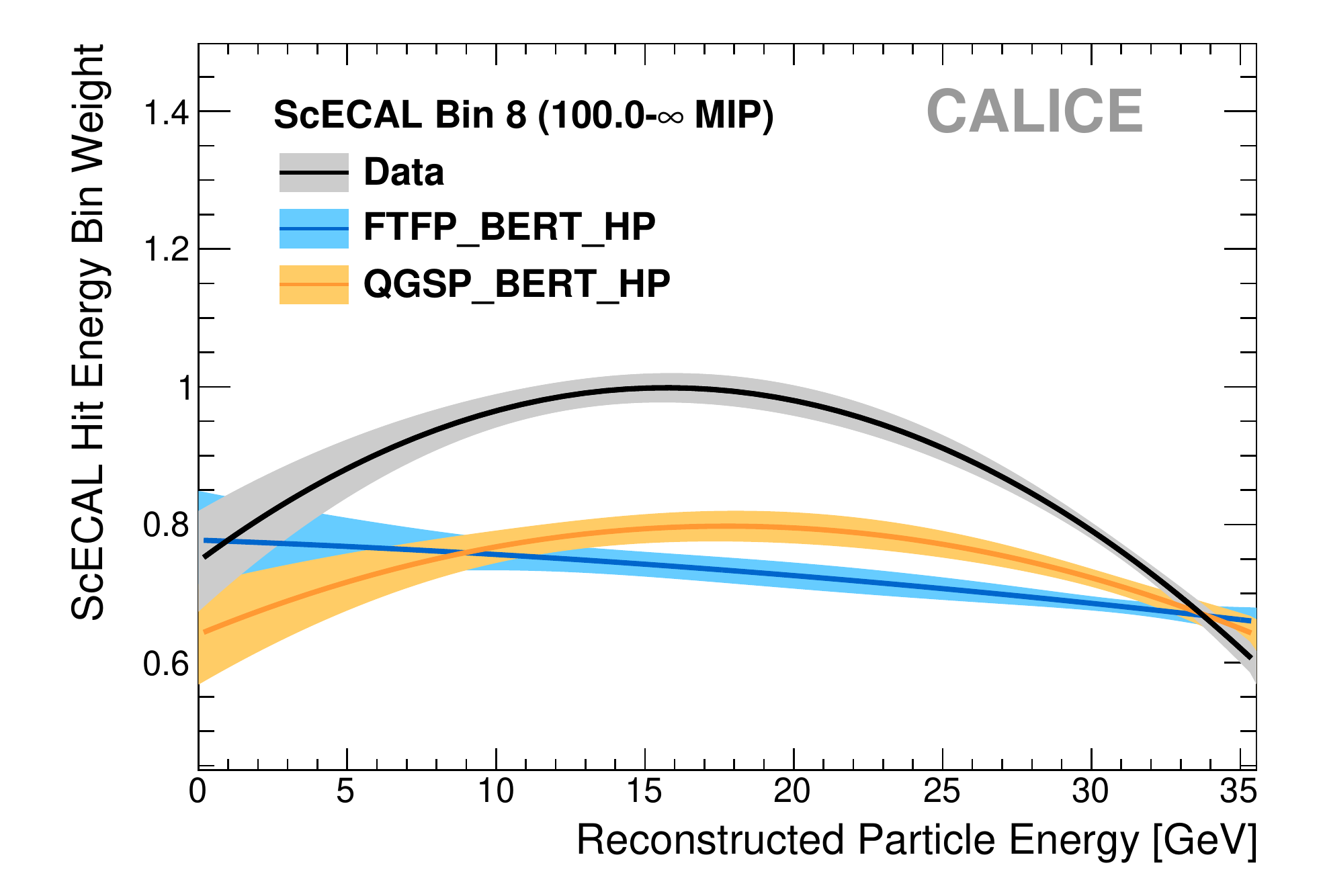}}\hfill
	{\includegraphics[width=0.5\textwidth]{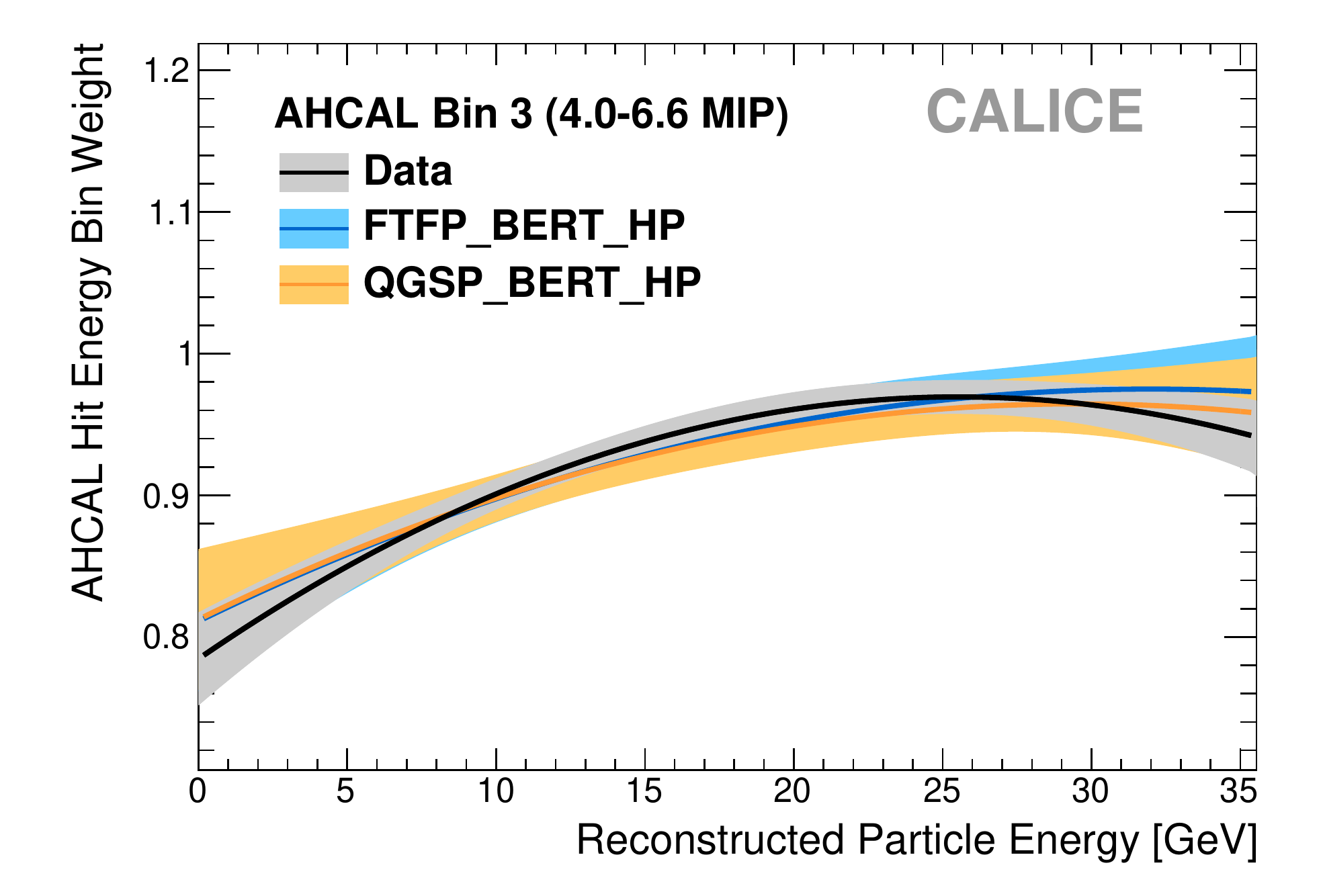}}
	
	\caption{Hit energy bin weights as a function of estimated particle energy for data events and different simulations. The width of each line indicates the weight uncertainty propagated from the parameter errors.
	}
	\label{fig:SCWeightsDataVsMC}
\end{figure}

The identical procedure of reconstructing energies and optimising the weight parameters is applied to simulated events. Individual bin weights as a function of estimated particle energy for data and simulation are shown in \autoref{fig:SCWeightsDataVsMC} for selected hit energy bins. The AHCAL shows reasonable agreement between weights derived from data and simulations in all hit energy bins. In the ScECAL discrepancies are seen especially in the two first hit energy bins and the highest hit energy bin, which also shows discrepancies between the used simulation physics lists. In most hit energy bins, both used simulation physics lists agree with each other within their parameter uncertainties. The TCMT weight also has a large discrepancy between data and simulations, although mostly for low beam momenta in which TCMT energy deposits are expected to be minimal.

\begin{figure}[htbp]
\begin{center}
	{\includegraphics[width=0.5\textwidth]{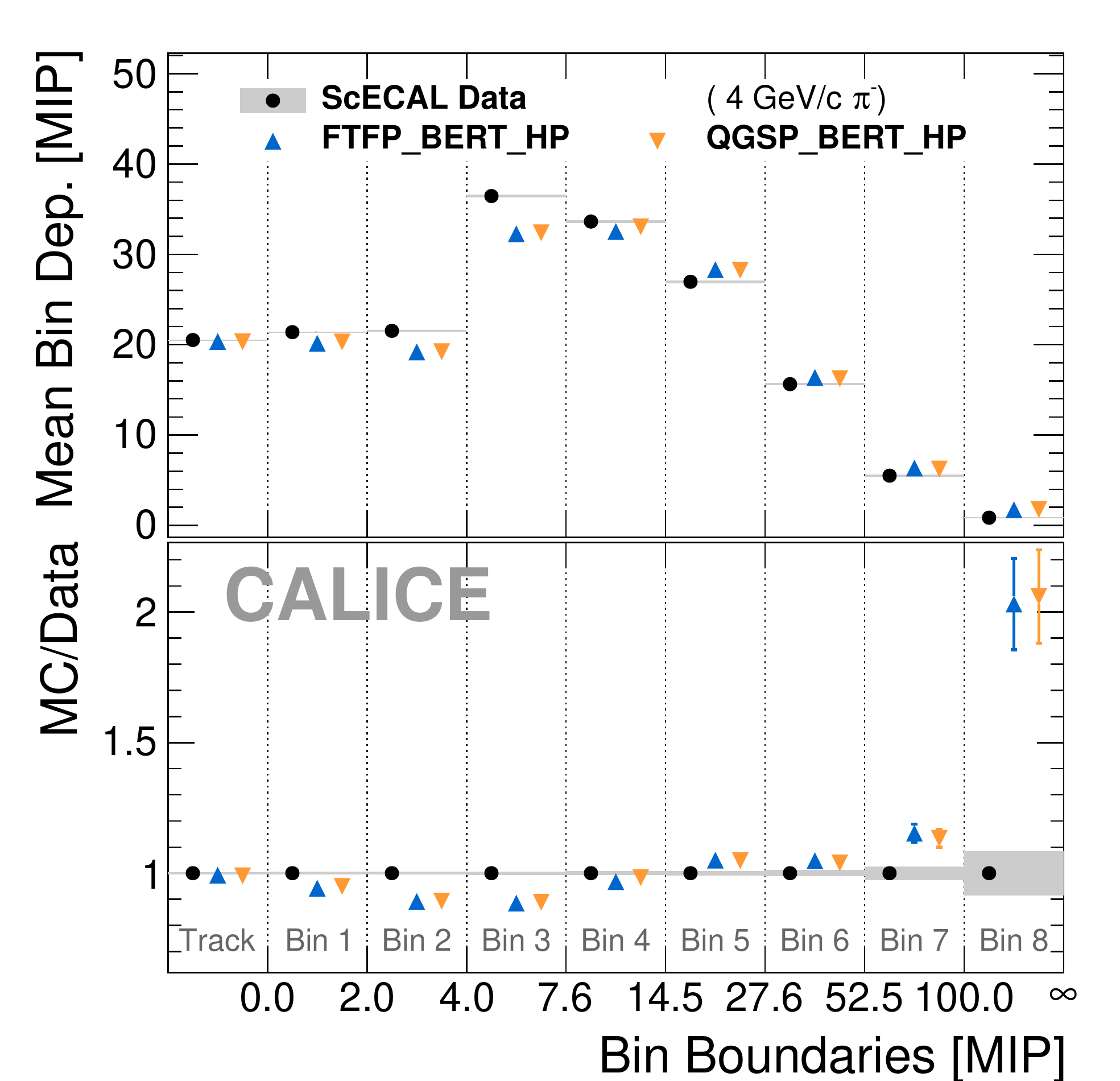}}\hfill
	{\includegraphics[width=0.5\textwidth]{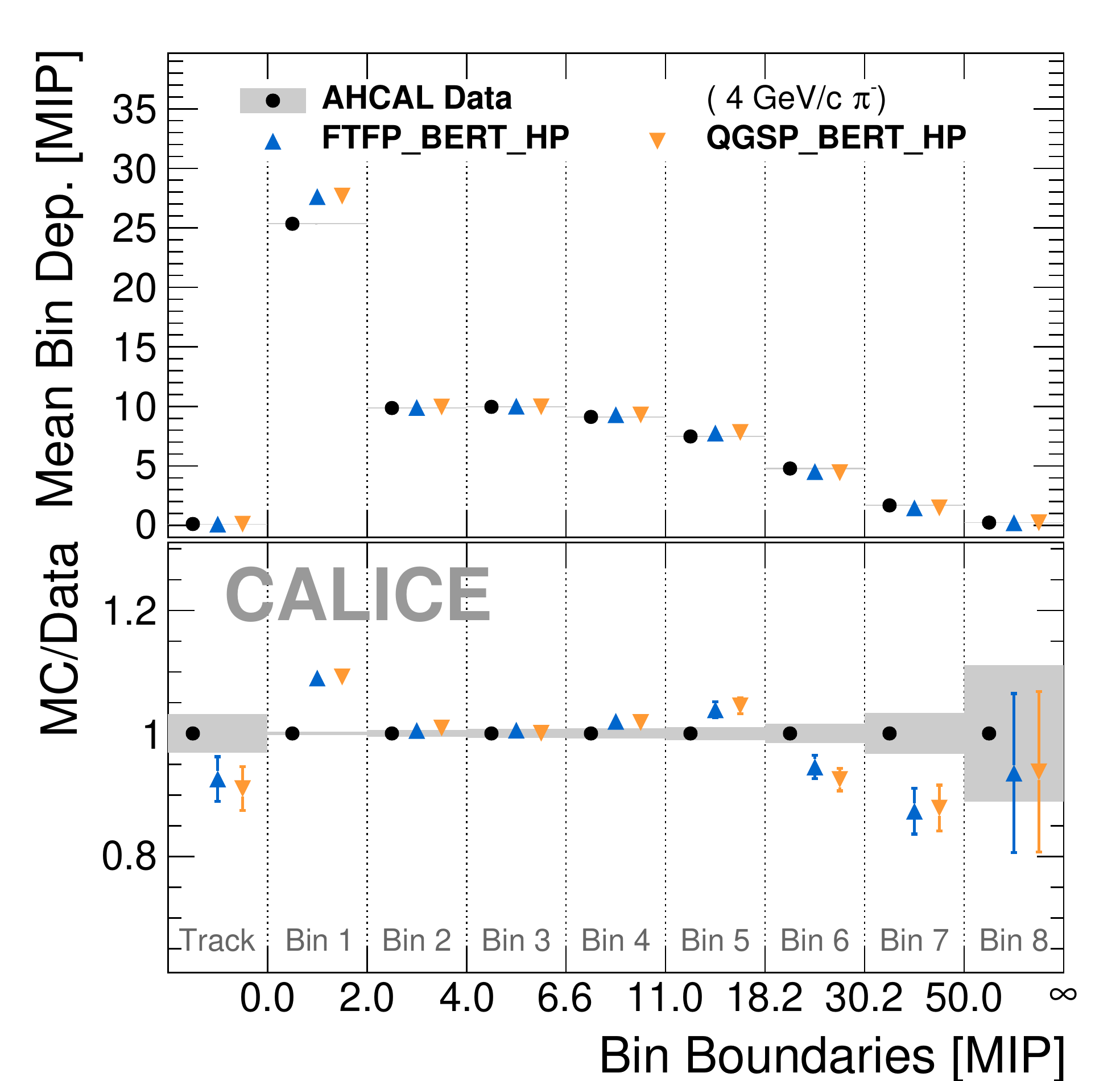}}
	{\includegraphics[width=0.5\textwidth]{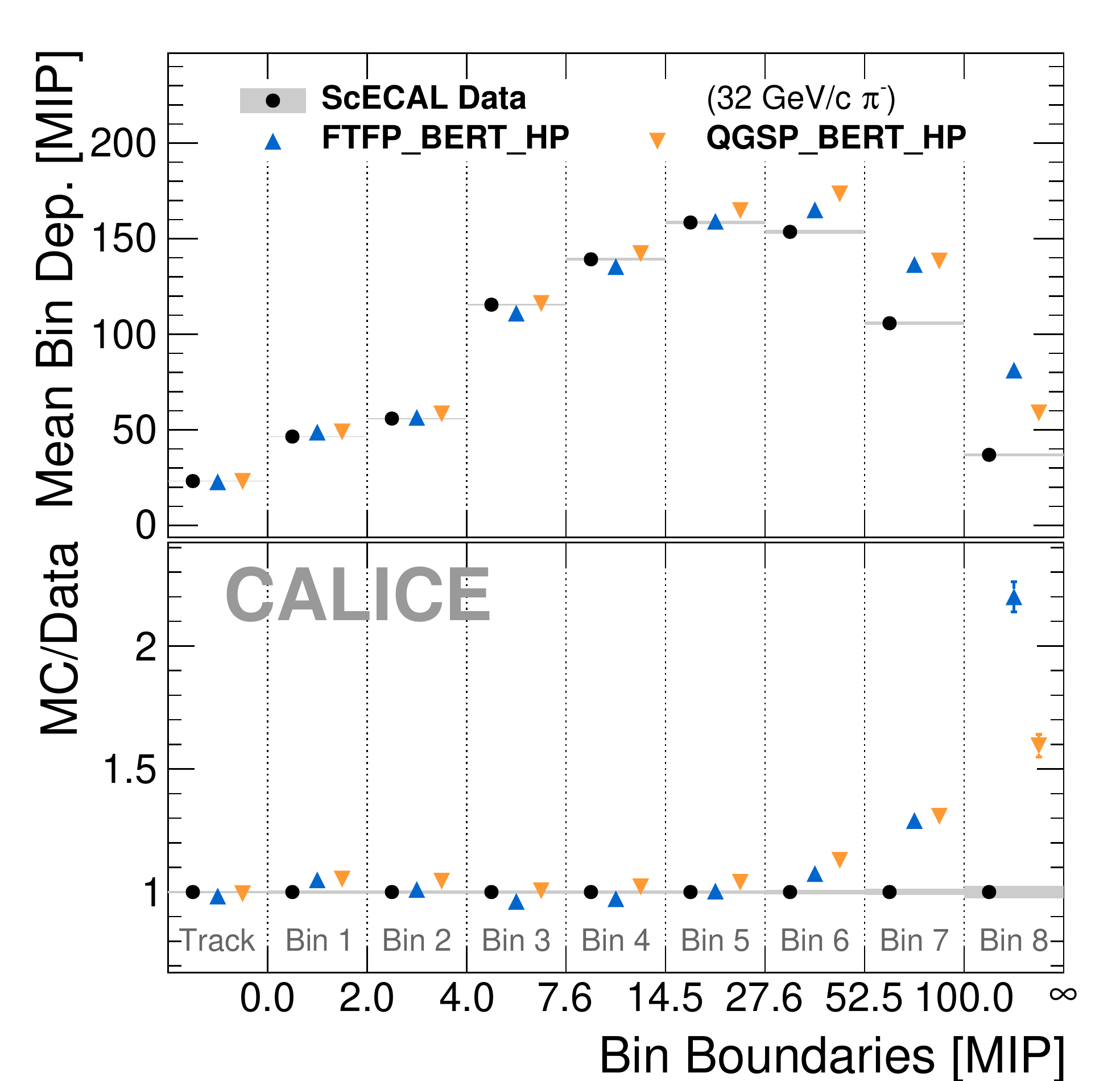}}\hfill
	{\includegraphics[width=0.5\textwidth]{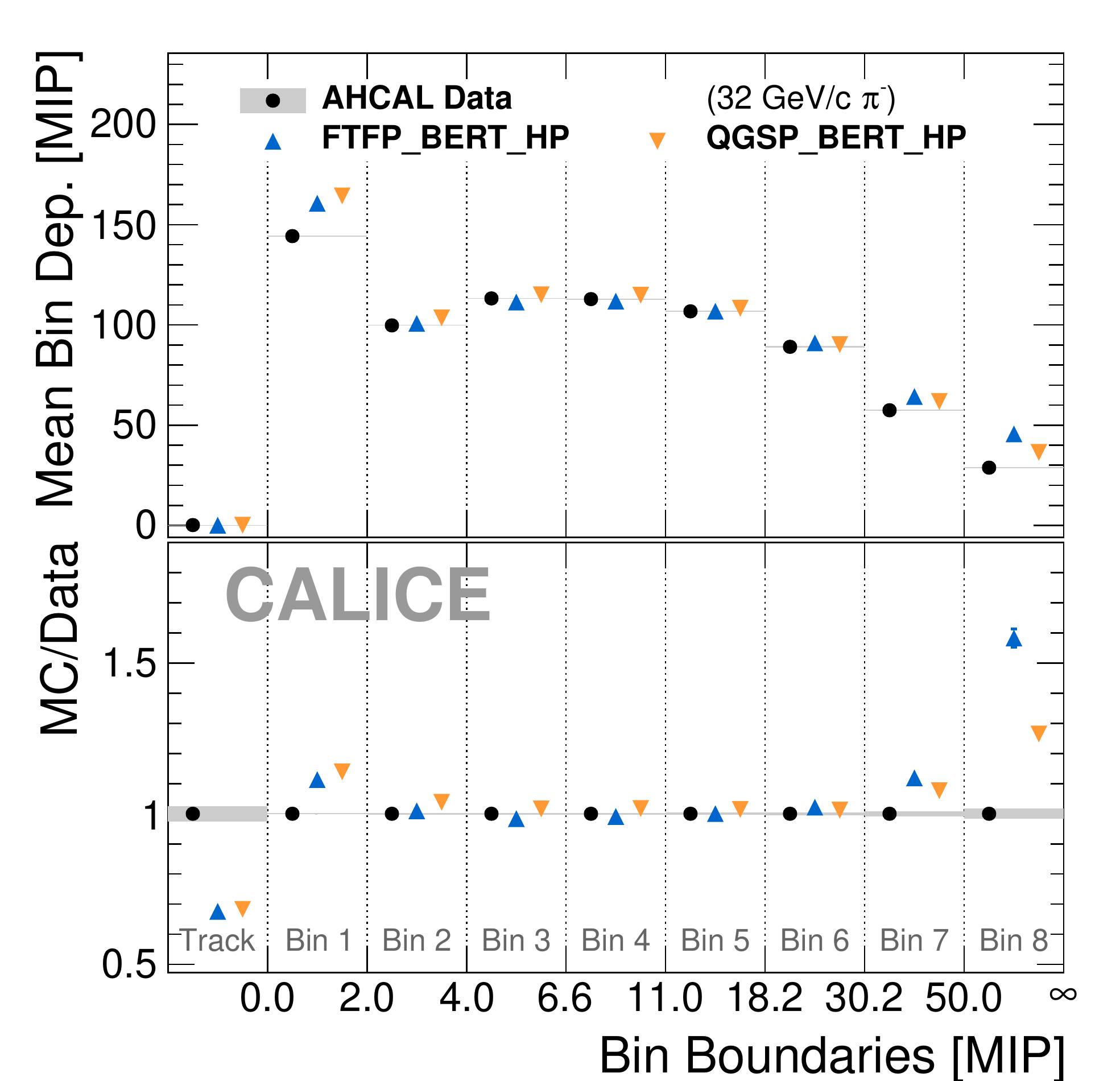}}
\end{center}
	\caption{Averaged energy sum per hit energy bin per event for data and simulated events in the ScECAL (left) and AHCAL (right) in \SI{4}{\GeV/c} (top) and \SI{32}{\GeV/c} (bottom) events. For most entries the statistical error is smaller than the used markers.}
	\label{fig:hitenergyBinsPion}
\end{figure}

The averaged summed energy deposit per event for each bin is investigated in order to better understand the observed differences between weights derived from data and simulated events, as shown in \autoref{fig:hitenergyBinsPion}. The highest hit energy bin in the ScECAL has around twice the mean energy deposit in the simulation compared to data, with significant differences evident from the sixth bin. This misdescription of the hit energy spectra especially for very high energy hits has also been observed in electromagnetic showers \cite{ScECALPaper}. For lower beam momenta, the ScECAL hit energy range around \SIrange{2}{8}{\mip} shows some underestimations in the hit energy spectra. In the AHCAL all bins show reasonable agreement between data and simulation, with overestimations up to \SI{50}{\percent} in the highest hit energy bin at the highest particle energy.

On their own, the observed differences in the mean hit energy per bin (and thus hit energy spectra) do not sufficiently explain the differences in the optimised software compensation weights. Although the highest ScECAL hit energy bin shows discrepancies between data and simulated events for all beam momenta, its bin weight is only different in the central part of the beam momenta. Likewise, the mean energy sum in the first ScECAL hit energy bin is well described for all energies, but the corresponding bin weights are not.

Shower simulation modelling effects could affect the optimised weights even in hit energy ranges where the observed hit energy spectra match reasonably well between data and simulation, as bin weights are necessarily anti-correlated to conserve the mean reconstructed energy. We thus assume the possibility that the observed discrepancies in the optimised software compensation weights is due to the remaining imperfect shower simulations, especially with the high-Z absorbers of the ScECAL and on the granularity scale studied here. However, the impact of these discrepancies on the resulting energy reconstruction is fairly small, as discussed further in \autoref{sec:sc_data_mc}.

\section{Systematic uncertainties}\label{sec:systematics}
We investigate systematic uncertainties from the following sources:
\paragraph{Energy calibration stability}{The calibration factors of individual detector cells depend on environmental conditions, which might change in between and also during individual runs. Imperfections in correcting for such factors typically shift the energy scale for the whole detector, as most cells will be influenced in the same way. Based on the results obtained from comparing muon calibration samples as shown in \autoref{fig:MIP}, as well as reconstructed MIP-like track segments within the studied data sets, we assign an uncorrelated systematic uncertainty of \SI{1}{\percent} to the energy scale for each separate beam momentum sample.}
\paragraph{Sensor saturation correction}{Varying the parameters used in the correction of the sensor saturation within their respective uncertainties does not significantly influence the optimised energy reconstruction weights and adds only negligible systematic uncertainties to the resulting energy resolution and response in both data and simulated events.}
\paragraph{Event selection}{Possible biases on the energy response and resolution introduced by the event selection as well as systematic uncertainties from varying the applied cuts are investigated by comparing a minimal selection and the full pion selection applied to simulated events. The minimal selection contains the lower and upper cuts on the FHI layer to roughly preserve sampling fractions and average containment between the selections. The bias on both the response and the energy resolution is $\lesssim1\%$ for all examined beam momenta and physics lists, as given in detail in \autoref{table:pion_selection_bias}. The resulting systematic uncertainties on the energy response are found to be negligible compared to the previously discussed energy scale uncertainties. The determined resolution bias is added to the corresponding systematic uncertainty of the simulated samples. In data, the arithmetic mean between the resolution biases extracted from both physics list is used as an additional systematic uncertainty.

\paragraph{Remaining data impurities}{The biggest systematic uncertainties on the results obtained from data stem from the remaining sample impurities as discussed in \autoref{sec:selection_impurities}. The resulting small biases are determined as described below, and corrected for. To account for the resulting additional uncertainty on the results, the magnitude of the correction is added to the respective data uncertainties in quadrature.

The influence of the remaining single electron event contamination is estimated from simulated single pion samples mixed with simulated single electron events according to the electron contamination fractions obtained from the template fits shown in \autoref{fig:contaminations} and \autoref{sec:selection_impurities}, and examining their influence on the fitted response and resolution. For \SI{4}{\GeV/c} pion events with a \SI{1.8}{\percent} electron contamination the mean bias on the fitted response is found to be \SI{+0.96}{\percent}, while the mean bias on the fitted resolution is determined as \SI{+1.65}{\percent} (relative to the fitted resolution). The low single electron event contamination of higher beam momentum samples does not lead to any significant systematic uncertainty.

In order to account for the impact of events with additional energy deposits as shown in \autoref{sec:selection_impurities}, we perform a toy-model study based on the simple signal and contamination model shown in \autoref{fig:contaminations}. As the true shape of the contamination contribution is not known, a range of contamination shapes is explored by scanning over a range of possible mean contamination energy values. For each fixed mean contamination energy, the combined toy model is fitted to the data points, with the width and relative fraction of the contamination contribution as free parameters. Reconstructed energy distributions are generated according to the combined model fit and their energy response and resolution extracted with the same fitting procedure as used in the full analysis. The biases introduced by the added contamination model are extracted by comparing the fit results to the true values of the assumed signal component. All biases obtained with different background shapes are averaged into a single number per beam momentum by weighting each iteration with the inverse reduced $\chi^2$ of the combined toy model fit, favouring models which result in a good agreement between the toy model and the data.

The resulting biases on the mean reconstructed energy are almost negligible, varying from \SI{0.20}{\percent} at \SI{12}{\GeV/c} to \SI{0.16}{\percent} in \SI{32}{\GeV/c} samples. The mean bias on the extracted energy resolution varies between \SIrange{0.5}{1.4}{\percent}. Since the method to determine the contamination fraction is not applicable to the \SI{4}{\GeV/c} data sample, the respective maximum values of all data samples are used as an estimate for the \SI{4}{\GeV/c} sample.}

\paragraph{ScECAL absorber material}{Pion event samples have been simulated using both ends of the range of possible ScECAL absorber compositions corresponding to effective radiation lengths of \SI{7.6}{\g/\cm^2} and \SI{7.9}{\g/\cm^2} \cite{ScECALPaper}. The difference is found to be negligible for all of the hadron shower observables discussed in this paper.}

\paragraph{ScECAL hit energy spectrum}{In order to assess the systematic uncertainties due to the mis-description in the tails of the hit energy spectra in simulated electron showers in the ScECAL \cite{ScECALPaper}, a full set of additional simulations with an  altered effective Moli\`ere radius in the ScECAL\footnote{The effective Moli\`ere radius is increased by artificially doubling the air gap between the ScECAL layers.} is used \cite{Hartbrich}. All software compensation weights are reoptimised on this set of simulated samples, yielding weights with small but significant differences to the default simulation\footnote{The actual difference in response and resolution between applying both sets of weights is negligible.}. Comparing fits to the resulting reconstructed energy spectra of the modified simulation events to the default simulation yields no significant difference in response and a small but significant difference in resolution in the range of \SIrange{2.0}{3.5}{\percent} for the standard energy reconstruction and \SIrange{0.1}{2.0}{\percent} with the software compensation energy reconstruction. The observed differences in resolution are used as the systematic uncertainties on the simulation results. The influence on the longitudinal pion shower profiles is found to be negligible.}

\paragraph{}{In summary, the most important sources of systematic uncertainties of the results presented here are the uncertainty in the run-to-run energy calibration, small biases due to the applied event selection, the data impurities of unclear origin remaining after the event selection, the remaining single electron contamination in the \SI{4}{\GeV/c} data sample as well as the general overestimation of the high energy tails in the ScECAL hit energy distributions in simulated events. A detailed breakdown of all numeric values is given in \autoref{table:systematics}.

\section{Results}
This section discusses longitudinal shower observables, energy reconstruction and linearity, and energy resolution of the full calorimeter system for standard reconstruction and software compensation reconstruction. The data are compared with simulations.
\subsection{Profiles}\label{sec:profiles}
An example of a distribution of the reconstructed layer of the first hadronic interaction in the ScECAL and AHCAL in pion selected data is shown in \autoref{fig:fhi_data_mc} for  data and both examined simulation models. No events with reconstructed FHI layer $\leq 4$ are present in the sample due to the applied pion event selection. The slight suppression of events with reconstructed FHI layer around \numrange{5}{8} is due to the clean primary track requirement in the event selection, the efficiency of which decreases for short primary tracks \cite{Hartbrich}.

The data are well described by the simulated samples within their statistical uncertainties, including the efficiency effect of the primary track selection step and the transition region between the ScECAL and the AHCAL. This demonstrates a good material description of the calorimeter setup, a similar performance of the event selection in data and simulated events, and no significant remaining single electron contamination in the dataset.

\begin{figure}[htbp]
\begin{center}
\includegraphics[width=0.8\textwidth]{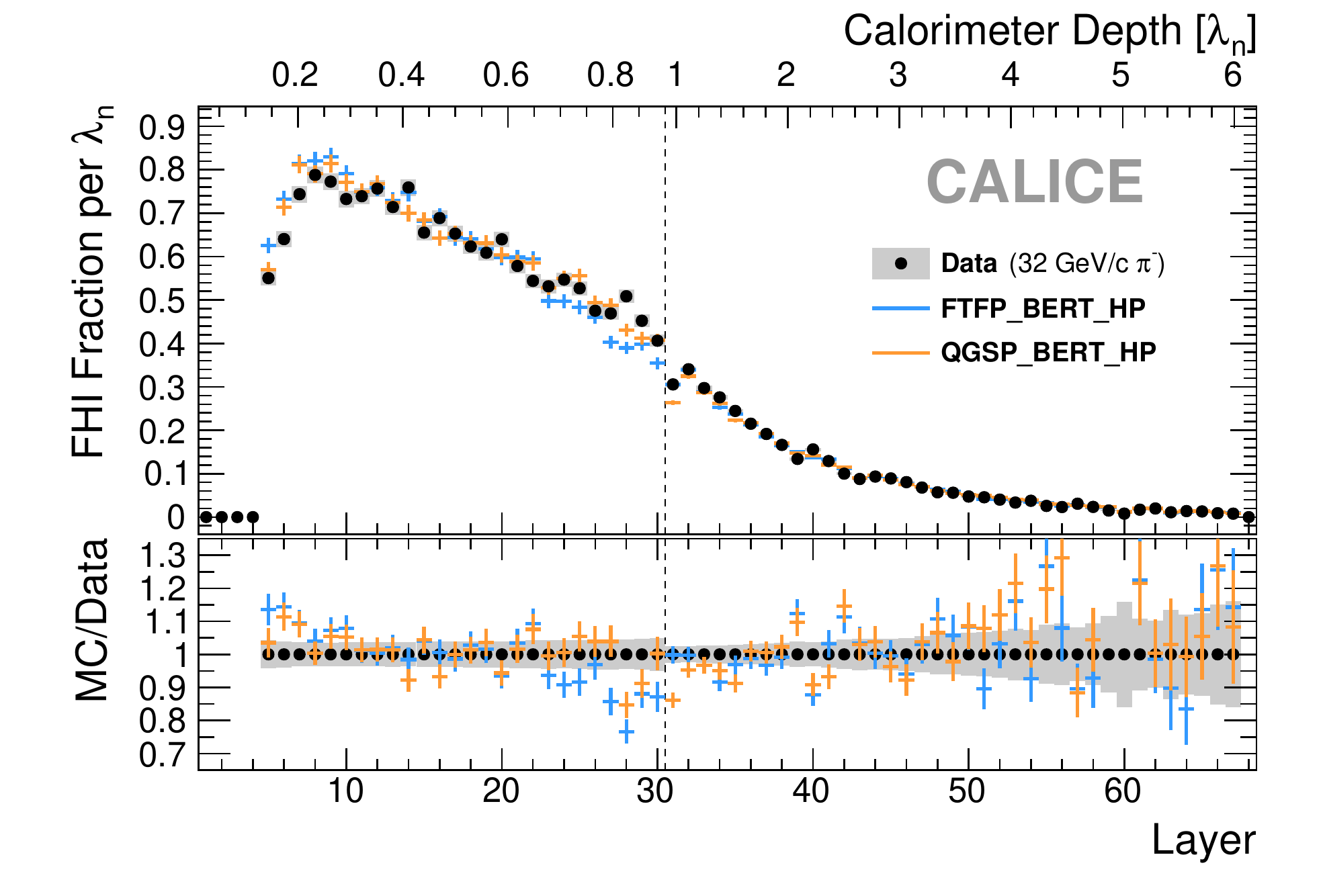}
\caption{Distribution of the reconstructed FHI layer for 32\,GeV/c \piminus events in data and different simulation physics lists. Energy deposits in the ScECAL are shown in layers 1 to 30, the layers 31 to 68 belong to the  AHCAL.}
\label{fig:fhi_data_mc}
\end{center}
\end{figure}

The mean longitudinal profile of all pion shower events passing the event selection for \SI{32}{\GeV/c} beam momentum is shown in \autoref{fig:profile_long_full}. The low mean energy deposit in the first five layers is due to the event selection, similar to the effect seen in \autoref{fig:fhi_data_mc}. The low mean energy deposit in the last AHCAL layers points to good average shower containment even without including the TCMT layers. The general shape of the profile is reasonably well described by all physics lists, including the dip in responses in the ScECAL around the transition region between the ScECAL and AHCAL. Both studied simulation models overestimate the mean energy deposits by around \SI{5}{\percent} regardless of the beam momentum, as already noted in \autoref{sec:energyrecoclassic}. This difference is larger than the systematic uncertainty on the MIP calibration or saturation effects. 

\begin{figure}[htbp]
\begin{center}
\includegraphics[width=0.8\textwidth]{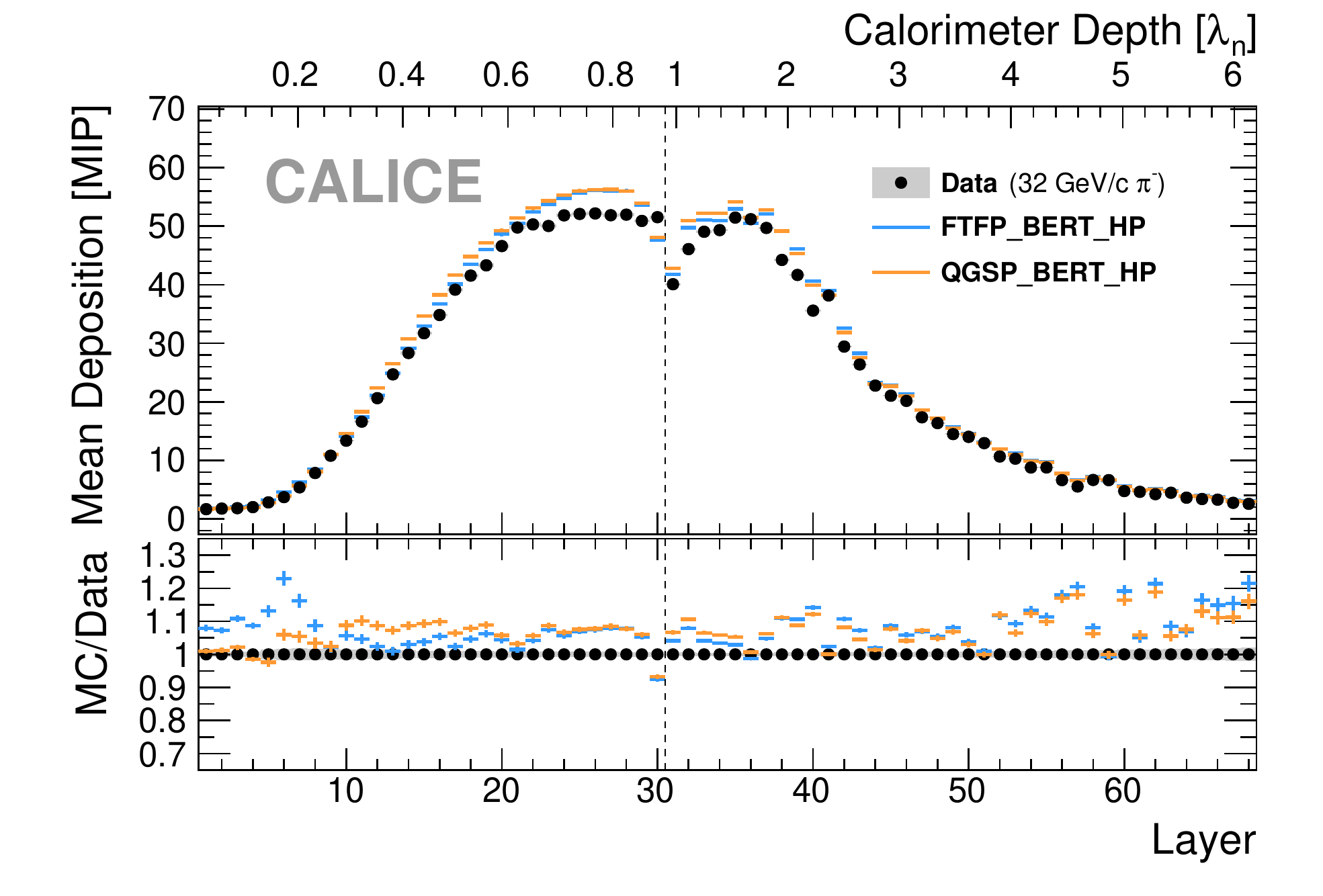}
\caption{Average longitudinal shower profile for 32\,GeV/c \piminus\ events in data and different simulation physics lists. Energy deposits in the ScECAL are shown in layers 1 to 30, the layers 31 to 68 belong to the  AHCAL.}
\label{fig:profile_long_full}
\end{center}
\end{figure}

\autoref{fig:profile_long_fhi} shows longitudinal shower profiles as a function of the distance from the shower starting point, taken as the reconstructed FHI layer, separately for showers starting in the ScECAL and in the AHCAL. For events with a reconstructed shower start in the ScECAL, the shower profile predictions by the FTFP\_BERT\_HP and QGSP\_BERT\_HP simulations are quite similar for the lower studied beam momenta, but vary by up to \SI{20}{\percent} at the higher beam momenta. Compared to data, energy deposits are under and overestimated in different layers for both tested physics lists. QGSP\_BERT\_HP simulations seem to generally produce more strongly peaked shower profiles, while FTFP\_BERT\_HP simulations generate slightly wider shower profiles than measured in data. Most of the data points lie between the predictions of these two physics lists used. The overestimation of energy deposits in the reconstructed shower start layer by the FTFP\_BERT\_HP simulations also explains the peaks in the energy deposits around layer 5, as seen for this physics list in \autoref{fig:profile_long_full}. 
For events in which the reconstructed shower start is located in the AHCAL the examined simulation models agree well with each other. 
Both simulation models show around \SI{10}{\percent} underestimated depositions in the first \SI{0.5}{\LambdaN}, with an overestimation of \SIrange{10}{20}{\percent} in the deeper layers. 
The agreement between data and simulation is on a similar level, with similar features in the MC/data ratio, as a comparable study performed on data taken with the AHCAL only \cite{MarinaDecomposition}.

\begin{figure}[htbp]
\begin{center}
	{\includegraphics[width=0.5\textwidth]{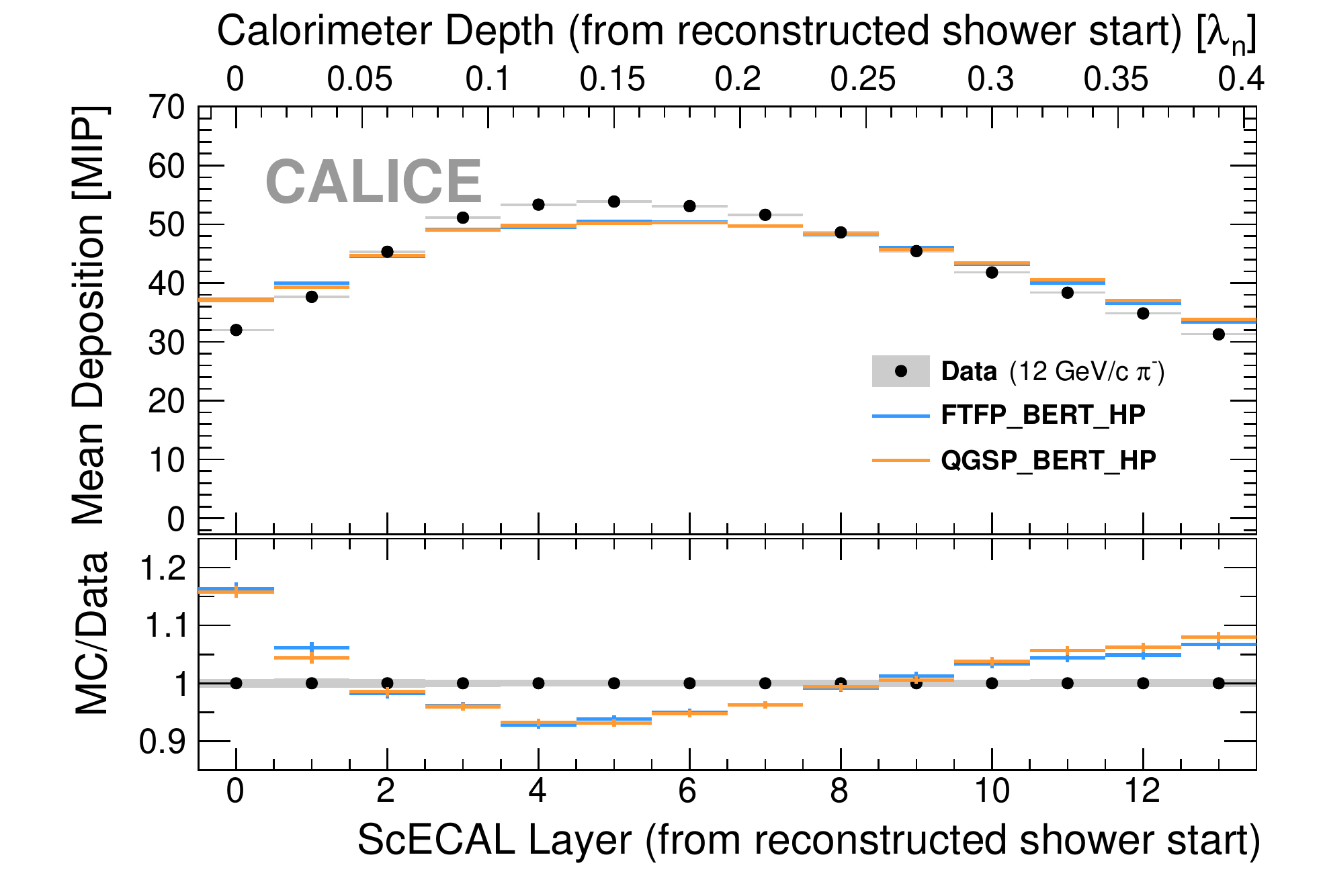}}\hfill
	{\includegraphics[width=0.5\textwidth]{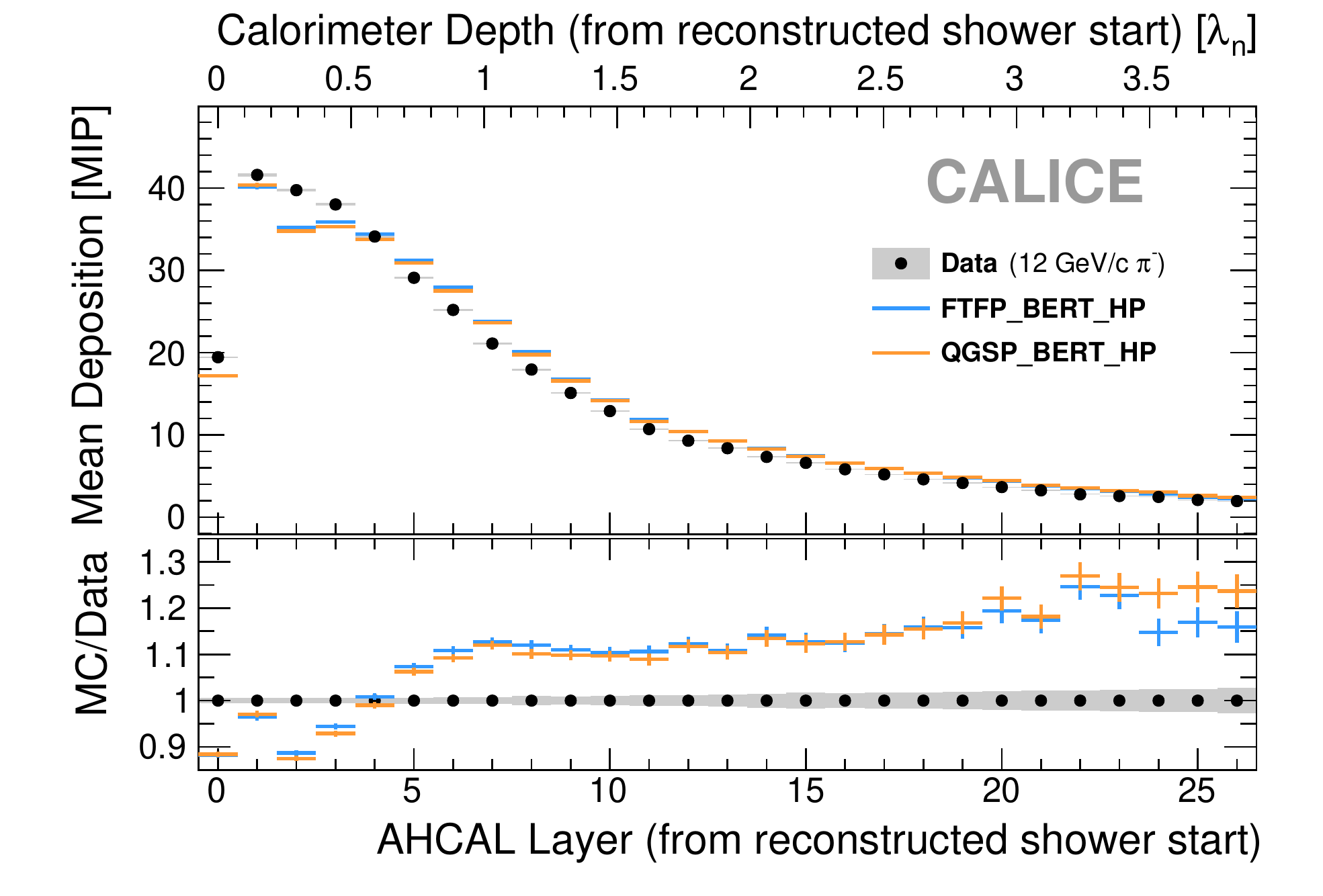}}
	{\includegraphics[width=0.5\textwidth]{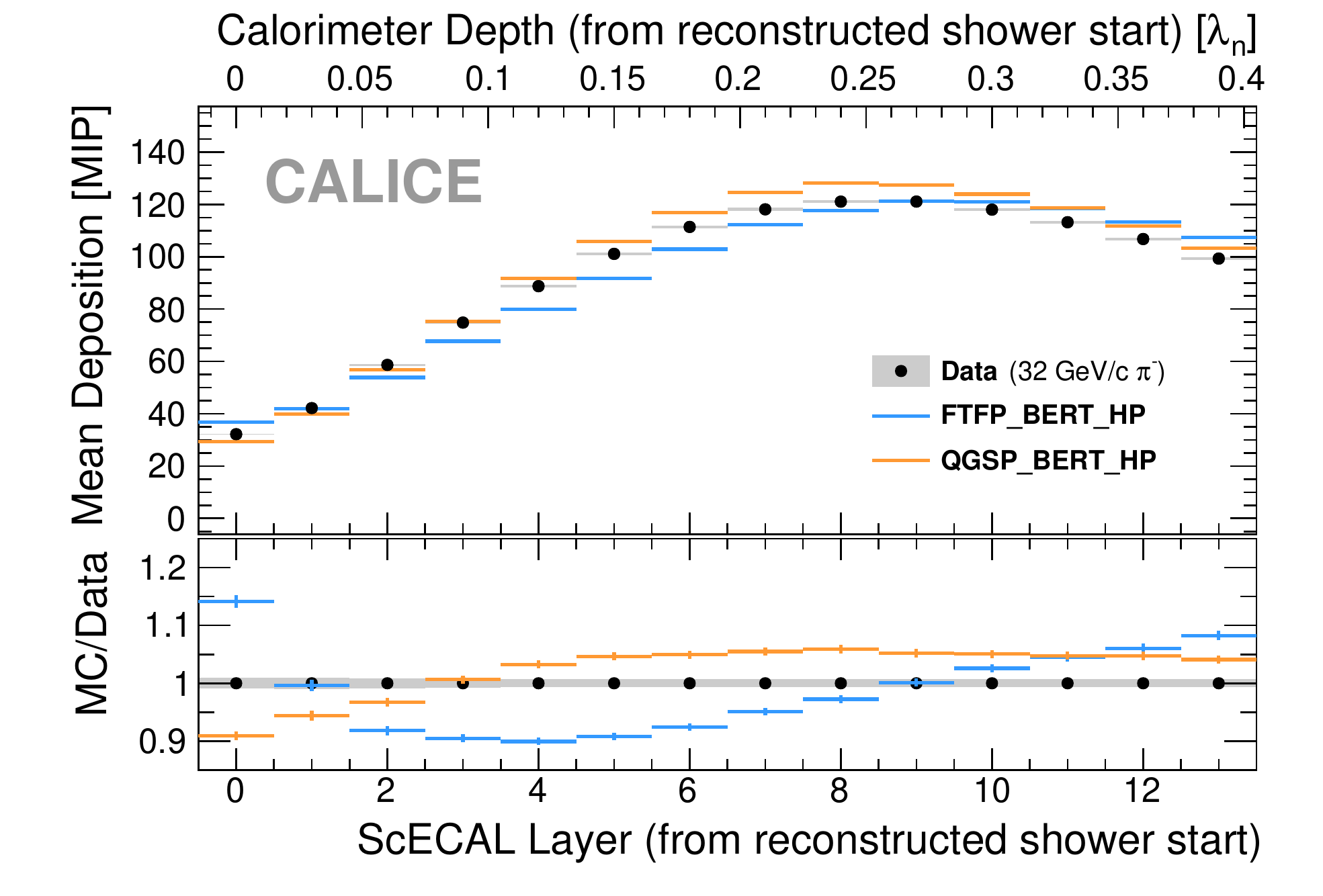}}\hfill
	{\includegraphics[width=0.5\textwidth]{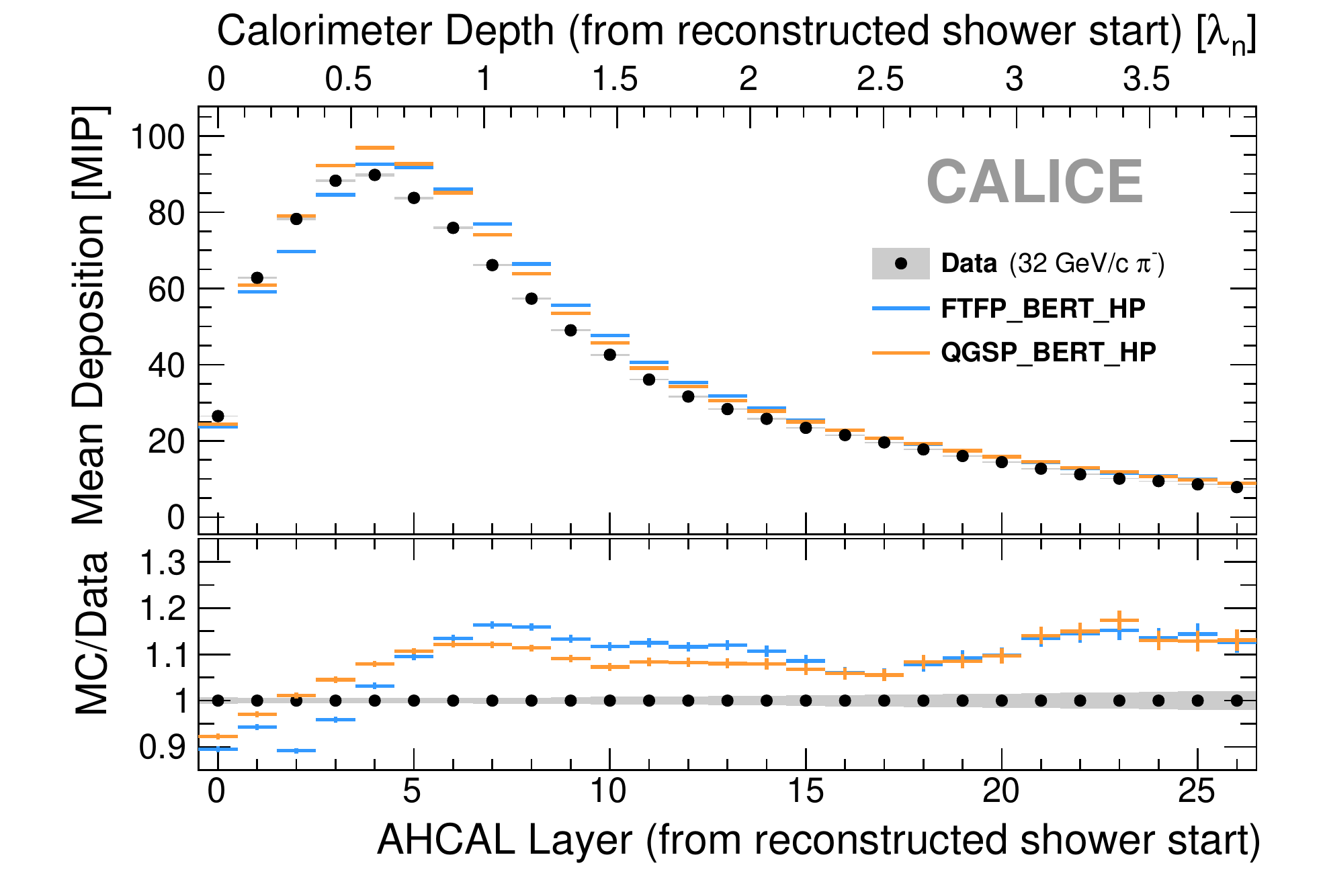}}
\end{center}
	\caption{Average longitudinal shower profiles for 12\,GeV/c (top) and 32\,GeV/c (bottom) \piminus\ events in data and different simulation physics lists for showers starting in ScECAL layer 5--16 (left) and AHCAL layer 0--10 (right). Energy deposits are plotted as function of distance to reconstructed FHI layer.}
	\label{fig:profile_long_fhi}
\end{figure}

\subsection{Energy reconstruction and linearity}
For each event the particle energy is reconstructed with the techniques described in \autoref{sec:energyreco}, without use of the known beam momentum.
The energy reconstruction of all data samples and simulations is performed using the weighting parameters optimised from their own datasets. 
The ratio of mean contributions to the reconstructed energy in the ScECAL and AHCAL $r = \nicefrac{E_\text{rec}^\text{AHCAL}}{E_\text{rec}^\text{ScECAL}}$ varies from around unity at \SI{4}{\GeV/c} to around three at \SI{32}{\GeV/c}. 

The spectrum of reconstructed energies for each sample is fitted with a Novosibirsk function \cite{Novosibirsk}, of which the mean response and resolution are calculated using Monte-Carlo integration \cite{Hartbrich, Neubueser}. 
The systematic uncertainties discussed in the previous section are added to the generally small statistical uncertainties on the fitted parameters. 

The mean standard reconstructed energies in data agree very well with the beam energy with all deviations, defined as $\frac{E_\text{rec}-E_\text{beam}}{E_\text{beam}}$, smaller than \SI{3}{\percent} as shown in \autoref{fig:pion_linearity_standard}. 
The reconstructed energy observed in simulations shows a small non-linearity, especially towards the lower beam momenta, with no deviation from the beam energy exceeding \SI{5}{\percent}, in qualitative agreement with the expectation of an undercompensating calorimeter. Even though the \SI{4}{\GeV/c} data point  has the largest assigned uncertainties overall, it does differ significantly from the simulated samples. 

When reconstructing the particle energy using the software compensation scheme described in \autoref{sec:energyrecosc}, the linearity of both data and simulated events is better than \SI{3}{\percent}, see \autoref{fig:pion_linearity_sc}. The \SI{4}{\GeV/c} data point is well aligned with the simulated results, due to the additional non-linear degrees of freedom in the software compensation reconstruction.

\begin{figure}[htbp]
\begin{center}
	{\label{fig:pion_linearity_standard}\includegraphics[width=0.5\textwidth]{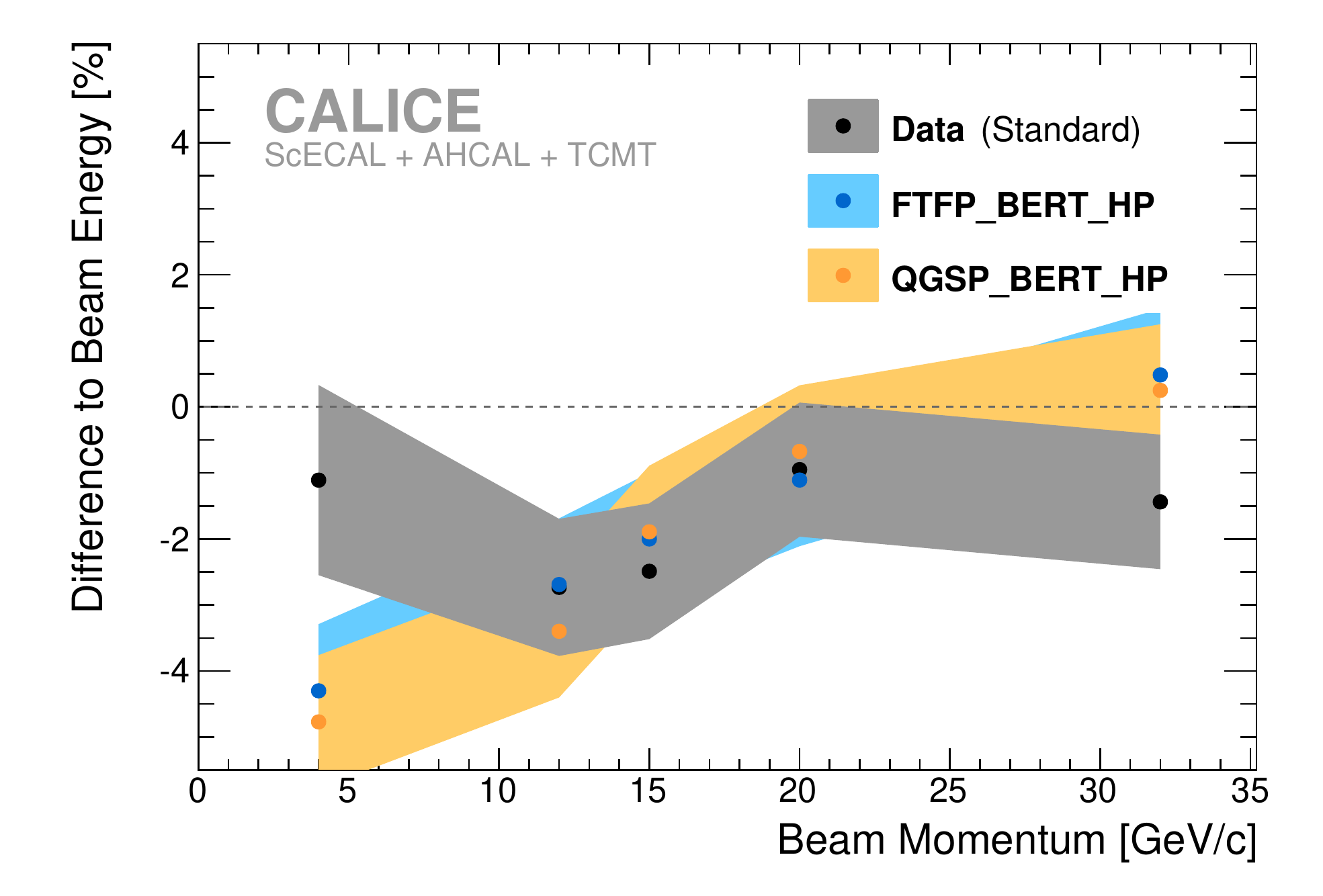}}\hfill
	{\label{fig:pion_linearity_sc}\includegraphics[width=0.5\textwidth]{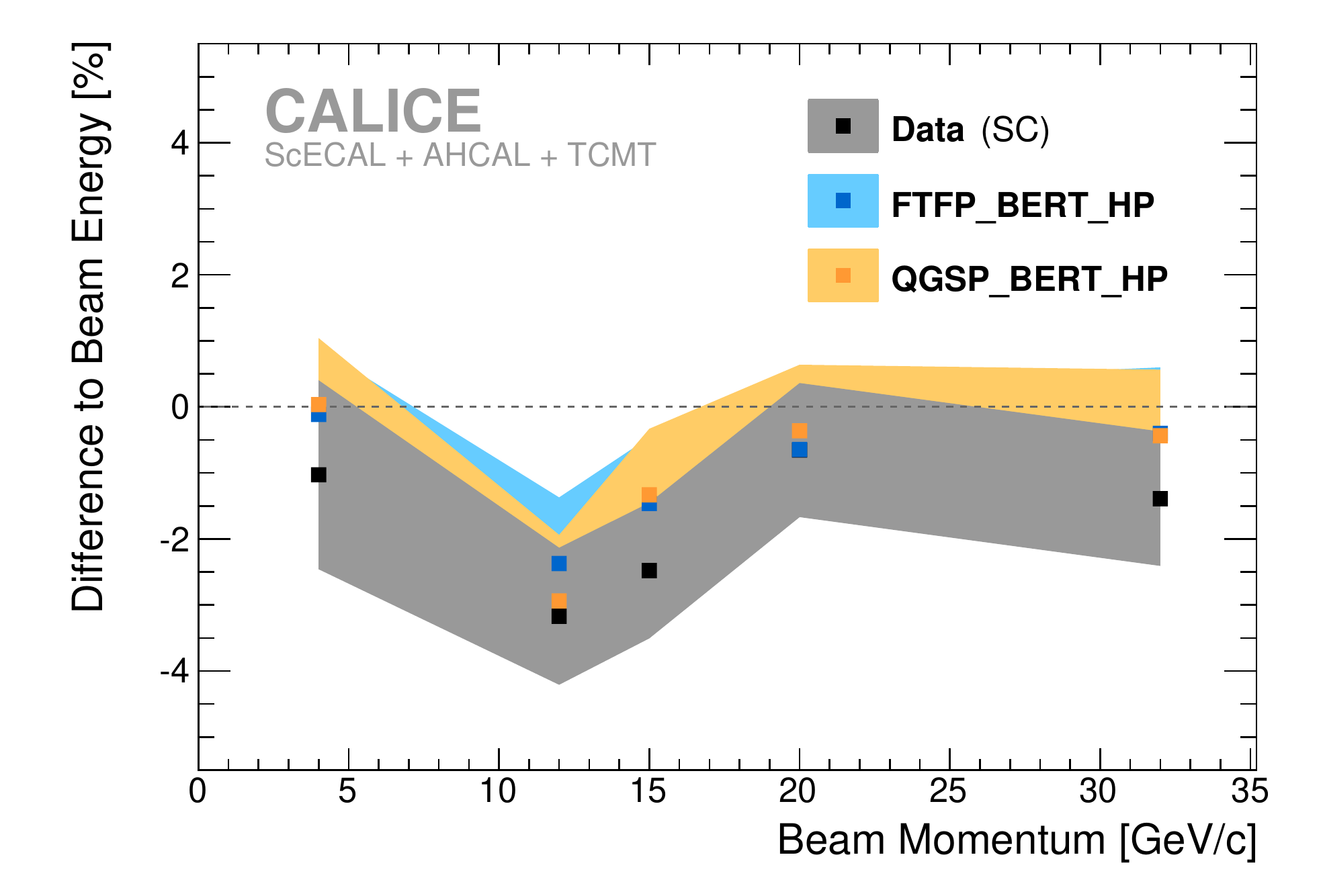}}
\end{center}
	\caption{Residual of mean fitted reconstructed energy over beam momentum in data and different simulation physics lists using the standard reconstruction (left) and software compensation reconstruction (right). The given residual is defined as $\frac{E_\text{rec}-E_\text{beam}}{E_\text{beam}} \times \SI{100}{\percent}$. }
	\label{fig:pion_linearity}
\end{figure}

\subsection{Energy resolution}
The energy resolution for each sample is calculated from the ratio of the width and mean of the Novosibirsk function fitted to the reconstructed energy spectra. For the standard energy reconstruction, the data energy resolutions are generally well described by the simulations as shown in \autoref{fig:pion_resolution_plots}. QGSP\_BERT\_HP produces a slightly better agreement with data compared to the FTFP\_BERT\_HP samples, which show a small deviation in the highest energy point. The software compensation reconstruction leads to a relative improvement of the data resolutions between \SI{10}{\percent} and \SI{20}{\percent} (\SI{2}{\percent} and \SI{3}{\percent} in terms of absolute resolution) depending on the beam momentum, see \autoref{fig:pion_resolution_plots}. The software compensation resolution obtained from simulated samples agrees between the physics lists, but significantly overestimates the achievable resolution improvements at all energies by \SIrange{5}{10}{\percent}.  All extracted resolutions, including the final combined uncertainties on each point are listed in \autoref{table:pion_reso_values}.

\begin{figure}[htbp]
\begin{center}
	{\includegraphics[width=0.5\textwidth]{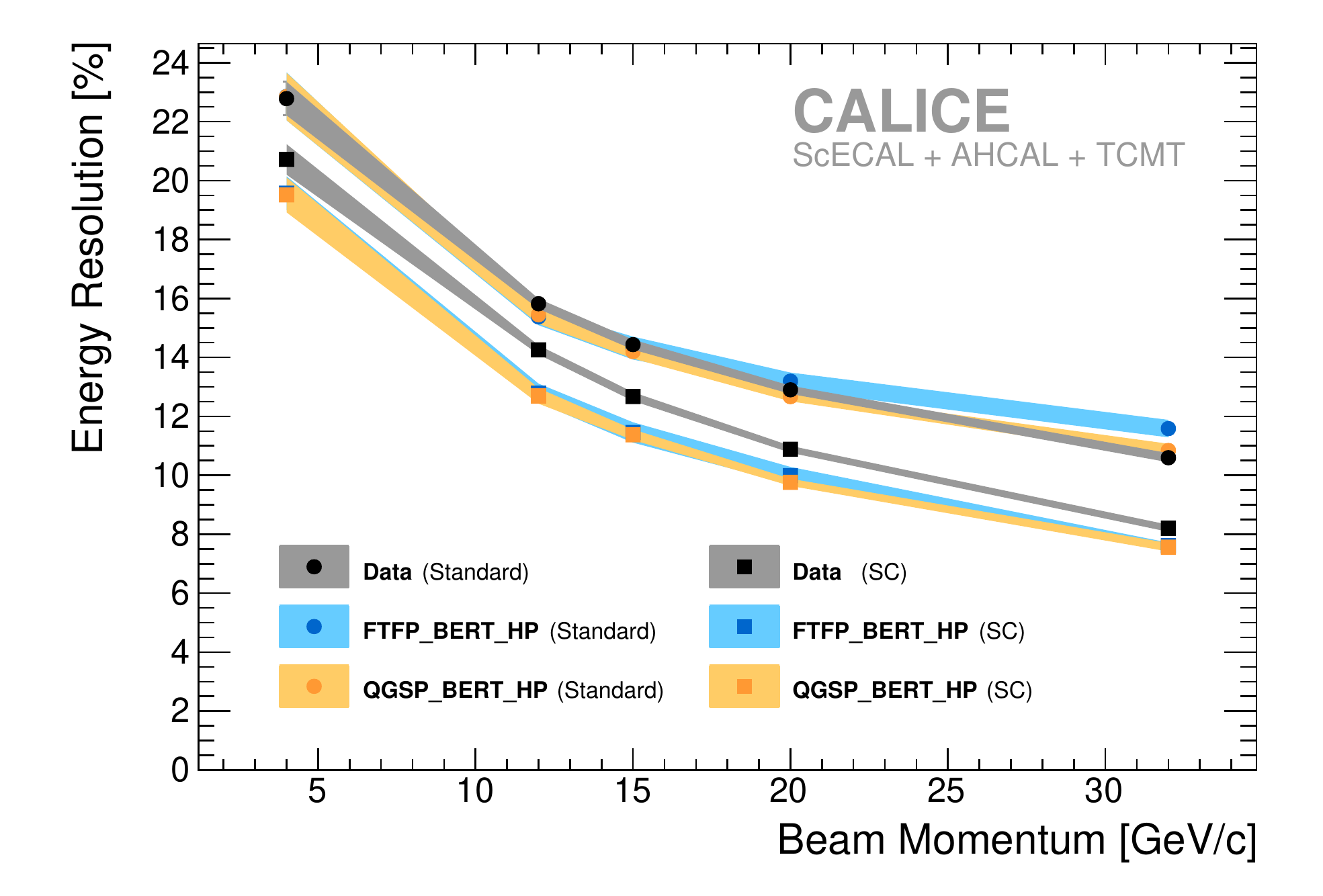}}\hfill
	{\includegraphics[width=0.5\textwidth]{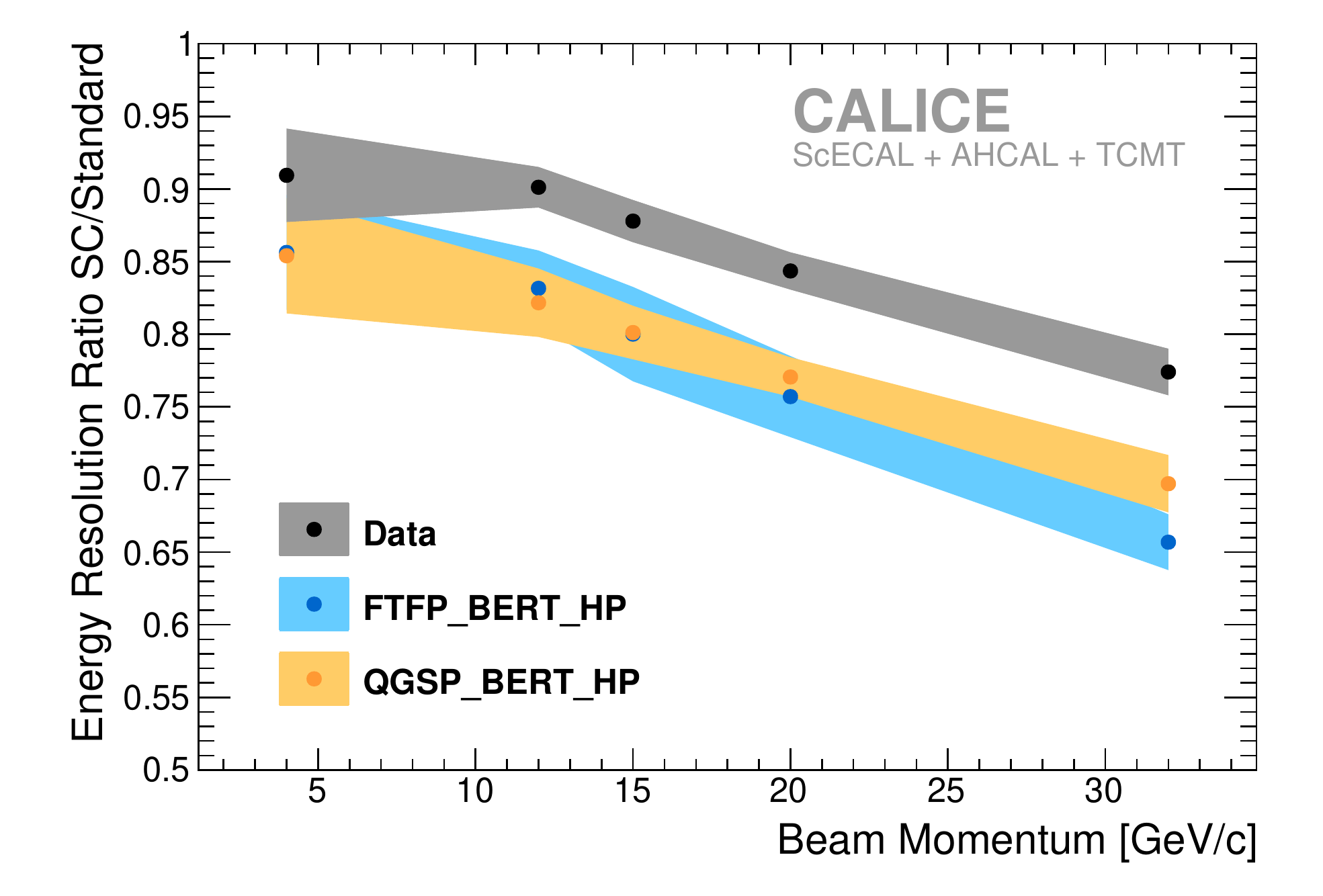}}
\end{center}
	\caption{Reconstructed energy resolution as a function of beam momentum in data and different simulation physics lists for standard and software compensation reconstruction (left), as well as the resolution improvement from the software compensation reconstruction (right). All plotted values are given in \autoref{table:pion_reso_values}.}
	\label{fig:pion_resolution_plots}
\end{figure}

\autoref{fig:pion_resolution_vs_ahcal} compares the energy resolution obtained from data samples in this analysis to the energy resolutions obtained from a previous analysis in which only pion showers with a reconstructed shower start in the AHCAL are considered \cite{SCPaper}. The resolutions obtained for both analyses are in reasonable agreement, indicating that the  combined calorimeter system with varying absorber materials and sampling fractions maintains the good single pion energy resolution of the standalone AHCAL.

\begin{figure}[htbp]
\begin{center}
\includegraphics[width=0.8\textwidth]{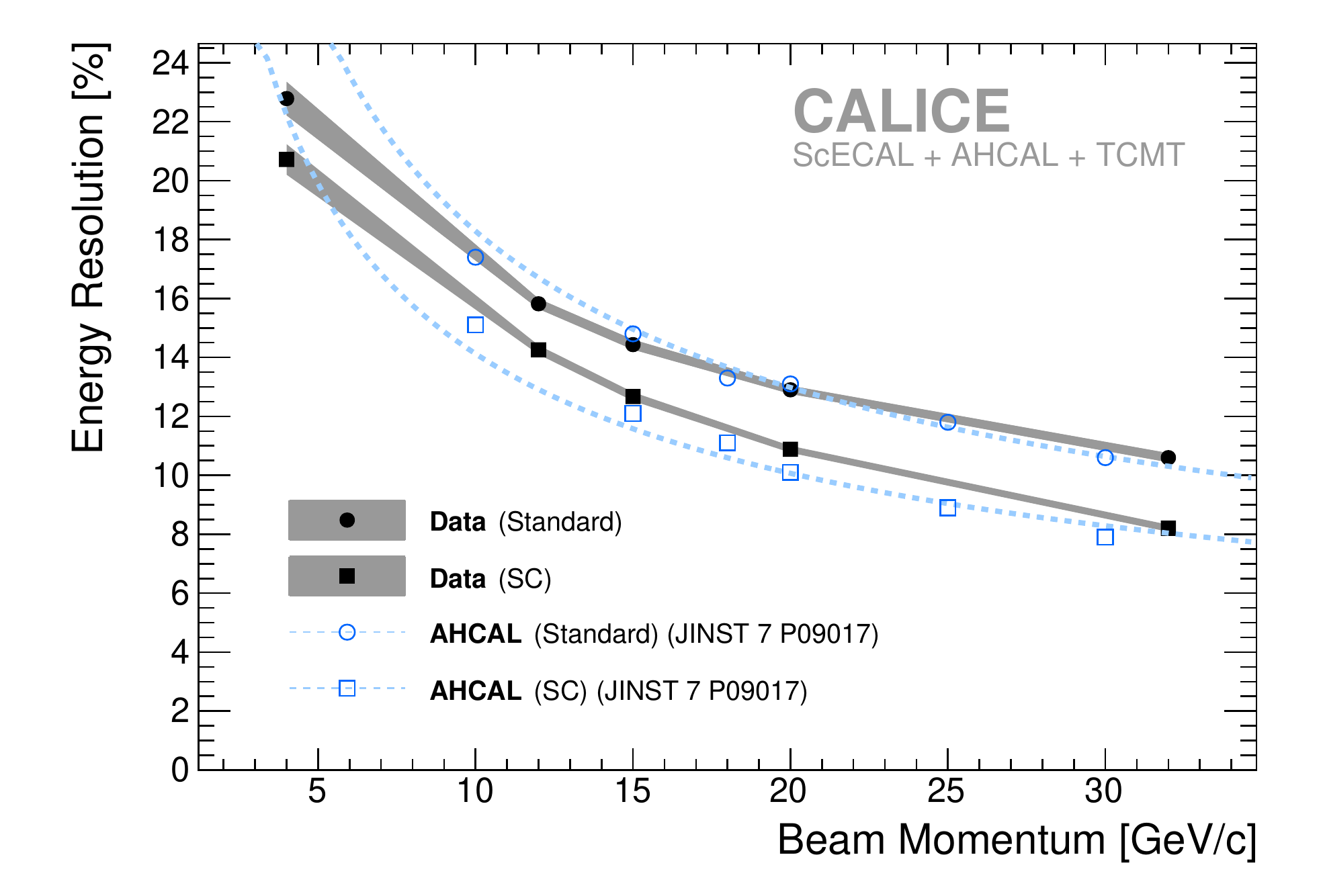}
\caption{Single pion energy resolutions with standard and software compensation reconstruction from the combined ScECAL+AHCAL+TCMT system compared to resolutions obtained from AHCAL+TCMT in \cite{SCPaper}.}
\label{fig:pion_resolution_vs_ahcal}
\end{center}
\end{figure}
\subsection{Application of software compensation weights from simulation to data}\label{sec:sc_data_mc}
The influence of the deviations observed in the software compensation weights between data and simulations was estimated by applying the software compensation weights obtained from simulated samples to the reconstruction of data events.  \autoref{fig:pion_data_mc_weights} shows the energy resolution and linearity of data samples reconstructed using weights optimised from both data and simulation. For this comparison the simulation weights optimised from QGSP\_BERT\_HP are used, as the difference between weights of different simulation physics lists is small. Furthermore only the software compensation specific weights $\alpha_i, \beta_i, \gamma$ are used as optimised from the simulation, while the standard reconstruction weights $w_{\text{ECAL}}, w_{\text{HCAL}}$ are used from data to set the correct energy reconstruction scale (see \autoref{sec:energyrecoclassic}).

Applying the software compensation weights obtained from simulations to data events actually improves the energy resolution slightly by a relative \SIrange{1}{3}{\percent}. However the achieved linearity is deteriorated, with additional deviations of magnitudes similar as seen in the resolution of \SIrange{1}{4}{\percent}. The \SI{4}{\GeV/c} point shows the biggest deviation when applying the simulation weights to data, in line with the previous observation that the simulated \SI{4}{\GeV/c} response profits most from the software compensation reconstruction.

Although there are significant differences between data and simulation in the first two and the last ScECAL hit energy bin weights, applying weights optimised from simulation onto data events data events yields similar performance to the use of weights optimised from data.
\begin{figure}[htbp]
\begin{center}
	{\includegraphics[width=0.5\textwidth]{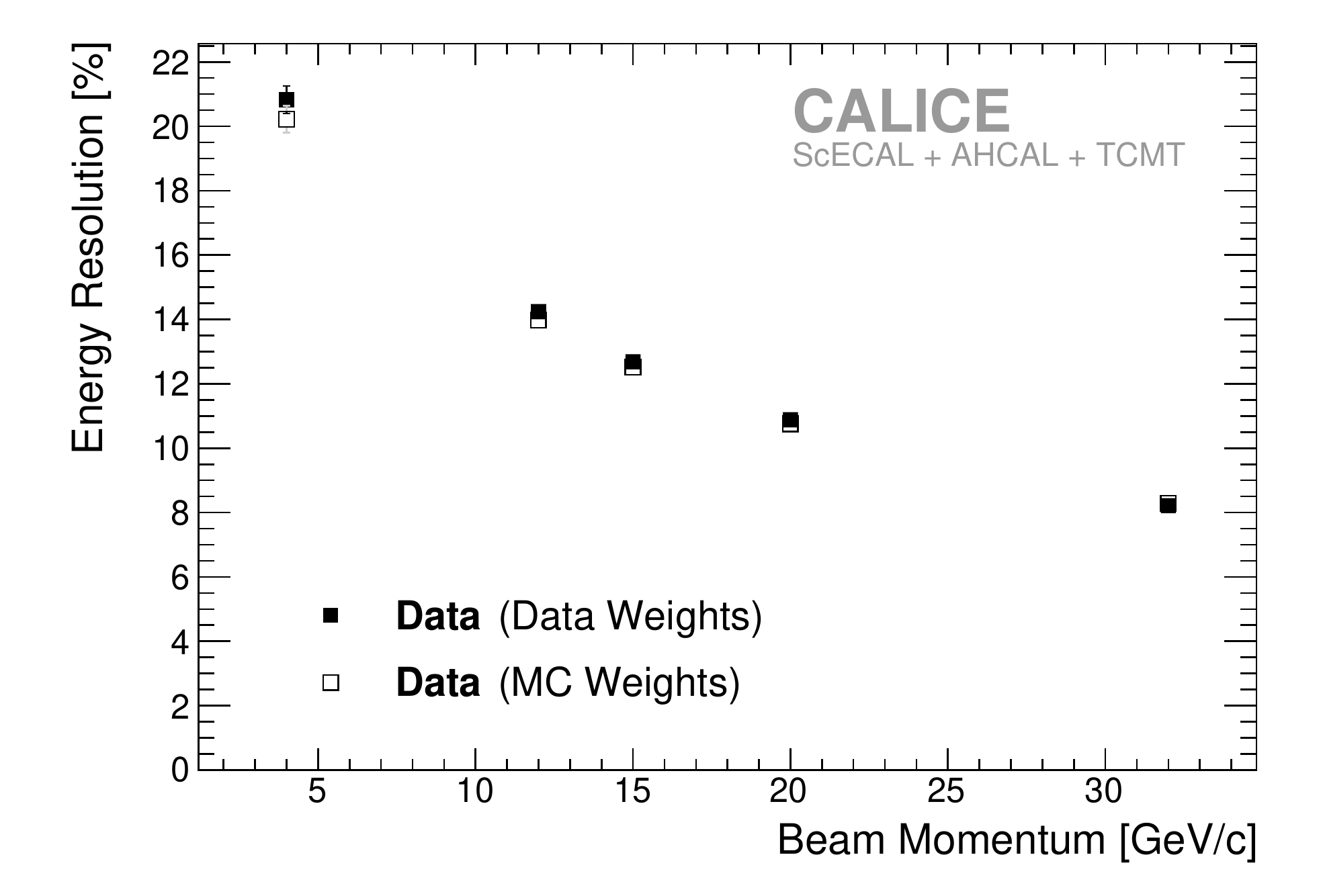}}\hfill
	{\includegraphics[width=0.5\textwidth]{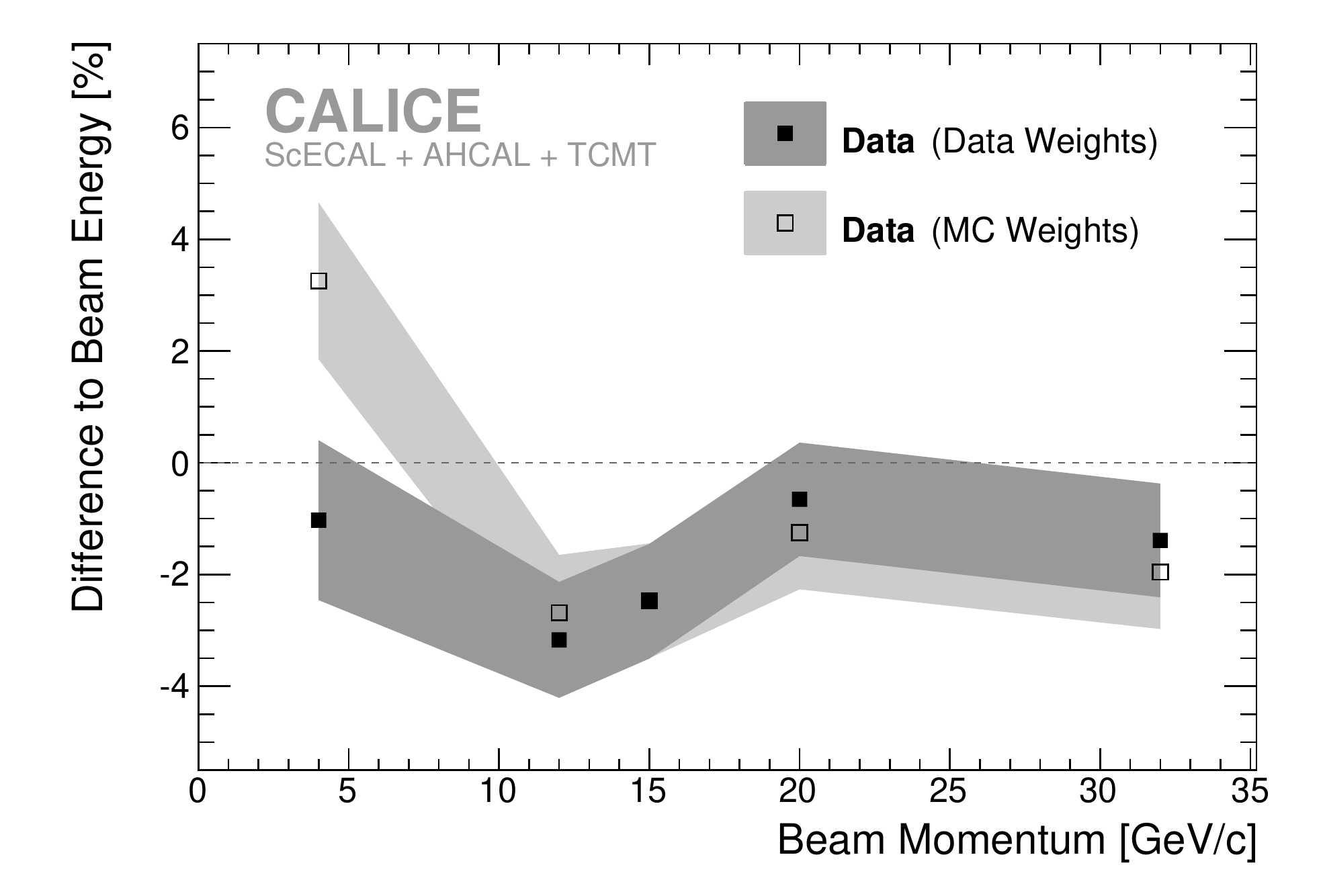}}
\end{center}
	\caption{Reconstructed energy resolution (left) and linearity (right) for the data sample when using software compensation weights derived from data and simulation (QGSP\_BERT\_HP in Geant\,4 10.2p1) for energy reconstruction.}
	\label{fig:pion_data_mc_weights}
\end{figure}
\section{Summary}
This paper presents results obtained with pion beams in the momentum range \SIrange{4}{32}{\GeV/c} at the FNAL testbeam facility and the combined CALICE scintillator-SiPM calorimeter system consisting of ScECAL, AHCAL and TCMT.
 The results are compared to data simulated with the physics lists QGSP\_BERT\_HP and FTFB\_BERT\_HP in Geant 4 version 10.1p2.

The longitudinal pion shower profiles show reasonable shape agreement between data and simulation. The longitudinal shower profile as a function of distance to the reconstructed shower start in the ScECAL shows up to \SI{20}{\percent} differences between simulation models, especially for high beam momenta.

Two separate schemes are used to reconstruct the primary particle energy. The linear standard energy reconstruction uses one constant weight per subdetector. These weights are in good agreement for data and simulated events apart from a general constant overestimate of energy deposits in the simulations. The deviation from linearity of the standard energy reconstruction is \SI{<5}{\percent} in simulation and \SI{<3}{\percent} in data. The energy resolution of the standard reconstruction is well reproduced by both simulation models. The second scheme, based on local software compensation, parametrises individual energy reconstruction weights for each bin in hit energy as a function of the standard reconstruction particle energy estimate. In the AHCAL, these weights are in good agreement between data and simulations. Significant discrepancies between data and simulation are observed in some ScECAL bins. The discrepancies may be partially explained by the observed deviations in the ScECAL hit energy spectra. The linearity deviation using software compensation reconstruction improves to \SI{<3}{\percent} for data and simulation. In data, the energy resolution improves by \SIrange{10}{20}{\percent} with the software compensation reconstruction. The improvements in energy resolution from the software compensation reconstruction are confirmed but overestimated in simulation by \SIrange{5}{10}{\percent} for all beam energies and simulation models.

The energy resolutions from data samples measured with the standard reconstruction and software compensation reconstruction are in good agreement with resolutions obtained from a previous analysis in which only showers starting in the AHCAL are considered. The good standalone single pion energy resolution of the AHCAL is thus maintained by adding the ScECAL in front.

\appendix
\begin{table}[htbp]
\ra{1.3}
\centering
\caption{Stepwise selection efficiencies of the single pion selection for data and different simulation physics lists and particle types. The event quality efficiency $\epsilon_\text{evQ}$ is normalised to the number of remaining events after the raw cut. Efficiencies marked with $^{\dagger}$ are normalised to the number of remaining events after the event quality cut to enable comparison of data and simulation results and are given as cumulative fractions of events passing the selections up to the respective step. The raw preselection efficiency of the 4\,GeV/c simulations is low, as the simulated beam profile is wider than the ScECAL geometry.}
\label{table:pion_selection_efficiency}
\begin{tabular}{p{1.8cm}llSSSSSS}
	\toprule
	Momentum & \multicolumn{2}{l}{Type}	& $\frac{n_\text{raw}}{n_\text{evts}}$ & $\epsilon_\text{evQ}$ & ${\epsilon_{\uppi\text{Sel}}}^{\dagger}$ & ${\epsilon_\text{mult}}^{\dagger}$  & ${\epsilon_\text{cont}}^{\dagger}$ & $\frac{n_\text{sel}}{n_\text{raw}}$\\
	\multicolumn{3}{c}{} 			& {[\%]}	& [\%]	& [\%]	& [\%]	& [\%]	& [\%]\\
	\midrule
	\multirow{4}{*}{\begin{minipage}{1.8cm} 4\,GeV/c\end{minipage} }	& \piminus	&\footnotesize{Data} 				& 87.1	& 66.9	& 66.4	& 52.8	& 38.5	& 25.8 \\
												& \piminus	&\footnotesize{FTFP\_BERT\_HP} 			& 54.0	& 96.9	& 74.0	& 66.7	& 47.5	& 46.1 \\
												& \piminus	&\footnotesize{QGSP\_BERT\_HP} 			& 54.0	& 96.9	& 74.1	& 66.6	& 47.4	& 45.9 \\
												& \eminus	&\footnotesize{QGSP\_BERT} 			& 51.9	& 99.9	& 2.4	& 0.3	& 0.2	& 0.2 \\
	\midrule
	\multirow{4}{*}{\begin{minipage}{1.8cm} 12\,GeV/c\end{minipage} }	& \piminus	&\footnotesize{Data} 				& 91.1	& 70.2	& 80.3	& 61.3	& 43.8	& 30.8 \\
												& \piminus	&\footnotesize{FTFP\_BERT\_HP} 			& 97.7	& 98.0	& 82.4	& 70.7	& 50.5	& 49.5 \\
												& \piminus	&\footnotesize{QGSP\_BERT\_HP} 			& 97.8	& 98.0	& 82.4	& 70.8	& 50.4	& 49.4 \\
												& \eminus	&\footnotesize{QGSP\_BERT} 			& 96.7	& 100.0	& 1.7	& {\textless 0.1} & {\textless 0.1}	& {\textless 0.1}\\
													\midrule
	\multirow{4}{*}{\begin{minipage}{1.8cm} 15\,GeV/c\end{minipage} }	& \piminus	&\footnotesize{Data} 				& 90.3	& 70.8	& 81.2	& 60.9	& 43.4	& 30.7 \\
												& \piminus	&\footnotesize{FTFP\_BERT\_HP} 			& 94.2	& 98.2	& 82.8	& 70.2	& 49.7	& 48.8 \\
												& \piminus	&\footnotesize{QGSP\_BERT\_HP} 			& 94.2	& 98.2	& 83.5	& 70.8	& 50.1	& 49.2 \\
												& \eminus	&\footnotesize{QGSP\_BERT} 			& 93.0	& 100.0	& 1.2	& {\textless0.1}	& {\textless0.1}	& {\textless0.1} \\
	\midrule
	\multirow{4}{*}{\begin{minipage}{1.8cm} 20\,GeV/c\end{minipage} }	& \piminus	&\footnotesize{Data} 				& 91.1	& 73.0	& 80.4	& 57.8	& 41.0	& 29.9 \\
												& \piminus	&\footnotesize{FTFP\_BERT\_HP} 			& 98.6	& 98.6	& 83.0	& 68.6	& 48.2	& 47.5 \\
												& \piminus	&\footnotesize{QGSP\_BERT\_HP} 			& 98.6	& 98.5	& 84.8	& 69.6	& 48.7	& 48.0 \\
												& \eminus	&\footnotesize{QGSP\_BERT} 			& 98.3	& 100.0	& 0.8	& {\textless0.1}	& {\textless0.1}	& {\textless0.1} \\
	\midrule
	\multirow{4}{*}{\begin{minipage}{1.8cm} 32\,GeV/c\end{minipage} }	& \piminus	&\footnotesize{Data} 				& 90.7	& 72.9	& 80.8	& 55.2	& 38.5	& 28.0 \\
												& \piminus	&\footnotesize{FTFP\_BERT\_HP} 			& 99.5	& 99.0	& 83.1	& 64.7	& 44.8	& 44.3 \\
												& \piminus	&\footnotesize{QGSP\_BERT\_HP} 			& 99.5	& 98.9	& 85.4	& 66.3	& 45.5	& 45.0 \\
												& \eminus	&\footnotesize{QGSP\_BERT} 			& 99.4	& 100.0	& 0.4	& {\textless0.1}	& {\textless0.1}	& {\textless0.1} \\

 	\bottomrule
\end{tabular}
\end{table}

\begin{table}[htbp]
\ra{1.3}
\centering
\caption{Biases on the response $\delta_\mu = 1 - \frac{\mu_\text{FHI}} {\mu_\text{sel}}$ and resolution $\delta_{\nicefrac{\sigma}{\mu}} = 1 - \frac{\nicefrac{\sigma_\text{FHI}}{\mu_\text{FHI}}}{\nicefrac{\sigma_\text{sel}}{\mu_\text{sel}}}$ for different physics lists. $\mu$ labels the extracted mean reconstructed response, $\sigma$ is the extracted width of the reconstructed energy distribution. The minimum bias FHI sample is selected for scintillator trigger and FHI layer cuts only, the sel sample contains the full pion selection. Statistical errors on the fit results are negligibly small.}
\label{table:pion_selection_bias}
\begin{tabular}{p{1.8cm}clSSSSSS}
	\toprule
	Momentum & \multicolumn{2}{l}{Type} & $\mu_\text{FHI}$ & $\nicefrac{\sigma_\text{FHI}}{\mu_\text{FHI}}$ & $\mu_\text{sel}$ & $\nicefrac{\sigma_\text{sel}}{\mu_\text{sel}}$ & $\delta_\mu$ & $\delta_{\nicefrac{\sigma}{\mu}}$\\
	\multicolumn{3}{c}{ } 			& [\si{\GeV}]	& [\%]	& [\si{\GeV}]	& [\%]	& [\%]	& [\%]	\\
	\midrule   
	\multirow{2}{*}{\begin{minipage}{1.8cm} \flushright 4\,GeV/c\end{minipage} }	& \piminus	&\footnotesize{FTFP\_BERT\_HP} 			& 3.82	& 22.92	& 3.83	& 22.86 & 0.3 & -0.3\\
												& \piminus	&\footnotesize{QGSP\_BERT\_HP} 			& 3.80	& 22.95	& 3.81	& 22.87 & 0.3 & -0.3 \\
	\midrule
	\multirow{2}{*}{\begin{minipage}{1.8cm} \flushright  12\,GeV/c\end{minipage} }	& \piminus	&\footnotesize{FTFP\_BERT\_HP} 			& 11.62	& 15.49	& 11.67	& 15.37 & 0.4 & -0.8\\
												& \piminus	&\footnotesize{QGSP\_BERT\_HP} 			& 11.54	& 15.54	& 11.59	& 15.45 & 0.4 & -0.6\\
	\midrule
	\multirow{2}{*}{\begin{minipage}{1.8cm}  \flushright 15\,GeV/c\end{minipage} }	& \piminus	&\footnotesize{FTFP\_BERT\_HP} 			& 14.64	& 14.43	& 14.70	& 14.30 & 0.4 & -0.9\\
												& \piminus	&\footnotesize{QGSP\_BERT\_HP} 			& 14.66	& 14.30	& 14.71	& 14.20 & 0.3 & -0.7\\
	\midrule
	\multirow{2}{*}{\begin{minipage}{1.8cm}  \flushright 20\,GeV/c\end{minipage} }	& \piminus	&\footnotesize{FTFP\_BERT\_HP} 			& 19.69	& 13.25	& 19.78	& 13.19 & 0.5 & -0.5\\
												& \piminus	&\footnotesize{QGSP\_BERT\_HP} 			& 19.78	& 12.74	& 19.86	& 12.66 & 0.4 & -0.6\\
	\midrule
	\multirow{2}{*}{\begin{minipage}{1.8cm}  \flushright 32\,GeV/c\end{minipage} }	& \piminus	&\footnotesize{FTFP\_BERT\_HP} 			& 31.97	& 11.59	& 32.15	& 11.58 & 0.6 & -0.1\\
												& \piminus	&\footnotesize{QGSP\_BERT\_HP} 			& 31.92	& 10.96	& 32.08	& 10.82 & 0.5 & -1.3\\
	\bottomrule
\end{tabular}
\end{table}

\begin{table}[htbp]
\ra{1.3}
\centering
\caption{Individual contributions to the relative systematic uncertainties on the energy resolutions obtained in this paper from the event selection (Evt. sel.), remaining electron contaminations in data after the selection (\eminus~cont.) and the modeling uncertainty in the simulation (Simu.), remaining impurities in data after the event selection (Impur.), as well as the total systematic uncertainty calculated from their addition in quadrature. Values marked with ${}^\dagger$ are uncertainties applied to data which are extracted from simulations. The value marked with ${}^\times$ is estimated from the uncertainties of different data samples, as the method to obtain the uncertainty is not applicable to that data point. Empty fields indicate systematic uncertainties that are not applicable to the specific sample, which are thus not included in the uncertainty calculation.}
\label{table:systematics}
\begin{tabular}{p{1.8cm}llSSSSS}
	\toprule
	Momentum & \multicolumn{2}{l}{Type}	& {Evt. sel.}  & {\eminus~cont.} & {Impur.} & {Simu.}  & {Total}\\
	\multicolumn{3}{c}{} 			& {[\%]}	& [\%]	& [\%]	& [\%]	& [\%]\\
	\midrule
	\multirow{5}{*}{\begin{minipage}{1.8cm} 4\,GeV/c\end{minipage} }	& \piminus	&\footnotesize{Data} 		& 0.3${}^\dagger$	& 1.7	& 1.4${}^\times$ 	& {-}	& 2.2 \\
												& \piminus	&\footnotesize{FTFP\_BERT\_HP (Standard)} 				& 0.3	& {-}		& {-}		& 3.5	& 3.5 \\
												& \piminus	&\footnotesize{FTFP\_BERT\_HP (SC)} 				& 0.3	& {-}		& {-}		& 2.9	& 2.9 \\
												& \piminus	&\footnotesize{QGSP\_BERT\_HP (Standard)} 				& 0.3	& {-}		& {-}		& 3.4	& 3.4 \\
												& \piminus	&\footnotesize{QGSP\_BERT\_HP (SC)} 				& 0.3	& {-}		& {-}		& 2.9	& 2.9 \\
	\midrule
	\multirow{5}{*}{\begin{minipage}{1.8cm} 12\,GeV/c\end{minipage} }	& \piminus	&\footnotesize{Data}	& 0.7${}^\dagger$	& 0.1	& 0.9	&  {-}	& 1.3 \\
												& \piminus	&\footnotesize{FTFP\_BERT\_HP (Standard)} 				& 0.8	&  {-}	&  {-}	& 1.8	& 2.0 \\
												& \piminus	&\footnotesize{FTFP\_BERT\_HP (SC)} 				& 0.8	&  {-}	&  {-}	& 2.4	& 2.5 \\
												& \piminus	&\footnotesize{QGSP\_BERT\_HP (Standard)} 				& 0.6	&  {-}	&  {-}	& 1.9	& 2.0 \\
												& \piminus	&\footnotesize{QGSP\_BERT\_HP (SC)} 				& 0.6	&  {-}	&  {-}	& 1.9	& 2.0 \\
													\midrule
	\multirow{5}{*}{\begin{minipage}{1.8cm} 15\,GeV/c\end{minipage} }	& \piminus	&\footnotesize{Data} 	& 0.8${}^\dagger$	& 0.0	& 0.9	&  {-}	& 1.3 \\
												& \piminus	&\footnotesize{FTFP\_BERT\_HP (Standard)} 				& 0.9	&  {-}	&  {-}	& 2.6	& 2.8 \\
												& \piminus	&\footnotesize{FTFP\_BERT\_HP (SC)} 				& 0.9	&  {-}	&  {-}	& 2.8	& 2.9 \\
												& \piminus	&\footnotesize{QGSP\_BERT\_HP (Standard)} 				& 0.7	&  {-}	&  {-}	& 1.4	& 1.6 \\
												& \piminus	&\footnotesize{QGSP\_BERT\_HP (SC)} 				& 0.7	&  {-}	&  {-}	& 1.0	& 1.2 \\
	\midrule
	\multirow{5}{*}{\begin{minipage}{1.8cm} 20\,GeV/c\end{minipage} }	& \piminus	&\footnotesize{Data} 	& 0.6${}^\dagger$	& 0.0	& 0.5	& {-}	& 0.8 \\
												& \piminus	&\footnotesize{FTFP\_BERT\_HP (Standard)} 				& 0.5	&  {-}	&  {-}	& 2.1	& 2.2 \\
												& \piminus	&\footnotesize{FTFP\_BERT\_HP (SC)} 				& 0.5	&  {-}	&  {-}	& 2.8	& 2.8 \\
												& \piminus	&\footnotesize{QGSP\_BERT\_HP (Standard)} 				& 0.6	&  {-}	&  {-}	& 1.0	& 1.2 \\
												& \piminus	&\footnotesize{QGSP\_BERT\_HP (SC)} 				& 0.6	&  {-}	&  {-}	& 0.8	& 1.0 \\
	\midrule
	\multirow{5}{*}{\begin{minipage}{1.8cm} 32\,GeV/c\end{minipage} }	& \piminus	&\footnotesize{Data} 	& 0.7${}^\dagger$ &  0.0	& 1.4	& {-} & 1.5 \\
												& \piminus	&\footnotesize{FTFP\_BERT\_HP (Standard)} 				& 0.1	&  {-}	&  {-}	& 2.4	& 2.4 \\
												& \piminus	&\footnotesize{FTFP\_BERT\_HP (SC)} 				& 0.1	&  {-}	&  {-}	& 0.1	& 0.1 \\
												& \piminus	&\footnotesize{QGSP\_BERT\_HP (Standard)} 				& 1.3	&  {-}	&  {-}	& 2.0	& 2.4 \\
												& \piminus	&\footnotesize{QGSP\_BERT\_HP (SC)} 				& 1.3	&  {-}	&  {-}	& 0.2	& 1.3 \\

 	\bottomrule
\end{tabular}
\end{table}

\begin{table}[htbp]
\ra{1.3}
\centering
\caption{Energy resolutions extracted from pion samples in data and simulations, using the standard energy reconstruction (Std.) and the software compensation reconstruction (SC) as plotted in \autoref{fig:pion_resolution_plots}. All errors include the full systematic uncertainties as discussed in \autoref{sec:systematics} and given in \autoref{table:systematics}.}
\label{table:pion_reso_values}
\begin{tabular}{p{1.8cm}llSSSSSS}
	\toprule
	Momentum & \multicolumn{2}{l}{Type}	& \multicolumn{2}{c}{$\left(\nicefrac{\sigma}{\mu}\right)_\text{Std.}$} & \multicolumn{2}{c}{$\left(\nicefrac{\sigma}{\mu}\right)_\text{SC}$}& \multicolumn{2}{c}{$\frac{\left(\nicefrac{\sigma}{\mu}\right)_\text{SC}}{\left(\nicefrac{\sigma}{\mu}\right)_\text{Std.}}$}\\
	\multicolumn{3}{c}{} 			& [\%]	& $\Delta$[\%]	& [\%]	& $\Delta$[\%]	& [\%]	& $\Delta$[\%]\\
	\midrule
	\multirow{3}{*}{\begin{minipage}{1.8cm} 4\,GeV/c\end{minipage} }	& \piminus	&\footnotesize{Data} 				 & 22.78 & 0.57 & 20.72 & 0.52 & 0.91 & 0.03 \\
												& \piminus	&\footnotesize{FTFP\_BERT\_HP} 			  & 22.86 & 0.82 & 19.57 & 0.57 & 0.86 & 0.04 \\
												& \piminus	&\footnotesize{QGSP\_BERT\_HP} 			& 22.85 & 0.80 & 19.51 & 0.59 & 0.85 & 0.04 \\
	\midrule
	\multirow{3}{*}{\begin{minipage}{1.8cm} 12\,GeV/c\end{minipage} }	& \piminus	&\footnotesize{Data} 				 & 15.82 & 0.17 & 14.26 & 0.16 & 0.90 & 0.01 \\
  	
												& \piminus	&\footnotesize{FTFP\_BERT\_HP} 			  & 15.38 & 0.28 & 12.79 & 0.32 & 0.83 & 0.03 \\
												& \piminus	&\footnotesize{QGSP\_BERT\_HP} 			 & 15.45 & 0.31 & 12.69 & 0.26 & 0.82 & 0.02 \\
	\midrule
	\multirow{3}{*}{\begin{minipage}{1.8cm} 15\,GeV/c\end{minipage} }	& \piminus	&\footnotesize{Data} 			
	& 14.43 & 0.17 & 12.67 & 0.15 & 0.88 & 0.01 \\
												& \piminus	&\footnotesize{FTFP\_BERT\_HP} 			 & 14.32 & 0.39 & 11.45 & 0.35 & 0.80 & 0.03 \\
												& \piminus	&\footnotesize{QGSP\_BERT\_HP} 			 & 14.20 & 0.24 & 11.38 & 0.18 & 0.80 & 0.02 \\
	\midrule
	\multirow{3}{*}{\begin{minipage}{1.8cm} 20\,GeV/c\end{minipage} }	& \piminus	&\footnotesize{Data} 			
	& 12.90 & 0.14 & 10.88 & 0.12 & 0.84 & 0.01 \\
												& \piminus	&\footnotesize{FTFP\_BERT\_HP} 			 & 13.19 & 0.29 & 9.99 & 0.30 & 0.76 & 0.03 \\
												& \piminus	&\footnotesize{QGSP\_BERT\_HP} 			 & 12.67 & 0.16 & 9.76 & 0.12 & 0.77 & 0.01 \\
	\midrule
	\multirow{3}{*}{\begin{minipage}{1.8cm} 32\,GeV/c\end{minipage} }	& \piminus	&\footnotesize{Data} 			
	& 10.62 & 0.16 & 8.24 & 0.12 & 0.78 & 0.02 \\
												& \piminus	&\footnotesize{FTFP\_BERT\_HP} 			  & 11.59 & 0.30 & 7.61 & 0.11 & 0.66 & 0.02 \\
												& \piminus	&\footnotesize{QGSP\_BERT\_HP} 			 & 10.84 & 0.23 & 7.56 & 0.14 & 0.70 & 0.02 \\
 	\bottomrule
\end{tabular}
\end{table}

\clearpage


\begin{figure}[htbp]
	{\includegraphics[width=0.50\textwidth]{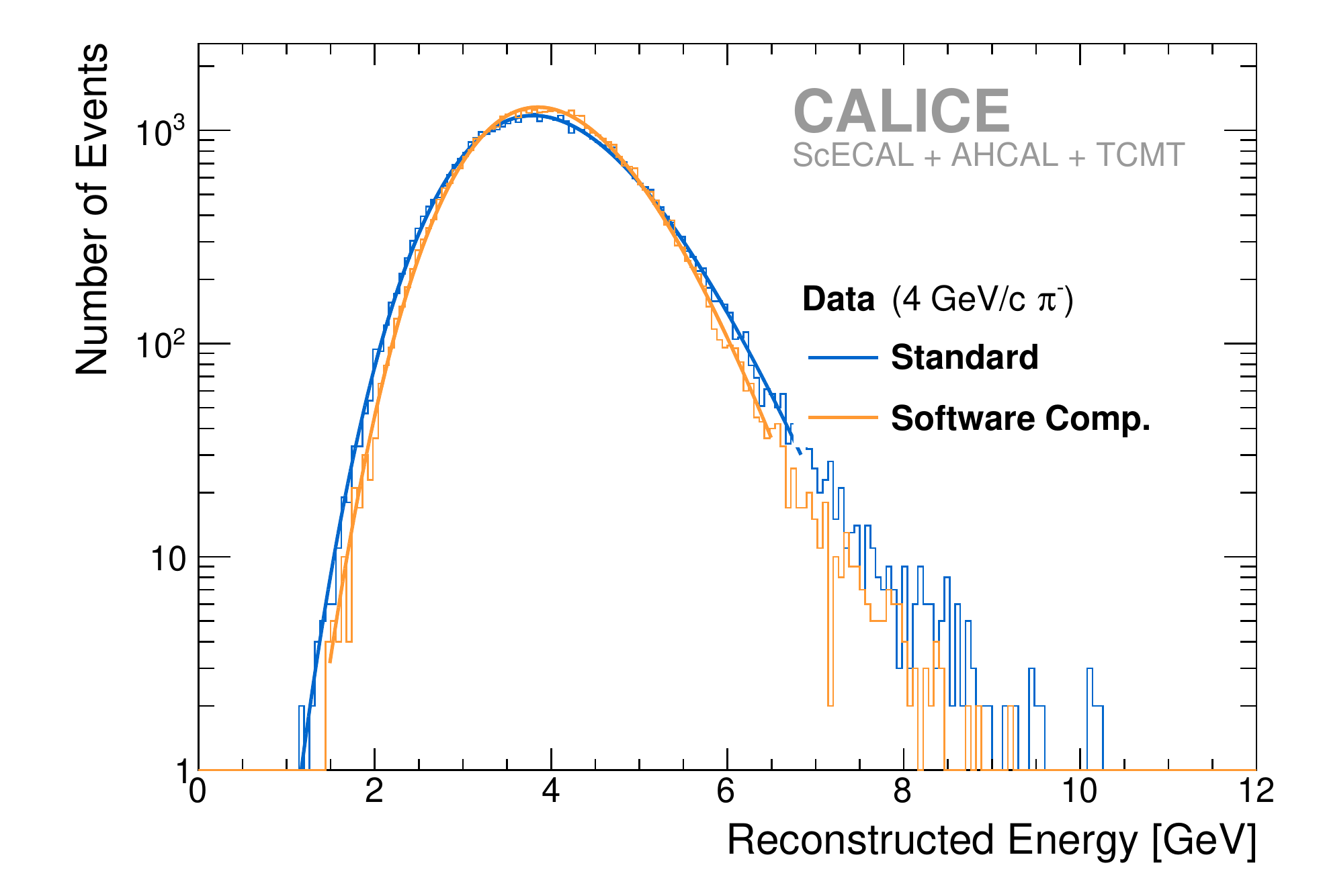}}\hfill
	{\includegraphics[width=0.50\textwidth]{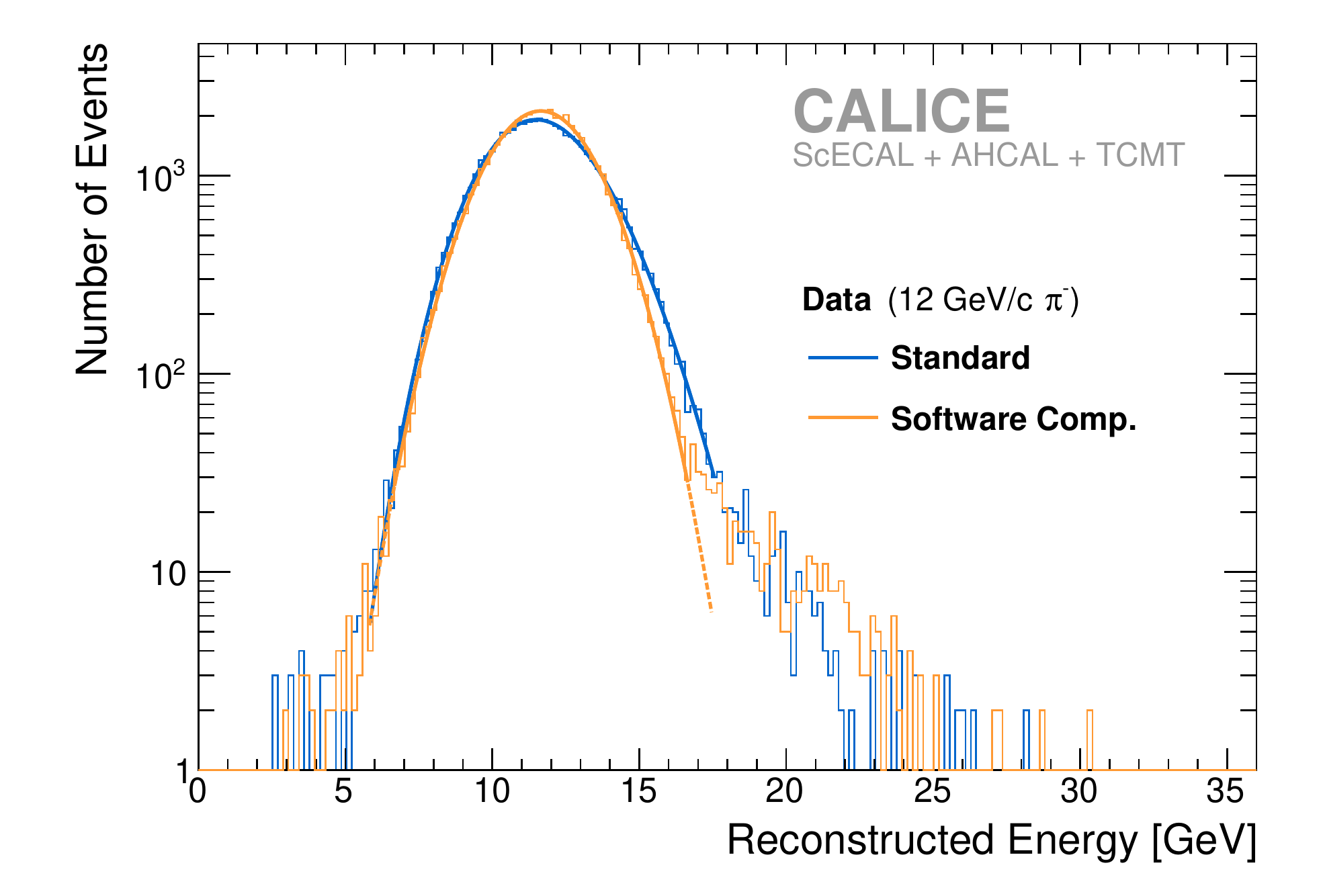}}\hfill
	{\includegraphics[width=0.50\textwidth]{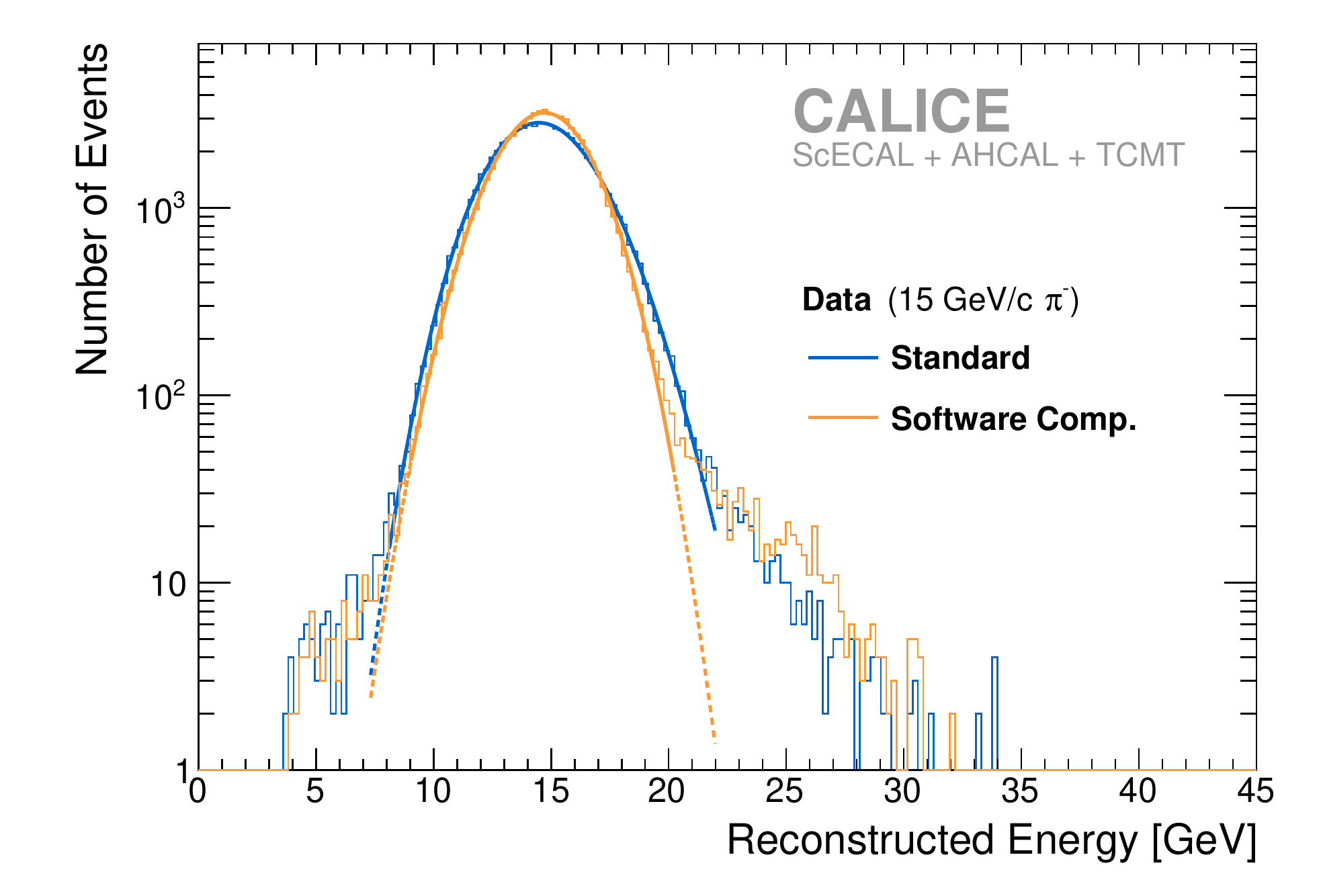}}\hfill
	{\includegraphics[width=0.50\textwidth]{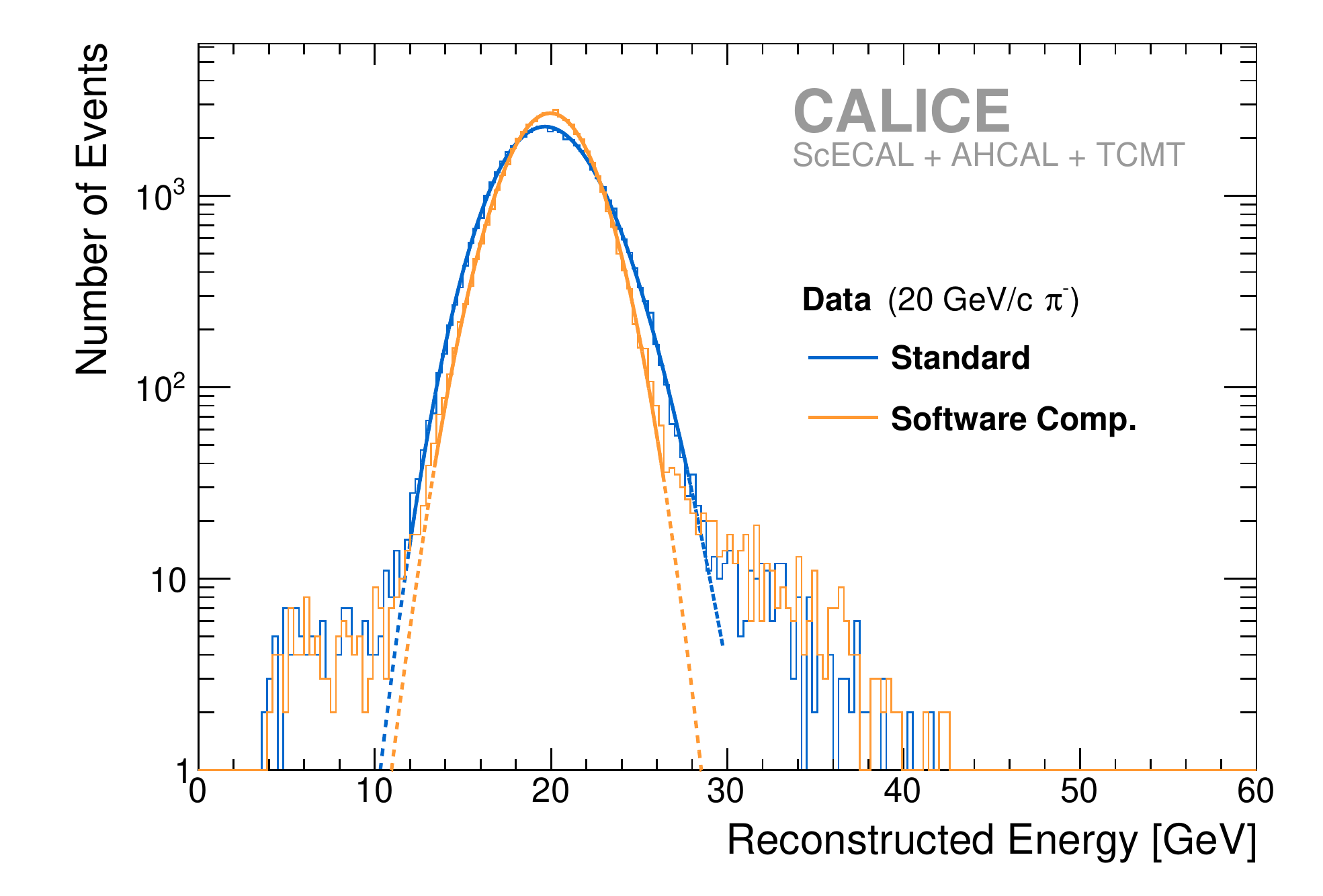}}\hfill
	
	\caption[]{Reconstructed energy spectra from data samples using the standard and software compensation reconstruction.}
	\label{fig:additionalspectra}
\end{figure}

\clearpage

\acknowledgments
We gratefully acknowledge the Fermilab management for their support and hospitality, and their accelerator staff for the reliable and efficient beam operation. 

This work was supported by the FWO, Belgium; by the Natural Sciences and Engineering Research Council of Canada; by the Ministry of Education, Youth and Sports of the Czech Republic; by the European Union's Horizon 2020 Research and Innovation programme under Grant Agreement 654168; by the European Commission within Framework Programme 7 Capacities, Grant Agreement 262025; by the Alexander von Humboldt Stiftung (AvH), Germany; by the Bundesministerium f\"ur Bildung und Forschung (BMBF), Germany; by the Deutsche Forschungsgemeinschaft (DFG), Germany; by the Helmholtz-Gemeinschaft (HGF), Germany; by the I-CORE Program of the Planning and Budgeting Committee, Israel; by the Nella and Leon Benoziyo Center for High Energy Physics, Israel; by the Israeli Science Foundation, Israel; by the JSPS KAKENHI Grant-in-Aid for Scientific Research (B) No. 17340071 and specially promoted research No.  223000002, Japan; by the National Research Foundation of Korea; by the Korea-EU cooperation programme of National Research Foundation of Korea, Grant Agreement 2014K1A3A7A03075053; by the Russian Ministry of Education and Science contracts 3.2989.2017 and 14.W03.31.0026; by the Spanish Ministry of Economy and Competitiveness 	FPA2014-53938-C3-R2 and Grant MDM-2015-0509; by the Science and Technology Facilities Council, UK; by the US Department of Energy; by the National Science Foundation of the United States of America, and by the Nuclear Physics, Particle Physics, Astrophysics and Cosmology Initiative, a Laboratory Directed Research, USA.


\end{document}